\pdfoutput=1
\documentclass[12pt]{article}

\usepackage{graphicx,psfrag,epsf,color}
\usepackage{xcolor}
\usepackage{amsmath,amssymb,amsfonts}
\usepackage{array}
\usepackage{cite}
\usepackage{rotating}
\usepackage{slashed,mathtools}
\usepackage{xparse}
\usepackage{stmaryrd}
\usepackage{cancel}

\usepackage{centernot}

\newcounter{MBQ}

\newcounter{RSQ}

\newcommand{\be}{\begin{equation}}
\newcommand{\ee}{\end{equation}}
\newcommand{\bea}{\begin{eqnarray}}
\newcommand{\eea}{\end{eqnarray}}
\newcommand{\bi}{\begin{itemize}}
\newcommand{\ei}{\end{itemize}}
\newcommand{\ben}{\begin{enumerate}}
\newcommand{\een}{\end{enumerate}}
\newcommand{\bt}{\begin{tabular}}
\newcommand{\et}{\end{tabular}}

\newcommand{\nn}{\nonumber}
\newcommand{\klammer}[1]{ \left( #1 \right)  }

\newcommand{\nm}{n_-}
\newcommand{\np}{n_+}

\newcommand{\as}{\alpha_s}
\newcommand{\eps}{\epsilon}
\newcommand{\ord}{\mathcal{O}}

\newcommand{\PS}{d\mbox{PS}}

\begin{document}
\allowdisplaybreaks

\begin{titlepage}

\begin{flushright}
{\small
TUM-HEP-1398/22\\
IPPP/22/25\\
09 May 2022
}
\end{flushright}

\vskip0.1cm
\begin{center}
{\Large \bf Next-to-leading power endpoint factorization and\\[0.1cm] 
resummation for off-diagonal ``gluon'' thrust}
\end{center}

\vspace{0.2cm}
\begin{center}
{\sc M.~Beneke,$^{a,b}$ M.~Garny,$^a$ S.~Jaskiewicz,$^c$ 
J.~Strohm,$^{a,b}$ R.~Szafron,$^d$ L.~Vernazza,$^{e,f}$ 
J.~Wang$^g$} \\[6mm]
{\it $^a$ Physik Department T31,\\
James-Franck-Stra\ss e~1, 
Technische Universit\"at M\"unchen,\\
D--85748 Garching, Germany
\\[0.2cm]
\it $^b$ Excellence Cluster ORIGINS,\\
Technische Universit\"at M\"unchen,
D--85748 Garching, Germany
\\[0.2cm]
\it $^c$ Institute for Particle Physics Phenomenology, Durham University\\
Durham, DH1 3LE, United Kingdom
\\[0.2cm]
 ${}^d$Department of Physics, Brookhaven National Laboratory,\\ 
Upton, N.Y., 11973, U.S.A.\\[0.2cm]
 ${}^e$ INFN, Sezione di Torino, Via P. Giuria 1, 
I-10125 Torino, Italy\\[0.2cm]
 ${}^f$ Nikhef, Science Park 105, NL-1098 XG Amsterdam, 
The Netherlands\\[0.2cm]
$^g$ School of Physics, Shandong University,\\ 
Jinan, Shandong 250100, China
}
\end{center}

\vspace{0.0cm}
\begin{abstract}
\vskip0.2cm\noindent
The lack of convergence of the convolution integrals appearing in 
next-to-leading-power (NLP) factorization theorems prevents the 
applications of existing methods to resum power-suppressed large 
logarithmic corrections in collider physics. We consider thrust 
distribution in the two-jet region for the flavour-nonsinglet 
off-diagonal 
contribution, where a gluon-initiated jet recoils against 
a quark-antiquark pair, which is power-suppressed. With the help 
of operatorial endpoint factorization conditions, we obtain  
a factorization formula, where the individual terms are free from 
endpoint divergences in convolutions and can be expressed in 
terms of renormalized hard, soft and collinear functions in four 
dimensions. This allows us to perform the first resummation of the 
endpoint-divergent SCET$_{\rm I}$ observables at the leading 
logarithmic accuracy using 
exclusively renormalization-group methods. The presented approach 
relies on universal properties of the soft and 
collinear limits and may serve as a paradigm for the systematic NLP 
resummation for other $1\to 2$ and $2\to 1$ collider physics processes.
\end{abstract}
\end{titlepage}


\section{Introduction}
\label{sec:introduction}

Hadronic event shape variables in the two-jet region have played a key 
role for the development of resummation techniques in 
QCD \cite{Catani:1991kz,Catani:1992ua}.\footnote{See \cite{Dokshitzer:1978dr,Parisi:1979se,Collins:1981uk,Kodaira:1981np} for other early work discussing the break-down of the expansion in the 
strong coupling for back-to-back jets and semi-inclusive particle production.} Later, the utility of soft-collinear effective theory (SCET) in 
extending the resummation order beyond what was achieved with
diagrammatic methods has first been demonstrated for the thrust 
variable $T$~\cite{Becher:2008cf}. In the present work we develop the
factorization of thrust for the particular contribution to the 
two-jet region, when a gluon recoils (at leading order) against a
quark-antiquark pair (``gluon thrust''). 
The motivation for this derives from 
the fact that this process is of next-to-leading power in $1-T\ll 1$ in 
the two-jet region, and thus not covered by the existing factorization
theorem. It represents the hadronic $e^+ e^-$ annihilation analogue of the off-diagonal $qg$ parton channels in deep-inelastic scattering (DIS) at
large Bjorken-$x$, or in Drell-Yan (DY) production near threshold. These and the power corrections to the corresponding diagonal channels  have 
recently received much attention \cite{Bonocore:2016awd,Moult:2018jjd,Beneke:2018gvs,Moult:2019mog,Bahjat-Abbas:2019fqa,Beneke:2019mua,Moult:2019uhz,Beneke:2019oqx,Moult:2019vou,Ajjath:2020ulr,Beneke:2020ibj,Ajjath:2020sjk,vanBeekveld:2021mxn}
in an effort to 
extend the classical leading-power (LP) factorization theorems for these observables to next-to-leading power (NLP) in the kinematic regime,
where large logarithms spoil the weak-coupling expansion.
We focus on gluon thrust here, since it presents the same
difficulties as the forementioned off-diagonal processes, but does
not require the factorization of parton distribution functions.

\begin{figure}
\begin{center}
\includegraphics[width=0.22\textwidth]{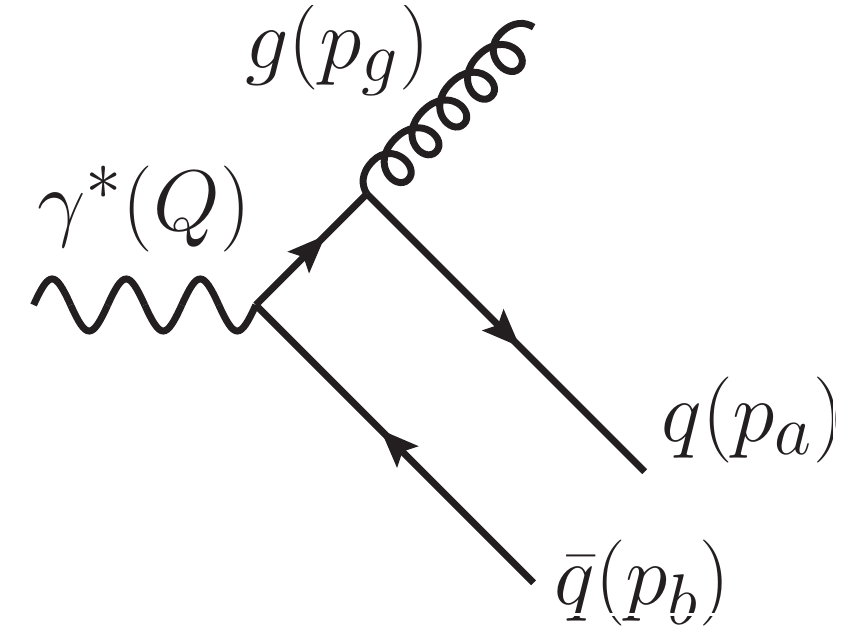}\qquad\qquad
\includegraphics[width=0.22\textwidth]{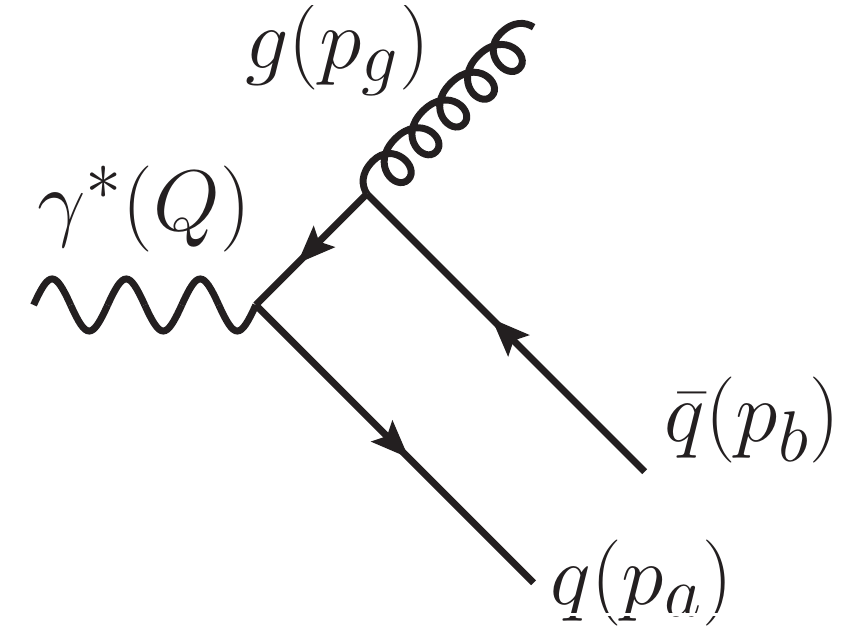}
\caption{\label{fig:QCD} Leading $\mathcal{O}(\alpha_s)$ QCD 
diagrams contributing to gluon thrust.}
\end{center}
\end{figure}

The thrust of a hadronic event in $e^+ e^-$ annihilation with 
center-of-mass energy $Q$ is defined as 
\begin{align}
T=\mbox{max}_{\vec n}\,\frac{\sum_{i}
\left|\vec{p_{i}}\cdot\vec{n}\right|}
{\sum_{i}\left|\vec{p_{i}}\right|},
\end{align}
where $i$ runs over all hadrons (partons) in the final state. 
The plane orthogonal to the thrust axis $\vec{n}$ divides 
space into a left and right hemisphere. The total invariant mass 
(squared) of the hadrons in these hemispheres will be denoted by 
$M_L^2$ and $M_R^2$, respectively. As 
\begin{align}
\label{eq:tau}
\tau=1-T \to 0,
\end{align}
the particles cluster into a pair of back-to-back jets, 
and perturbation theory in the strong coupling $\alpha_s(Q)$ 
breaks down due to large logarithms $\ln\tau$ 
at every order. This has been extensively studied for 
the leading quark-antiquark two-jet process. All-order resummations 
are essential to provide reliable calculations of thrust 
and jet masses in the two-jet regions.
In this work we consider instead the phase-space region
\begin{equation}
e^+ e^-\to\gamma^*\to [g]_{c} + [q\bar q]_{\bar c}\,,
\end{equation}
where the direction of gluon jet is defined to be the 
``collinear direction'' and the direction of the recoiling 
$q\bar q$-jet the ``anti-collinear'' one. This process 
starts at order $\alpha_s$ only (see Fig.~\ref{fig:QCD}). 
More importantly, as $\tau\to 0$, it does not have the 
leading-power $[\ln\tau/\tau]_+$ behaviour from
soft-gluon emission. Instead the leading term is 
$\alpha_s\ln \tau$, and the entire process is of 
next-to-leading power in $\tau$,\footnote{For clarification, 
we note that we do not consider here non-perturbative 
power corrections in the strong 
interaction scale $\Lambda_{\rm QCD}$, which are  
expected to be  $\mathcal{O}(\Lambda_{\rm QCD}/(\tau Q))$, that is, 
we assume the strong hierarchy $\Lambda_{\rm QCD}/Q\ll \tau\ll1$.} 
similar to the off-diagonal DIS and 
DY processes related by crossing.

``Gluon thrust'' and its unconventional all-order logarithmic 
structure was first studied in \cite{Moult:2019uhz} and an 
expression for the leading double logarithms was written down, 
which was subsequently derived from $d$-dimensional consistency 
relations \cite{Beneke:2020ibj}. The result 
of \cite{Moult:2019uhz} exhibits an unconventional ``quark'' 
Sudakov form factor, related to an enhancement when 
either the quark or anti-quark in the 
$q\bar q$-jet becomes soft. In this case 
the octet colour charge of the gluon jet is balanced by 
anti-collinear particles in a colour-triplet or anti-triplet. 
The colour mismatch of the back-to-back energetic particles causes 
a new type of double logarithms proportional to the difference 
$C_A-C_F$ of the colour charges at the leading double-logarithmic 
order. Despite these insights, and sketches of the form of 
a factorization theorem within SCET 
\cite{Moult:2019uhz,Moult:2019mog,Beneke:2020ibj}, it is not yet known 
how to proceed beyond the double-logarithmic order. The 
issue is an endpoint-divergence in the convolution 
integrals that link the hard, (anti-) collinear and soft 
functions in the factorization theorem. In the present 
work we build on refactorization ideas developed for 
DIS \cite{Beneke:2020ibj}, and Higgs decay to two photons 
through bottom-quark loops \cite{Liu:2019oav,Liu:2020wbn} 
to derive a factorization theorem suitable for systematic 
resummation. To this 
end we combine standard SCET factorization with endpoint 
factorization.

The outline of this paper is as follows: In Sec.~\ref{sec:heuristic} 
we provide a heuristic discussion of the factorization formula, 
which motivates the proposed endpoint subtraction. In 
Sec.~\ref{sec:bare} we derive the ``bare'' factorization theorem 
using standard SCET methods. For endpoint factorization to 
work, the hard, collinear and soft functions must satisfy 
certain asymptotic conditions, which are derived in 
Sec.~\ref{sec:refactorization}, and subsequently employed to 
rearrange the factorization formula such that the convolution 
integrals are finite. The renormalization-group equations required 
for the various functions are given in Sec.~\ref{sec:consistency} and 
solved with leading-logarithmic (LL) accuracy. Finally, in 
Sec.~\ref{sec:resummation}, we obtain the resummed gluon thrust 
distribution and display the numerical effect of resummation. 

Throughout the paper, we employ dimensional regularization with 
$d=4-2\epsilon$ space-time dimensions, and the 
$\overline{\rm MS}$ subtraction 
scheme whenever renormalization is required.


\section{Heuristic discussion}
\label{sec:heuristic}

At leading order (LO) a virtual photon couples to 
a quark-antiquark pair producing two back-to-back jets 
in the center-of-mass frame. In this work, we consider 
instead the different two-jet situation, where one of the jets 
is initiated by a gluon. This process is not only suppressed by 
one power of $\alpha_s$ (see Fig.~\ref{fig:QCD}) relative to the LO 
process, but also by one power of $\tau$.

We define the hemisphere of the gluon jet to be in 
the ``collinear'' direction, the other jet is then in the 
``anti-collinear'' direction. There are two possibilities to 
generate the gluon jet at $\mathcal{O}(\alpha_s)$:
\begin{enumerate}
\item[I] Quark and anti-quark have both large  anti-collinear 
momentum and 
form a single jet, which recoils against the gluon.
\item[II] The quark or anti-quark is anti-collinear and balances
the gluon momentum, while the other of the two is soft.
\end{enumerate}
Both configurations are suppressed by a factor of $\tau$. 
Their further evolution is determined by the standard leading-power 
collinear splittings and soft emissions. The above separation 
introduces an ambiguity from the precise meaning  of ``soft'' 
and ``anti-collinear''. We can start with situation I, and 
decrease the large anti-collinear momentum component of the 
quark or anti-quark. 
At some point, it becomes soft (and anti-collinear) and 
should be counted as part of contribution II. In perturbative 
computations in powers of $\alpha_s$, this ambiguity is not a 
problem as long as the jet definition is infrared-safe. If, 
however,  one wants to perform resummation, the effective field 
theory approach requires the separation of soft and collinear 
modes to split the large logarithms into single-scale quantities. 
This conflict between physical and mathematical mode separation 
leads to a so-called endpoint divergence in the convolution 
integrals appearing in the factorization theorems. 

\begin{figure}
\begin{center}
\includegraphics[width=0.28\textwidth]{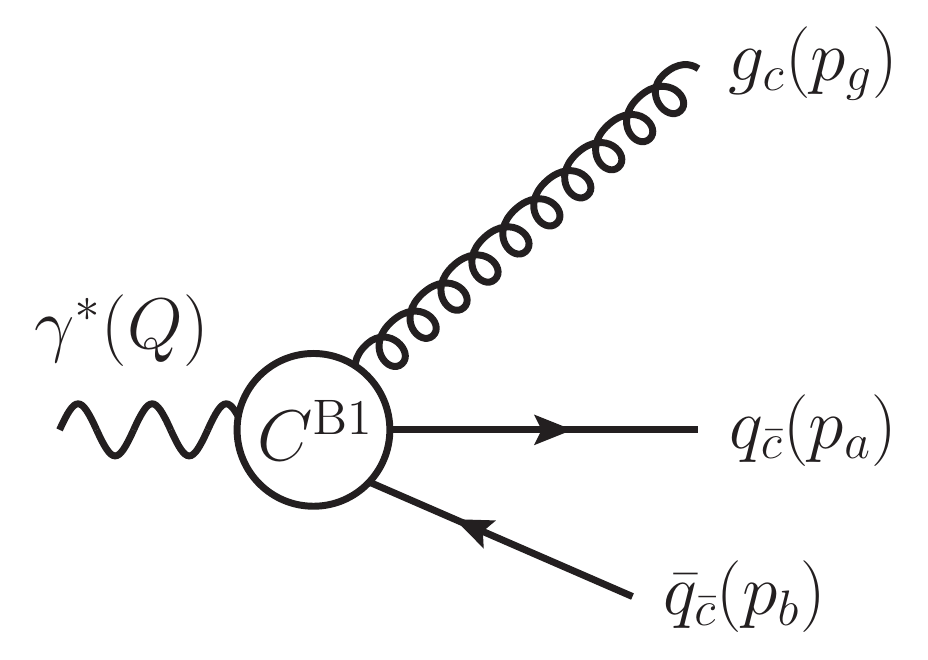}\hspace{0.15cm}
\includegraphics[width=0.28\textwidth]{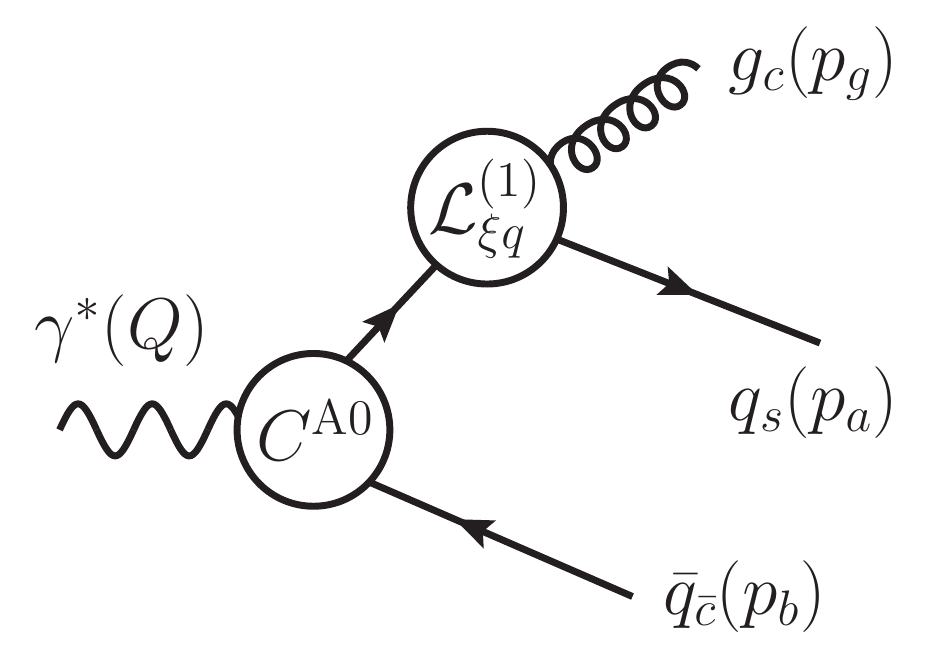}\hspace{0.15cm}
\includegraphics[width=0.28\textwidth]{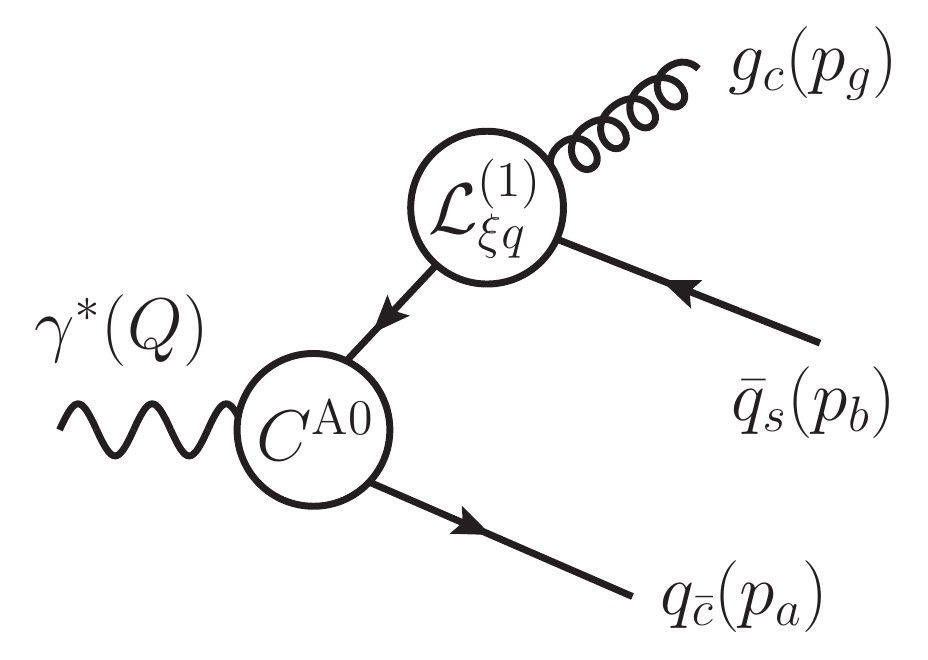}
\caption{\label{fig:SCET} SCET representation of the 
``off-diagonal'' gluon-thrust amplitude in the two-jet region. 
The diagrams show the NLP structure 
of the process with all LP interactions removed.}
\end{center}
\end{figure}

Let us analyze these two possibilities from the SCET point of view 
by looking first at the Feynman diagrams shown in 
Fig.~\ref{fig:QCD}. We begin with I (later called ``B-type'' 
contribution, following the SCET notation for the corresponding 
hard vertex). The intermediate propagator in the diagram on the 
left carries momentum $q_a=p_g + p_a$. Since $p_g$ is 
assumed to be collinear and $p_a$ is anti-collinear, $q_a$ is a 
hard momentum with virtuality $q_a^2\sim Q^2$. The $q_a$-internal 
line is integrated out, and the primary hard SCET vertex produces 
a $q\bar{q} g$ state directly, see the first diagram in 
Fig.~\ref{fig:SCET}. Since there are two partons ($q\bar q$) in the 
anti-collinear direction, only their total momentum is fixed and the 
amplitude depends on the fraction of anti-collinear momentum 
carried by each parton. When one of these fractions tends to zero, 
the parton becomes effectively soft and one moves to 
possibility II (called ``A-type'' later). The 
$q_a$-internal line (or $q_b$, if the anti-quark becomes soft) 
is no longer hard. Hence the primary vertex is the standard 
$\gamma^*\to q\bar {q}$ process. The energetic quark or anti-quark 
subsequently transfers its entire momentum to the gluon, while 
becoming soft. This process is represented by a power-suppressed 
SCET Lagrangian term $\mathcal{L}^{(1)}_{\xi q}$ describing 
soft (anti-) quark emission, see the second and third diagram in  
Fig.~\ref{fig:SCET}.  

The fact that in some intermediate region the same physical process 
can be represented by the limits of two different mathematical 
expressions is the key to the idea of endpoint factorization. 
To anticipate what will be discussed in technical terms further 
below, we state the factorization theorem for the two-hemisphere 
invariant mass distribution of gluon thrust in 
schematic form in Laplace space:
\begin{align}
\frac{1}{\sigma_{0}}\frac{\widetilde{d\sigma}}{ds_R ds_L}
&=
\int_{0}^{\Lambda} d\omega d\omega'\,
\left|C^{A0}\right |^{2}\times
\mathcal{J}^{(\bar{q})}_{\bar{c}}\times 
\mathcal{J}_{c}\left(\omega,\omega'\right)
\otimes S_{\rm NLP} \left(\omega,\omega'\right)
\nonumber \\&+
\int_{\Lambda/Q}^{1-\Lambda/Q} drdr'\,
C^{B1}(r) C^{B1}(r')^{*}\otimes
\mathcal{J}^{q\bar q}_{\bar{c}}\left(r,r'\right)\times 
\mathcal{J}_{c}^{(g)}\times S^{(g)}\,.
\label{eq:ffschematic}
\end{align}
Here $C$ are hard matching coefficients, $\mathcal{J}$ are jet 
functions and $S$ are soft functions.  The precise definitions 
will be given in subsequent sections. We dropped all function 
arguments except for the convolution variables, which have 
divergent integrals. The soft-collinear $\omega$, $\omega'$ 
convolution integrals are logarithmically divergent for 
$\omega,\omega'\to \infty$, 
while the hard-anti-collinear integrals over $r$, $r'$ diverge 
logarithmically when $r,r'\to 0,1$. However, both integrals are 
well-defined in dimensional regularization as long as the 
limit to four space-time dimensions is not taken.

In the intermediate region, the soft quark in the soft function 
$S_{\rm NLP}$ that appears in possibility II has {\em large} soft 
momentum $\omega$ and can equally well be regarded as part of 
the anti-collinear function $\mathcal{J}^{q\bar q}_{\bar{c}}$ 
in the second line with {\em small} 
anti-collinear momentum fraction $r$. Removing the  
quark from $S_{\rm NLP}$ leaves the single-particle soft 
function $S^{(g)}$, hence $S_{\rm NLP}\to S^{(g)}$ and 
$\mathcal{J}^{(\bar{q})}_{\bar{c}} \to 
\mathcal{J}^{q\bar q}_{\bar{c}}$. At the same time, the hard 
process changes from A0-type to B1-type. 
Thus, the {\em integrands} of the two terms in 
\eqref{eq:ffschematic} should become identical in the 
respective limits. A rearrangement at the integrand level 
can then be performed in the singular limits, which 
introduces the factorization parameter $\Lambda$, such that 
both integrals are separately finite. At this point 
one can remove the dimensional regulator and use standard 
renormalization-group techniques to sum the 
logarithms in the hard, jet and soft functions. These 
involve only LP interactions. NLP interactions occur 
only once (at amplitude level), leading to the 
two terms in \eqref{eq:ffschematic}, and therefore provide 
only one additional logarithmic, but finite integration from the 
convolutions.

\section{Bare factorization theorem}
\label{sec:bare}

The distribution of an event shape $e$ in $e^+ e^-$ annihilation 
into hadrons through a photon with virtuality $Q^2=q^2$ is
\begin{align}
\frac{d\sigma}{de} =
\sum_{X}\int \!\PS_{X}\left(2\pi\right)^{d}\delta^{(d)}\left(q-p_{X}\right)L_{\mu\nu}\left(q\right)
\delta(e-\hat{e}(X))\left\langle 0|J^{\dagger\nu}(0)|X\right\rangle \left\langle X|J^{\mu}(0)|0\right\rangle ,
\label{eq:dsigma}
\end{align}
with the leptonic tensor
\begin{align}
L_{\mu\nu}\left(q\right)=\frac{8\pi^{2}\alpha_{\rm em}^{2}}
{3Q^{4}}\left(-g_{\mu\nu}+\frac{q_{\mu}q_{\nu}}{Q^{2}}\right)\,
\label{eq:leptontensor}
\end{align}
and the electromagnetic current $J^\mu=\bar\psi\gamma^\mu\psi$, 
for simplicity of a single quark flavour with electric charge 
$e_\psi=+1$.
For a given final state $X$, $\hat{e}(X)$ returns the value of the 
event shape variable. For small $\tau$, 
\begin{equation}
\tau = \frac{M_R^2+M_L^2}{Q^2}+\mathcal{O}(\tau^2),
\label{eq:tauMRML}
\end{equation}
and the thrust distribution is related to the two-hemisphere 
invariant-mass distribution by
\begin{equation}
\frac{d\sigma}{d\tau} =
\int dM_R^2dM_L^2 \,\delta\!\left(\tau-\frac{M_R^2+M_L^2}{Q^2}
\right)\frac{d\sigma}{dM_R^2 dM_L^2}\,.
\label{eq:taudist}
\end{equation}

Eq.~(\ref{eq:dsigma}) is the starting point of the perturbative 
computation in terms of partonic final states $X$. The LP 
factorization theorem for the resummed event shape in the two-jet 
limit has been known for a 
long time \cite{Catani:1991kz,Catani:1992ua}. The resummation of NLP 
double-logarithmic corrections to the quark-antiquark 
back-to-back jet configuration, which is present already at LP,
 has been derived recently 
\cite{Moult:2018jjd}. Defining the left hemisphere to be the 
collinear one and employing the SCET framework, the hemisphere 
invariant-mass distribution for $M_R^2,M_L^2\ll Q^2$ is expressed 
at LP as 
\begin{equation}
\frac{1}{\sigma_0}\frac{d\sigma}{dM_R^2 dM_L^2} = 
|C^{\rm A0}(Q^2)|^2 \int dl_+ dl_-\,
\mathcal{J}_c^{(q)}(M_L^2-Q l_-)\mathcal{J}_{\bar c}^{(\bar q)}(M_R^2-Q l_+)\,
S_{\rm LP}(l_+,l_-)
\end{equation}
in terms of the convolution of the jet (anti-) collinear functions 
with the hemisphere-soft function, multiplied by a 
universal hard function\cite{Fleming:2007qr,Schwartz:2007ib,Bauer:2008dt,Becher:2008cf}.\footnote{By charge conjugation symmetry, the 
quark and anti-quark jet functions coincide, $\mathcal{J}_{\bar c}^{(\bar q)}(p^2) = \mathcal{J}_c^{(q)}(p^2)$.} The distribution is normalized to 
the leading-order total cross section
\begin{equation}
\sigma_0=\frac{4\pi N_c\alpha_{\rm em}^{2}}
{3Q^2}\,,
\end{equation} 
with the number of colours $N_c=3$.
Factorization theorems at NLP are more complicated due to 
multi-local convolutions, which further exhibit endpoint 
divergences, when expressed in terms of renormalized hard, 
collinear and soft functions \cite{Beneke:2019oqx}.

Before turning to endpoint factorization, we provide in this 
section the SCET derivation of the factorization formula for  
off-diagonal gluon thrust, which has already been outlined 
in \cite{Moult:2019uhz}. We introduce two light-like vectors 
$n_\pm^\mu$, satisfying $\np\nm=2$ and pointing into the 
directions of the back-to-back jets. The vector $\nm^\mu$ defines 
the direction of the collinear modes of the gluon jet. (Anti-) 
collinear modes therefore have large components $\np p_c\sim Q$ 
($\nm p_{\bar c}\sim Q$). 
The transverse momenta in the jet are of order 
$p_\perp \sim \lambda Q\sim\sqrt{\tau} Q \ll Q$, where 
$\lambda\sim\sqrt{\tau}$ denotes the SCET power-counting 
parameter. The momenta of soft particles scale uniformly, 
$p_s\sim \tau Q$. In the adopted reference frame, $q_\perp^\mu=0.$
We further note that since gluon thrust vanishes at LP, there 
are no kinematic NLP corrections, and \eqref{eq:tauMRML}, 
\eqref{eq:taudist} continue to hold.

After integrating out the hard modes, the electromagnetic current 
matches to 
\begin{eqnarray}
\bar\psi\gamma_\perp^\mu\psi(0) &=& \int dtd\bar{t}\,
\widetilde{C}^{\rm A0}(t,\bar{t})\,
\bar{\chi}_c(t\np)\gamma_\perp^\mu\chi_{\bar c}(\bar t\nm) 
+ (c\leftrightarrow \bar{c})
\nonumber\\
&&+\,\sum_{i=1,2}\int dt d\bar{t}_1d\bar{t}_2\,
\widetilde{C}^{\rm B1}_i(t,\bar{t}_1,\bar{t}_2)\,
\bar{\chi}_{\bar c}(\bar{t}_1\nm)\Gamma_i^{\mu\nu}\mathcal{A}_{c\perp\nu}
(t\np)\chi_{\bar c}(\bar{t}_2\nm)
+\ldots \qquad
\label{eq:hardmatching}
\end{eqnarray}
in SCET. Here $\mathcal{A}_{c\perp\nu}$ denotes the gauge-invariant 
gluon field operator dressed with Wilson lines, which in light-cone gauge is related to the gluon field by $\mathcal{A}_{c\perp\nu}=
g_s A_{c\perp\nu}$. The omitted terms are higher than NLP 
corrections and purely gluonic ``flavour-singlet'' operators. 
For simplicity, we neglect the mixing 
of $q\bar{q}$ into $gg$ operators in the second line. 
Singlet terms cause well-understood 
technical complications but do not touch the essence of the 
factorization discussed in this work.

The originally local current is now represented by 
light-ray operators. The first line corresponds to the LP hard 
vertex, which emits a back-to-back quark-antiquark pair. The 
second line of \eqref{eq:hardmatching} 
represents the $\mathcal{O}(\lambda)$ suppressed 
``B-type'' SCET operator, which directly produces the 
configuration relevant to gluon thrust, a collinear gluon and an 
anti-collinear quark-antiquark pair. The first line can 
nevertheless contribute to gluon thrust through a time-ordered 
product with the  $\mathcal{O}(\lambda)$ suppressed SCET 
interaction  \cite{Beneke:2002ph,Beneke:2002ni}
\begin{align}
\mathcal{L}_{\xi q}(x)=\bar{q}_s(x_{-})\slashed{\mathcal{A}}_{c\perp}(x)\chi_{c}(x)+\rm{h.c.}\,,
\label{eq:Lxiq}
\end{align} 
which converts a collinear quark into a collinear gluon and 
a soft quark. Here
\begin{equation}
x_\mp^\mu = (n_\pm\cdot x)\frac{n_\mp^\mu}{2}\,,
\label{eq:xminusdef}
\end{equation}
and we further define the scalar quantity 
$x_\mp \equiv n_\pm\cdot x/2$. 
The gluon jet now recoils against a quark-antiquark 
pair in a highly asymmetric configuration, in which the 
antiquark carries almost all the momentum of the 
jet.\footnote{The hermitian conjugate in \eqref{eq:Lxiq} provides 
a second contribution, in which the antiquark turns into the 
collinear gluon and a soft antiquark.} The Dirac structures in the 
power-suppressed B-type operator are given by
\begin{equation}
\Gamma_1^{\mu\nu}=\frac{\slashed n_{-}}{2}\gamma_{\perp}^{\nu}\gamma_{\perp}^{\mu}\,,
\qquad\quad
\Gamma_2^{\mu\nu}=\frac{\slashed n_{-}}{2}\gamma_{\perp}^{\mu}\gamma_{\perp}^{\nu}\,.
\label{eq:BtypeDirac}
\end{equation}
We note that boost invariance together with the projection 
property $\slashed n_{-} \chi_c=0$, 
$\slashed n_+ \chi_{\bar c}=0$ implies that only the transverse 
components of the vector current contribute.

Eq.~\eqref{eq:hardmatching} is the starting point for deriving 
the factorization formula and the main steps are fairly standard.
After the collinear field redefinition with the 
soft Wilson line \cite{Bauer:2001yt} $Y_{n_-}(x)=\mathcal{P} \exp\left[i g_s \int_0^\infty ds n_-A_s(x+sn_-) \right]$ 
($Y_{n_+}(x)$ for anti-collinear fields), the collinear, 
anti-collinear and soft fields are decoupled at LP. The final 
state  $|X\rangle = |X_c\rangle|X_{\bar c}\rangle 
|X_s\rangle$ consists of a collection of corresponding modes, 
and hence the matrix element factorizes into suitably 
defined collinear, anti-collinear and soft functions. 
Performing these steps on the two terms in 
\eqref{eq:hardmatching}, more precisely the time-ordered 
product with \eqref{eq:Lxiq} for the first, gives rise 
to an ``A-type'' and a ``B-type'' term in the factorization formula, 
which we discuss separately next. Note that due to the different 
field/mode content in the final state (soft quark or not), 
there is no interference between the two terms when 
squaring the amplitude.

\subsection{A-type term (soft quark)}

This term stems from the square of the matrix element
\begin{equation}
\langle X_c|\langle X_{\bar c}|\langle X_s| \int d^d x \,T\,[\bar{\chi}_c(t\np)\gamma_\perp^\mu\chi_{\bar c}
(\bar t\nm), i\mathcal{L}_{\xi q}(x)]\,|0\rangle\,,
\label{eq:atypetproduct}
\end{equation}
summed and integrated over all possible final states. As mentioned 
above, there are two A-type terms, because either the 
quark or the anti-quark in the current can transfer its momentum 
to the collinear gluon. Due to the different final state, the 
two terms do not interfere. 
In \eqref{eq:atypetproduct} we consider explicitly only the 
first contribution. We comment 
on the second at the end. To separate the above matrix 
element into collinear, anti-collinear and soft factors, 
we perform the field redefinitions \cite{Bauer:2001yt}
\begin{eqnarray}
\chi_c(x) \to Y_{n_-}(x_-)\chi_c(x),&\qquad& 
\chi_{\bar c}(x) \to Y_{n_+}(x_+)\chi_{\bar c}(x), 
\\[0.1cm]
&&\hspace*{-4cm}
\mathcal{A}_{c\perp}^\mu(x)\to Y_{n_-}(x_-)\mathcal{A}_{c\perp}^\mu(x)
Y^\dagger_{n_-}(x_-)\,.
\end{eqnarray}
Inspection of \eqref{eq:hardmatching}, 
\eqref{eq:atypetproduct} shows that the following hard, (anti-) 
collinear and soft functions appear.

\paragraph{Hard function} This term involves the hard function 
$\widetilde{C}^{\rm A0}(t,\bar{t})$ of 
the LP A-type current. We define the momentum-space 
coefficient
\begin{equation}
C^{\rm A0}(n_+p_c, n_-p_{\bar{c}}) = 
\int dtd\bar{t}\,e^{it n_+ p_c+i \bar{t}\nm p_{\bar c}}\,
\widetilde{C}^{\rm A0}(t,\bar{t})\,.
\label{eq:CA0momentumspace}
\end{equation}
It depends only on the product $Q^2 = n_+p_c n_-p_{\bar{c}}$  of 
the large components of the collinear and anti-collinear momenta. 
At LO, we find $C^{\rm A0}(Q^2)=1+\mathcal{O}(\alpha_s)$. 
The two-loop result can be found in \cite{Becher:2007ty}.

\paragraph{Anti-collinear function}

The $\mathcal{L}_{\xi q}$ Lagrangian insertion acts in the collinear sector, hence the anti-collinear sector is unaffected. After squaring the amplitude and integrating over the anti-collinear final state, we obtain the LP anti-quark  jet function 
\begin{eqnarray}
&&\frac{1}{2\pi}\sum_{X_{\bar{c}}}\int \!\PS_{X_{\bar{c}}}
\langle 0|\bar{\chi}_{\bar{c}}(x)_{b\beta}|X_{\bar{c}}\rangle 
\langle X_{\bar{c}}|\chi_{\bar{c}}(0)_{a\alpha}|0\rangle 
\nonumber\\
&& \equiv\,\delta_{ab}\int\!\frac{d^d p}{(2\pi)^d}\,n_{-}p\,e^{-ipx}
\mathcal{J}^{(\bar{q})}_{\bar{c}}(p^2)\left(\frac{\slashed n_{+}}{2}\right)_{\alpha\beta}.
\label{eq:LPjet}
\end{eqnarray}
At LO, we have $\mathcal{J}^{(\bar{q})}_{\bar{c}}(p^2)=\delta^{+}(p^2)
\equiv \theta(p^0)\delta(p^2)$. This object is well understood and computed up to the third order in $\alpha_s$ \cite{Bruser:2018rad}. Here and below we use small Latin 
letters for colour and Greek letters for Dirac spinor indices. 

\paragraph{Collinear function} This function picks up the collinear fields from the 
$\mathcal{L}_{\xi q}$ insertion and represents a new non-local 
jet function that appears at NLP. Defining the non-local operator
\begin{equation}
\mathcal{O}_{a\alpha;b\beta}(\omega,x) = 
\int d^dy \,e^{i y_-\omega}\,T\big\{\bar{\chi}_{c,b\beta}(x),
\left[\,\not\hspace*{-1.5mm}\mathcal{A}_{\perp c}\chi_c\right]_{a\alpha}(x+y)\big\}\,,
\label{eq:nonlocaljetfield}
\end{equation}
where $y_-=\np\cdot y/2$, its jet function defined as in the LP 
case \eqref{eq:LPjet} by 
\begin{eqnarray}
&&\frac{1}{2\pi}\sum_{X_{c}}\int \PS_{X_{c}}\,
\frac{1}{g_{s}^{2}}\left\langle 0\right|
\mathcal{O}^\dagger_{b'\beta';a'\alpha'}(\omega^\prime,x)
\left|X_{c}\right\rangle 
\,\left[\frac{\slashed n_{+}}{2}\right]_{\alpha' \alpha} \!\!
\left\langle X_{c}\right|\mathcal{O}_{a\alpha;b\beta}(\omega,0)\left|0\right\rangle 	
\nn\\&& \hspace*{0.5cm} = \,(d-2)\,
\left[\frac{\slashed n_{-}}{2}\right]_{\beta'\beta}\,
\int\frac{d^dp}{(2\pi)^d}\,e^{-ipx}\,\Big\{
\left[t^{A}\right]_{ab}\left[t^{A}\right]_{b'a'}
\,\mathcal{J}_{c}(p^2,\omega,\omega^\prime)
\nn\\
&& \hspace*{1cm}+
\left[t^{A}\right]_{aa'}\left[t^{A}\right]_{b'b}
\,\widehat{\mathcal{J}}_{c}(p^2,\omega,\omega^\prime)
\Big\}\,.
\label{eq:defjetNLP}
\end{eqnarray}
The factor $1/g_s^2$ has been introduced for convenience, such 
that the leading term is $\mathcal{O}(\alpha_s^0)$. The Dirac matrix 
$\left[\slashed n_{+}/2\right]_{\alpha' \alpha}$ on the left-hand 
side ensures that only a single Dirac structure can appear 
on the right-hand side, and avoids a discussion of evanescent 
structures that would otherwise be present. In the derivation of 
the A-type term, this factor will arise from the decomposition of  
the soft function, see \eqref{eq:NLPsoftfndef} below.\footnote{This 
would remain true even if we 
had allowed for flavour-singlet contractions of the quark fields. 
These do not exist for the A-type term, since the contraction of 
different modes, that is the collinear quark field with the 
soft quark field from the 
soft function, is not allowed in SCET.} At 
LO in $\alpha_s$, the collinear final state $X_c$ 
consists of a single gluon,\footnote{The $q\bar{q} g$ final state 
at this order corresponds to a disconnected diagram, which does 
not contribute to the scattering amplitude.} and only the 
unhatted jet function is non-vanishing:
\begin{align}
\mathcal{J}_{c}(p^2,\omega,\omega')&= 
 \frac{1}{\omega\omega'}\,\delta^+(p^{2})\,,\\
\widehat{\mathcal{J}}_{c}(p^2,\omega,\omega')&=0\,.
\label{eq:AtypeJctree}
\end{align}

\paragraph{Soft function} 

Collecting the soft Wilson lines from 
the soft-decoupling field redefinition, as well as the 
soft quark field from the Lagrangian insertion \eqref{eq:Lxiq} 
leads to the following definition of the NLP soft quark function:
\begin{eqnarray}
&&g_s^2
\int\frac{dx_-}{2\pi}\frac{dx_-^\prime}{2\pi}
\,e^{-i (x_-\omega-x_-^\prime\omega^\prime)}	
\left\langle 0\right|\overline{T}\left\{
\left[Y_{n_+}^{\dagger}(0)Y_{n_-}(0)\right]_{cb'}
\left[Y^\dagger_{n_-}q_s\right]_{\alpha'a'}(x_-^\prime)
\right\}
\nn \\ 
&&\hspace*{1cm}\times\,
\mathcal{P}_s(l_+,l_-)\,
T\left\{\left[\bar{q}_sY_{n_-}\right]_{\alpha a}(x_-)\left[Y_{n_-}^{\dagger}\left(0\right)Y_{n_+}
\left(0\right)\right]_{bc}\right\}
\left|0\right\rangle 	\nn\\
&& = \, \left(\frac{\slashed n_{+}}{2}\right)_{\alpha'\alpha}\!
\left\{
\delta_{a'a}\delta_{bb'} \,
S_{\text{NLP}}(l_+,l_-,\omega,\omega^\prime) +
\delta_{ba}\delta_{a'b'} \,
\widehat{S}_{\text{NLP}}(l_+,l_-,\omega,\omega^\prime)
\right\} +\ldots\,,\qquad\quad
\label{eq:NLPsoftfndef}
\end{eqnarray}
where $x_-$ is defined below \eqref{eq:xminusdef}, and 
\begin{equation}
\mathcal{P}_s(l_+,l_-) \equiv \sum_{X_s}\int \PS_{X_s}\,
\delta(l_+-\np p_{X_s}^R)\delta(l_--\nm p_{X_s}^L)\,|X_s\rangle
\langle X_s|\,,
\label{eq:softmeasurement}
\end{equation}
contains the measurement function on the soft final state and 
the final-state sum pertaining to the two-hemisphere invariant 
mass distribution. With this definition, the left hemisphere 
contains the collinear gluon jet, and the right hemisphere 
is on the anti-collinear side. 
The dots in \eqref{eq:NLPsoftfndef} denote terms proportional to 
$\slashed n_{-}$, which do not contribute due to the projection 
property of collinear fields in other parts of the process.
The pre\-factor is chosen for convenience, such that 
the soft function starts at $\mathcal{O}(\alpha_s)$. 
The soft function for thrust is obtained from
\begin{equation}
S_{\text{NLP}}^{T}(k,\omega,\omega') = \int_0^\infty 
dl_+d l_-\,\delta(k-l_+ -l_-) \, 
S_{\text{NLP}}(l_+,l_-,\omega,\omega')\,.
\end{equation}
There is an analogous definition with obvious modifications 
for the NLP soft anti-quark 
function relevant to the second A-type contribution. The 
colour decomposition is chosen such that at leading $\mathcal{O}(\alpha_s)$, 
$\widehat{S}_{\text{NLP}}(l_+,l_-,\omega,\omega^\prime)=0$, 
while
\begin{eqnarray}
S_{\text{NLP}}(l_+,l_-,\omega,\omega^\prime) &=&
\frac{\alpha_s}{4\pi}\frac{\theta(\omega)\delta(\omega-\omega')}
{e^{-\epsilon\gamma_E}\Gamma(1-\epsilon)}\,
\bigg[\delta(l_-)\theta(\omega-l_+)\theta(l_+)\,
\omega\left(\frac{l_+\omega}{\mu^2}\right)^{\!-\epsilon}
\nonumber\\
&& -\,\frac{\omega^2}{1-\epsilon}\,\delta(l_+)\delta(l_--\omega)
\left(\frac{\omega^2}{\mu^2}\right)^{\!-\epsilon}\,\bigg]\,.
\label{eq:NLPsofttree}
\end{eqnarray}
The two-hemisphere NLP soft-quark function is introduced here 
for the first time, but we take note of the calculation of the NLP 
soft function that appears  for soft gluon emission in 
the Drell-Yan process \cite{Beneke:2018gvs}, which is already 
known to $\mathcal{O}(\alpha_s^2)$ 
\cite{Beneke:2018gvs,Broggio:2021fnr}.

\paragraph{Factorization formula}
With these definitions, we obtain the soft-quark contribution 
to gluon thrust in the two-jet region in the form
\begin{eqnarray}
\frac{1}{\sigma_0}\frac{d\sigma}{dM_R^2 dM_L^2}|_{\rm A-type} &=& 
\frac{2 C_F}{Q}\,f(\epsilon)\,
|C^{\rm A0}(Q^2)|^2 \int_0^\infty dl_+ dl_-\,
\int d\omega d\omega'\,
\mathcal{J}_{\bar c}^{(\bar q)}(M_R^2-Q l_+)
\nonumber\\
&&\hspace*{-2cm}\times\,
\bigg\{\,\mathcal{J}_{c}(M_{L}^{2}-Ql_-,\omega,\omega')
\,S_{\rm NLP}(l_+,l_-,\omega,\omega')
\nonumber\\
&&\hspace*{-1.2cm}
+\,\widehat{\mathcal{J}}_{c}(M_{L}^{2}-Ql_-,\omega,\omega')\,
\widehat{S}_{\rm NLP}(l_+,l_-,\omega,\omega')\,\bigg\}\,,
\label{eq:Atypefactformula}
\end{eqnarray}
where 
\begin{eqnarray}
f(\epsilon) = \bigg( \frac{Q^{2}}{4\pi} \bigg)^{-\epsilon} \frac{ (1-\epsilon)^2 \Gamma(1-\epsilon)}{ \Gamma(2- 2\epsilon)}
\end{eqnarray}
is a $d$-dimensional factor defined such that 
$f(0)=1$ in four dimensions.

To turn the convolution in $l_+, l_-$ into a product, it is 
convenient to consider the Laplace transform 
\begin{equation}
\frac{\widetilde{d\sigma}}{ds_R ds_L} = 
\int_0^\infty dM_R^2 dM_L^2\,\,e^{-s_R M_R^2/Q}\,
e^{-s_L M_L^2/Q}\,\frac{d\sigma}{dM_R^2 dM_L^2}
\label{eq:laplacetrafo}
\end{equation}
of the two-hemisphere invariant-mass distribution. In general, 
the Laplace transform of collinear functions with variable 
$p^2$ are defined with $\int_0^\infty dp^2\,e^{-s p^2/Q}$, while 
for soft functions we use $\int_0^\infty dl_+ dl_-\,
e^{-s_R l_+-s_L l_-}$. Then
\begin{eqnarray}
\frac{1}{\sigma_0}\frac{\widetilde{d\sigma}}{ds_R ds_L}|_{\rm A-type} &=& 
\frac{2 C_F}{Q} \,f(\epsilon)\,|C^{\rm A0}(Q^2)|^2 \,
\widetilde{\mathcal{J}}_{\bar c}^{(\bar q)}(s_R)
\,\int d\omega d\omega'\,
\nonumber\\
&&\hspace*{-3cm}\times\,
\bigg\{\,\widetilde{\mathcal{J}}_{c}(s_{L},\omega,\omega')
\,\widetilde{S}_{\rm NLP}(s_R,s_L,\omega,\omega')
+\;\,\widetilde{\!\!\widehat{\mathcal{J}}}_{\!c}(s_{L},
\omega,\omega')
\,\,\widetilde{\!\widehat{S}}_{\rm NLP}(s_R,s_L,\omega,\omega')\,\bigg\}\,. \quad
\label{eq:AtypeLaplace}
\end{eqnarray}
The total A-type term is twice the above \eqref{eq:Atypefactformula}, 
\eqref{eq:AtypeLaplace}, since there is 
an equal contribution from the soft anti-quark phase-space 
region. 

\subsection{B-type term}

The B-type term is obtained under the assumption that the 
intermediate propagator in Fig.~\ref{fig:QCD} is hard, which 
leads to the B1 operator in the matching equation 
\eqref{eq:hardmatching}. We then need the square of   
the matrix elements
\begin{equation}
\langle X_c|\langle X_{\bar c}|\langle X_s|\,\bar{\chi}_{\bar c}(\bar{t}_1\nm)\Gamma_i^{\mu\nu}
\mathcal{A}_{c\perp\nu}(t\np)\chi_{\bar c}
(\bar{t}_2\nm) \, |0\rangle\,,
\label{eq:btypeME}
\end{equation}
summed and integrated over all possible final states.
Inspection of \eqref{eq:hardmatching}, 
\eqref{eq:btypeME} shows that the following hard, (anti-) 
collinear and soft functions appear.

\paragraph{Hard function} 
There are two relevant operators, $i=1,2$, 
which differ by their Dirac structures \eqref{eq:BtypeDirac}. We 
introduce the Fourier transforms
\begin{equation}
C^{\rm B1}_i(n_+p_c, r n_-p_{\bar{c}},\bar{r} n_-p_{\bar{c}}) = 
\nm p_{\bar c} \int dtd\bar{t}_1d\bar{t}_2\,e^{it n_+ p_c
+i (\bar{t}_1 r + \bar{t}_2 \bar{r})\nm p_{\bar c}}\,
\widetilde{C}^{\rm B1}_i(t,\bar{t}_1,\bar{t}_2)
\label{eq:CB1momentumspace}
\end{equation}
of the corresponding hard functions. The B-type operators contain 
two fields (quark and anti-quark) in the same collinear direction. 
The momentum-space coefficient depends on the product 
$Q^2 = n_+p_c n_-p_{\bar{c}}$  of 
the large components of the collinear and total 
anti-collinear momentum, and $r$, which denotes the fraction 
of total anti-collinear momentum carried by the quark. 
Accordingly, $\bar{r}=1-r$ is the momentum fraction of the 
anti-quark. At the leading order in $\alpha_s$, we find\footnote{We 
recall that the definition of $\mathcal{A}_{c\perp}^\mu$ absorbs 
the factor $g_s$ from the tree-level diagram.}
\begin{align}
C_1^{\rm B1}(Q^2,r) &=\frac{1}{r}\,,
\label{eq:CB11tree}\\
C_2^{\rm B1}(Q^2,r) &=-\frac{1}{\bar{r}}\,.
\end{align}
From the transformation of \eqref{eq:hardmatching} under charge 
conjugation, we find that charge conjugation invariance implies that 
$C_2^{\rm B1}(Q^2,r) = -C_1^{\rm B1}(Q^2,\bar{r})$ to 
all orders in the strong coupling. 
We notice that the matching coefficient $C_1^{\rm B1}$ ($C_2^{\rm B1}$) 
is singular in the limit $r\to0$ ($r\to1$), in which  
the quark (anti-quark) becomes soft.  
From this observation, we can already anticipate a relation to the 
A-type contributions.

Further insight can be obtained by a helicity consideration. To this 
end, we separate the electromagnetic current in 
\eqref{eq:hardmatching} into a left- and right-handed piece 
by inserting $1=P_L+P_R$. In four space-time 
dimensions we have\footnote{Convention: $\epsilon^{0123} = +1$}
\begin{eqnarray}
\label{eq:b1helicity1}
\Gamma^{\mu\nu}_1 P_L &=& g^{\mu\nu}_{L} \slashed n_{-} P_L\,,
\\
\Gamma^{\mu\nu}_2 P_L &=& g^{\mu\nu}_{R} \slashed n_{-} P_L\,,
\\
\Gamma^{\mu\nu}_1 P_R &=& g^{\mu\nu}_{R} \slashed n_{-} P_R\,,
\\
\Gamma^{\mu\nu}_2 P_R &=& g^{\mu\nu}_{L} \slashed n_{-} P_R\,,
\label{eq:b1helicity2}
\end{eqnarray}
where 
\begin{equation}
g_{L/R}^{\mu\nu} = \frac{1}{2}\left(g_\perp^{\mu\nu} 
\pm\frac{i}{2}\epsilon^{\rho\sigma\mu\nu} n_{-\rho} n_{+\sigma}\right)
\label{eq:gLR}
\end{equation}
projects on the left $(-1)$ / right $(+1)$ 
helicity of the collinear gluon in the 
B1 operator \eqref{eq:hardmatching}. 
Since helicity is conserved and amplitudes with 
gluons of different helicity do not interfer (see 
\eqref{eq:LPgluonjet} below), the virtual photon and gluon 
always have the same helicity. Focusing on the negative  
helicity case for definiteness, we conclude from 
\eqref{eq:b1helicity1}~--~\eqref{eq:b1helicity2} that 
the operator with Dirac structure $\Gamma^{\mu\nu}_1$ 
produces a left-handed out-going quark, while for 
$\Gamma^{\mu\nu}_2$ it is right-handed. It follows that 
the two operators cannot interfer in the non-singlet channel, 
where the $\bar{\chi}_{\bar c}(\bar{t}_1\nm)
\chi_{\bar c}(\bar{t}_2\nm)$ fields in the operator in 
\eqref{eq:hardmatching} cannot be contracted among 
themselves. This strictly holds only in four space-time 
dimensions. In other words, a non-singlet interference 
term must be ``evanescent'' and vanish as $\epsilon\to 0$. 

The limit $r\to 0$ in the B1 matching coefficient implies 
that the momentum of the out-going quark in the left 
diagram of Fig.~\ref{fig:QCD} goes to zero, and the 
intermediate quark propagator goes on-shell. The helicity 
argument above implies that only $C_1^{\rm B1}(Q^2,r)$ 
corresponding to $\Gamma^{\mu\nu}_1$ can be singular 
in this limit. For definiteness, consider again the 
case of the negative helicity gluon and virtual photon. 
The nearly on-shell intermediate propagator implies 
that the B1 operator factorizes into $\gamma^* \to 
q+\bar{q}$ and $q\to q+g$. As $r\to 0$ the outgoing quark 
in the latter process has zero momentum, and the incoming 
quark and outgoing gluon have the same large collinear 
momentum. Since $g$ has negative helicity, the 
incoming quark must be left-handed, since a 
helicity change by $3/2$ is not possible. By helicity 
conservation of the strong interaction, the final soft 
quark must then also be left-handed. 
As shown above, this singles 
out  $\Gamma^{\mu\nu}_1$. The situation is reversed 
in the soft anti-quark limit 
$r\to 1$ and implies right-handedness, which 
singles out  $\Gamma^{\mu\nu}_2$. We conclude 
that, to all orders in perturbation theory, only 
$C_1^{\rm B1}(Q^2,r)$ ($C_2^{\rm B1}(Q^2,r)$) can be  
singular in the endpoint $r\to 0$ ($r\to 1$). 
  
\paragraph{Collinear function}

The collinear function is simply the standard LP gluon jet 
function, defined as  
\begin{eqnarray}
&&\frac{1}{2\pi}\frac{1}{g_{s}^2}
\sum_{X_c}\int \!\PS_{X_c}\,
\langle 0|\mathcal{A}_{c\perp\mu}^B(x)|X_{c}\rangle 
\langle X_c|\mathcal{A}_{c\perp\nu}^C(0)|0\rangle 
\nonumber\\
&& \equiv\,\delta^{BC}\,(-g_{\mu\nu}^\perp)
\int\!\frac{d^{d}p}{(2\pi)^d}\,e^{-ipx}
\mathcal{J}^{(g)}_{c}(p^2).
\label{eq:LPgluonjet}
\end{eqnarray}
At LO, we have $\mathcal{J}^{(g)}_{c}(p^2)=
\delta^+(p^2)$. This object is well understood and known up 
to the third order in $\alpha_s$ \cite{Banerjee:2018ozf}. 

\paragraph{Anti-collinear function}

The anti-collinear fields consist of the quark-anti-quark pair, 
which recoils against the gluon and is therefore in a colour-octet 
state. The corresponding jet function can be defined as for 
the gluon, but in terms of the composite field
\begin{equation}
\mathcal{Q}_i^{A\mu\nu}(x,r) = 
\frac{1}{2\pi}\int_0^\infty \!\!d\bar{t}\,
e^{-i r \bar{t} \nm\cdot p_{\bar{c}}}\,
\bar{\chi}_{\bar{c}}(x+\bar{t} \nm) t^A\,\Gamma_i^{\mu\nu}
\chi_{\bar{c}}(x)\,.
\end{equation}
At leading order, $r$ can be interpreted as the momentum fraction 
of the out-going quark. The Dirac structures refer to 
\eqref{eq:BtypeDirac}. 
Except for the colour and Dirac matrix, the operator coincides 
with the one that enters the definition of light-cone distribution 
amplitudes of mesons. However, instead of the vacuum-to-meson  
matrix element, here we need the inclusive jet function 
\begin{eqnarray}
&&\frac{g_s^2}{2\pi}
\sum_{X_{\bar{c}}}\int \!\PS_{X_{\bar{c}}}\,
\langle 0|\mathcal{Q}_{i'\mu\nu}^{\dagger B}(x,r')
|X_{\bar{c}}\rangle 
\langle X_{\bar{c}}|\mathcal{Q}_{i}^{A\mu\nu}(0,r)|0\rangle 
\nonumber\\
&&\hspace*{0cm}=\,
\delta^{AB}\,(d-2)^2\int\!\frac{d^{d}p}{(2\pi)^d}\,e^{-ipx}
\,\bigg\{
\delta_{i i'} \mathcal{J}_{\bar{c}}^{q\bar{q}(8)}(p^2,r,r')
+(1-\delta_{i i'})\,
\mathcal{\widehat{J}}_{\bar{c}}^{q\bar{q}(8)}(p^2,r,r^\prime)
\bigg\}\,,\quad
\label{eq:NLPqqbarjet}
\end{eqnarray}
which is a NLP object. It is easy to see that the 
four possible values of $i,i'=1,2$ give rise to only
two independent jet functions, which we decompose as 
above into a diagonal and off-diagonal term in $ii'$. 
At lowest non-vanishing order $\mathcal{O}(\alpha_s)$, 
the $X_{\bar{c}} = q\bar{q}$ final state results in 
\begin{align}
\mathcal{J}_{\bar{c}}^{q\bar{q}(8)}(p^2,r,r')=\frac{\alpha_{s}}{4\pi}
\,\frac{1}{\Gamma\left(1-\epsilon\right)}\,\delta(r-r')\,r\bar{r}\left(\frac{\mu^{2}e^{\gamma_E}}{p^{2}r \bar{r}}\right)^{\!\epsilon},
\end{align}
where $\bar r =1-r$. 

The hatted function, which describes the 
interference of the two B1 operators, is given as this order by 
$\mathcal{\widehat{J}}_{\bar{c}}^{q\bar{q}(8)}(p^2,r,r') = 
(-\epsilon)/(1-\epsilon)\,\mathcal{J}_{\bar{c}}^{q\bar{q}(8)}(p^2,r,r')$, 
and vanishes in four dimensions, as expected from the helicity 
discussion above. In principle, it should be possible to remove 
this evanescent anti-collinear function by a finite, non-minimal 
counterterm related to its mixing into the ``physical'' jet 
function. This relies crucially on a renormalized factorization 
theorem, i.e. being able to take $\epsilon\to 0$ at the 
level of the individual collinear, soft etc. functions. 
Since for the time being, we continue to work with $d$-dimensional 
quantities, we keep 
$\mathcal{\widehat{J}}_{\bar{c}}^{q\bar{q}(8)}(p^2,r,r^\prime)$ 
in the following. We will see later that it is irrelevant for 
the discussion of endpoint divergences.

\paragraph{Soft function} 
Performing the soft decoupling field redefinition on 
\eqref{eq:btypeME} results in the field product
\begin{equation}
\bar{\chi}_{\bar{c}a} [Y_{\np}^\dagger Y_{\nm}]_{ac} 
\big[t^B\big]_{cd}\, 
[Y_{\nm}^\dagger Y_{\np}]_{db} \chi_{\bar{c}b}\,
\mathcal{A}^B_{c\perp\nu}\,.
\end{equation}
The Wilson lines can be combined into the adjoint Wilson 
lines $\mathcal{Y}_{n_\mp}$ using 
\begin{equation}
\big[Y_{n_\mp} t^A Y_{n_\mp}^\dagger\big]_{ab}  = 
\big[Y_{n_\mp}^\dagger t^A Y_{n_\mp}\big]_{ab}  = 
\mathcal{Y}_{n_\mp}^{AB} \big[t^B\big]_{ab}\,.
\end{equation}
After squaring the matrix element and summing over the soft 
final state, we obtain the LP two-hemisphere soft 
function in the adjoint representation:
\begin{equation}
S^{(g)}(l_+,l_-)=\frac{1}{N_c^2-1}\langle 0|\overline{T}\left\{
\mathcal{Y}_{n_+}^{BD}(0)\mathcal{Y}_{n_-}^{DA}(0)\right\}
\,\mathcal{P}_s(l_+,l_-)\,
T\left\{\mathcal{Y}_{n_-}^{AC}(0)\mathcal{Y}_{n_+}^{CB}(0)
\right\}|0\rangle \,.
\end{equation}
The LO result is $S^{(g)}(l_+,l_-)=\delta(l_+)\delta(l_-)$.

\paragraph{Factorization formula}

Inserting these matrix element definitions, we 
obtain the factorization formula for the B-type term 
\begin{eqnarray}
\frac{1}{\sigma_0}\frac{d\sigma}{dM_R^2 dM_L^2}|_{\rm B-type} &=& 
\frac{2 C_F}{Q^2} \,f(\epsilon)\int_0^\infty dl_+ dl_-\,
\sum_{i,i'=1,2} \int dr dr'\,
C^{\rm B1*}_{i'}(Q^2,r')C^{\rm B1}_{i}(Q^2,r) 
\nonumber\\
&&\hspace*{-2cm}\times\,
\bigg\{\,
\delta_{i i'}\,\mathcal{J}_{\bar{c}}^{q\bar{q}(8)}(M_R^2-Ql_+,r,r^\prime)
+\,(1-\delta_{i i'})\,\mathcal{\widehat{J}}_{\bar{c}}^{q\bar{q}(8)}(M_R^2-Ql_+,r,r^\prime)
\bigg\}
\nonumber\\[0.2cm]
&&\hspace*{-2cm}\times\,
\mathcal{J}_{c}^{(g)}(M_L^2-Q l_-)\,S^{(g)}(l_+,l_-)\,. 
\label{eq:Btypefactformula}
\end{eqnarray}
In Laplace space:
\begin{eqnarray}
\frac{1}{\sigma_0}\frac{\widetilde{d\sigma}}{ds_R ds_L}|_{\rm B-type} 
&=& 
\frac{2 C_F}{Q^2}  \,f(\epsilon)\sum_{i,i'=1,2} \int dr dr'\,
C^{\rm B1*}_{i'}(Q^2,r')C^{\rm B1}_{i}(Q^2,r) 
\nonumber\\
&&\hspace*{-2.5cm}\times\,
\bigg\{\,
\delta_{i i'}\,\widetilde{\mathcal{J}}_{\bar{c}}^{q\bar{q}(8)}(s_R,r,r^\prime)
+\,(1-\delta_{i i'})\;\;\widetilde{\!\!\mathcal{\widehat{J}}}{}_{\bar{c}}^{q\bar{q}(8)}(s_R,r,r^\prime)
\bigg\}\,
\widetilde{\mathcal{J}}_{c}^{(g)}(s_L)\,\widetilde{S}^{(g)}(s_R,s_L)\,. \quad
\label{eq:BtypeLaplace}
\end{eqnarray}
Only the first term in the curly brackets proportional to 
$\delta_{i i'}$ is non-vanishing at leading 
$\mathcal{O}(\alpha_s)$.\footnote{Since the {\em sum} of 
the A-type and B-type term are finite in the limit $\epsilon\to 0$, 
the common prefactor $f(\epsilon)$ could be set to its 
four-dimensional value 1 everywhere at this point.} 

\subsection{Tree-level evaluation}
\label{sec:treelevelevaulation}

Inserting tree-level results for all functions given above, the
two terms in equations \eqref{eq:Atypefactformula} and
\eqref{eq:Btypefactformula} reduce to the following expressions. The
A-type soft-quark term is given by 
\begin{eqnarray} 
\label{eq:AtypesqTree}
\frac{1}{\sigma_0}\frac{d^2\sigma}{dM^2_R\,dM^2_L}|_{
\scriptsize \begin{array}{l}
$\rm A--type$,\\[-0.1cm] \rm tree \end{array}} 
&=&\frac{2C_F}{Q}\, f(\epsilon) \int dl_+\,dl_-
\, \int d\omega d\omega' \, \delta^+(M^2_R-Ql_+ )\,
\nonumber\\[0ex] 
&&\hspace*{-3cm} 
\times \,\frac{1}{\omega\omega'}\,\delta^+(M^2_L-Ql_-) 
\nonumber \,\frac{\alpha_s}{4\pi} \,\frac{\theta(\omega)
\delta(\omega-\omega') }{e^{-\epsilon\gamma_E}\Gamma(1-\epsilon)}
\,\bigg[ \delta(l_- ) \,\theta(\omega-l_+ ) \theta(l_+ )\,
\omega \left(\frac{l_+ \omega}{\mu^2}\right)^{\!-\epsilon}
\\[0ex] 
&& \hspace{-2cm}- \frac{ \omega^{2}}{1-\epsilon} 
\, \delta(l_+ )\,\delta(l_- -\omega)
\left(\frac{ \omega^2}{\mu^2}\right)^{\!-\epsilon}\bigg]
\nonumber\\
&&\hspace{-4cm} =
\frac{\alpha_s}{4\pi}\,
\frac{2C_F}{Q^2} \,\frac{f(\epsilon)}{e^{-\epsilon\gamma_E}\Gamma(1-\epsilon)}
\bigg[ \frac{1}{\epsilon}\delta^+(M^2_L)
\left(\frac{({M^2_R})^2}{{Q}^2\mu^2}\right)^{\!-\epsilon}
- \,\delta^+(M^2_R )\,\frac{ 1}{1-\epsilon}
\left(\frac{ (M^2_L)^2}{Q^2\mu^2}\right)^{\!-\epsilon}\bigg],\quad\;
\end{eqnarray}
where the pole in $\epsilon$ arises from the logarithmic divergence
of the integral $\int_{l_+}^\infty d\omega/\omega$ for large values
of the soft function variable. The total A-type contribution is
twice the above after adding the soft-antiquark term.
Similarly, the B-type term is given by
\begin{eqnarray}
\label{eq:BtypeTree}
\frac{1}{\sigma_0}\frac{d^2\sigma }{dM^2_L\,dM^2_R}
|_{\scriptsize \begin{array}{l}
$\rm B--type$,\\[-0.1cm] \rm tree \end{array}}
&=& \frac{2C_F}{Q^2 } \,f(\epsilon) \int dl_+\,dl_- \int_0^1\, 
dr dr'\,\delta^+(M^2_L-Ql_-) \delta(l_+)\delta(l_-)
\nonumber\\
&&\hspace{-3.5cm} \times\,\bigg\{
\left(\frac{1}{r}\frac{1}{r'}+\frac{1}{\bar{r}}\frac{1}{\bar{r}'} 
\right) 
+
\left(\frac{1}{r}\frac{1}{\bar{r}'}+\frac{1}{\bar{r}}\frac{1}{r'} 
\right)\frac{\epsilon}{1-\epsilon}
\bigg\}
\,\delta(r -r' ) 
\,\frac{\alpha_s}{4\pi}
\frac{ r \bar{r} }{\Gamma(1-\epsilon)}
\bigg(\frac{(M^2_R-Ql_+)\, {r} \bar{r} }{\mu^2e^{\gamma_E}} 
\bigg)^{\!-\epsilon}
\nonumber\\[0.2cm]
&&\hspace{-3.5cm} = \frac{\alpha_s}{4\pi}\,
\frac{4C_F}{Q^2 }\,f(\epsilon)\,\bigg\{
-\frac{1}{\epsilon} + 
\frac{\epsilon}{(1-\epsilon)^2 }\,\bigg\} \, \delta^+(M^2_L)
 \bigg(\frac{M^2_R}{\mu^2e^{ \gamma_E}} \bigg)^{\!-\epsilon}
\frac{\Gamma (2-\epsilon)}{\Gamma (2-2 \epsilon)}\,.
\end{eqnarray}
In this case, the singularity arises from a logarithmic divergence
in the integral over the momentum fraction $r$ as $r\to 0$ and 
$r\to 1$. Adding both terms and taking the limit $\epsilon \to 0$, 
we find
\begin{eqnarray}
\label{eq:treeresultA+B}
\frac{1}{\sigma_0}\frac{d^2\sigma }{dM^2_L\,dM^2_R} &=&
\frac{\alpha_s C_F}{\pi}\,\frac{1}{Q^2 }
\,\bigg\{\delta^+(M^2_L) \left[\ln\frac{Q^2}{M_R^2} -1\right] 
- \delta^+(M^2_R)\bigg\}\,.
\end{eqnarray}
The poles cancel in the sum of both the A and B-type contributions
in equations \eqref{eq:AtypesqTree} and \eqref{eq:BtypeTree}
(once we account for the soft-antiquark contribution to the A-type term)
as it should be, as the off-diagonal gluon-thrust is infrared safe.
After converting the two-hemisphere invariant mass distribution to the
thrust distribution according to \eqref{eq:taudist}, we reproduce
the coefficient $-\alpha_s C_F/\pi$ of the 
$\ln \tau$ term in \cite{Moult:2016fqy}.

The single logarithm arises from the dimensionally regulated 
convolution integrals, which are logarithmically 
divergent in $d=4$. For resummation we would like to define renormalized
hard, soft and (anti) collinear functions and take the limit
$\epsilon \to 0$ before performing the convolution integrals.
Proceeding in this way, however, the convolution integrals are
ill-defined. In the following sections we explain
how this problem can be solved. 

To prepare this discussion we make the following observation. 
The integrand of the two-dimensional convolution in $\omega, \omega'$
in \eqref{eq:AtypesqTree} in the limit of 
large $\omega, \omega'$ (where the divergence arises) is given by
\begin{eqnarray} 
\label{eq:AtypesqTreeAsy}
&&\frac{\alpha_s}{4\pi}\,\frac{2 C_F}{Q}\, f(\epsilon)\int dl_+dl_-\,
\delta^+(M^2_R-Ql_+ )\delta^+(M^2_L-Ql_-)
\delta(l_- ) \,\theta(l_+ )\, \,
\nonumber\\ 
&& \hspace*{-0cm} \times\frac{\delta(\omega-\omega')}{\omega\omega'}\,
\nonumber  \,\frac{1}{e^{-\epsilon\gamma_E}\Gamma(1-\epsilon)}
\,
\omega \left(\frac{l_+ \omega}{\mu^2}\right)^{\!-\epsilon}
\nonumber\\ 
&&= \,\frac{\alpha_s}{4\pi}\,\frac{2 C_F}{Q^2}\,\delta^+(M^2_L)\, 
\frac{\delta(\omega-\omega')}{\omega\omega'}\,
\frac{f(\epsilon)}{e^{-\epsilon\gamma_E}\Gamma(1-\epsilon)}
\,
\omega \left(\frac{M_R^2\omega}{Q\mu^2}\right)^{\!-\epsilon}.
\end{eqnarray}
In the B-type term \eqref{eq:BtypeTree}, the divergence of the 
convolution arises from small momentum fraction. The 
integrand of the $r, r'$ integrals in the limit of small $r$ 
reads 
\begin{eqnarray}\label{eq:BtypeTreeAsy}
&&\frac{\alpha_s}{4\pi}
\,\frac{2 C_F}{Q^2 } f(\epsilon)\int dl_+dl_-\,
\,\delta^+(M^2_L-Ql_-) \delta(l_+)\delta(l_-)
\nonumber\\
&&\times\,\frac{\delta(r -r' )}{r r'} 
\frac{1}{\Gamma(1-\epsilon)}\,r\,
\bigg(\frac{(M^2_R-Ql_+)\, {r} }{\mu^2e^{\gamma_E}} 
\bigg)^{\!-\epsilon}
\nonumber\\ 
&&= \frac{\alpha_s}{4\pi}
\,\frac{2 C_F}{Q^2 }\,\delta^+(M^2_L)\,
\frac{\delta(r -r' )}{r r'} 
\frac{ f(\epsilon)}{\Gamma(1-\epsilon)}\,r\,
\bigg(\frac{M^2_Rr }{\mu^2e^{\gamma_E}} 
\bigg)^{\!-\epsilon}\,,
\end{eqnarray}
which is identical to the previous expression provided we 
identify $r=\omega/Q$, $r'=\omega'/Q$. A similar agreement 
holds for the identical soft anti-quark contribution to the 
A-type term and the $r\to 1$ limit of the B-type 
term \eqref{eq:BtypeTree}. From the heuristic discussion of 
Sec.~\ref{sec:heuristic} it is evident that this agreement is 
not a coincidence, but holds to all orders in the expansion 
in $\alpha_s$.

\section{Endpoint factorization}
\label{sec:refactorization}

Eqs.~\eqref{eq:Atypefactformula}, \eqref{eq:Btypefactformula} 
provide factorized expressions for the two-hemisphere 
mass distribution in the two-jet limit 
in the gluon-thrust region from 
which the NLP logarithms can be computed 
to any logarithmic accuracy order by order in $\alpha_s$ 
by evaluating the convolutions of the $d$-dimensional 
hard, (anti-)collinear and soft function. The limit 
$\epsilon\to 0$ must be taken at the end, because the 
convolutions in the A-type term are divergent when 
$\omega$ and $\omega^\prime$ both tend to infinity. In the B-type 
term this happens when $r$ and $r^\prime$ both approach 0 or 1.

This form is not quite sufficient to sum the logarithms 
to all orders in $\alpha_s$. To apply the standard SCET 
renormalization-group method to the hard, (anti-)collinear and 
soft factors, one must renormalize them and take the  
limit $\epsilon\to 0$ before the convolution. In this section 
we employ the coincidence of the integrands in certain 
asymptotic limits, which should hold on physics grounds 
and has been demonstrated above at lowest order, to factorize 
and rearrange the endpoint contributions such that the 
convolution integrals are finite. Similar methods to derive 
endpoint factorization at the level of factorization formulas 
with functions defined by matrix elements of operators have 
been derived before only for exclusive $B$ decays to P-wave 
charmonia \cite{Beneke:2008pi}, which involves SCET and 
non-relativistic QCD, and Higgs decay to two photons 
\cite{Liu:2020wbn}, which is a SCET$_{\rm II}$ process. 
These are amplitude-factorization problems. The present 
case of gluon thrust is somewhat different as it is a 
SCET$_{\rm I}$ problem, and involves inclusive collinear and 
soft functions defined at cross-section level, similar 
to \cite{Beneke:2020ibj}. Nevertheless, the main mechanism 
that achieves endpoint factorization is similar to the 
above-mentioned cases.  
Interestingly, it partly involves functions, which are the 
non-abelian generalization of some that appear in 
$H\to\gamma\gamma$. 

\subsection{Soft-collinear limit of the B1 matching 
coefficients}

An important ingredient in this discussion 
is the factorization property of the coefficient function of 
$q\bar q g$ SCET B1 operators, when the light-cone momentum 
fraction carried by the quark or anti-quark becomes small. 
The endpoint divergence in the B-type contribution originates 
from the singular behaviour of the $C_i^{\rm B1}$ matching 
coefficients in this limit, which in turn arises because the 
intermediate quark or anti-quark goes on-shell, see 
Fig.~\ref{fig:QCD}. 

\begin{figure}
\begin{center}
\includegraphics[width=0.7\textwidth]{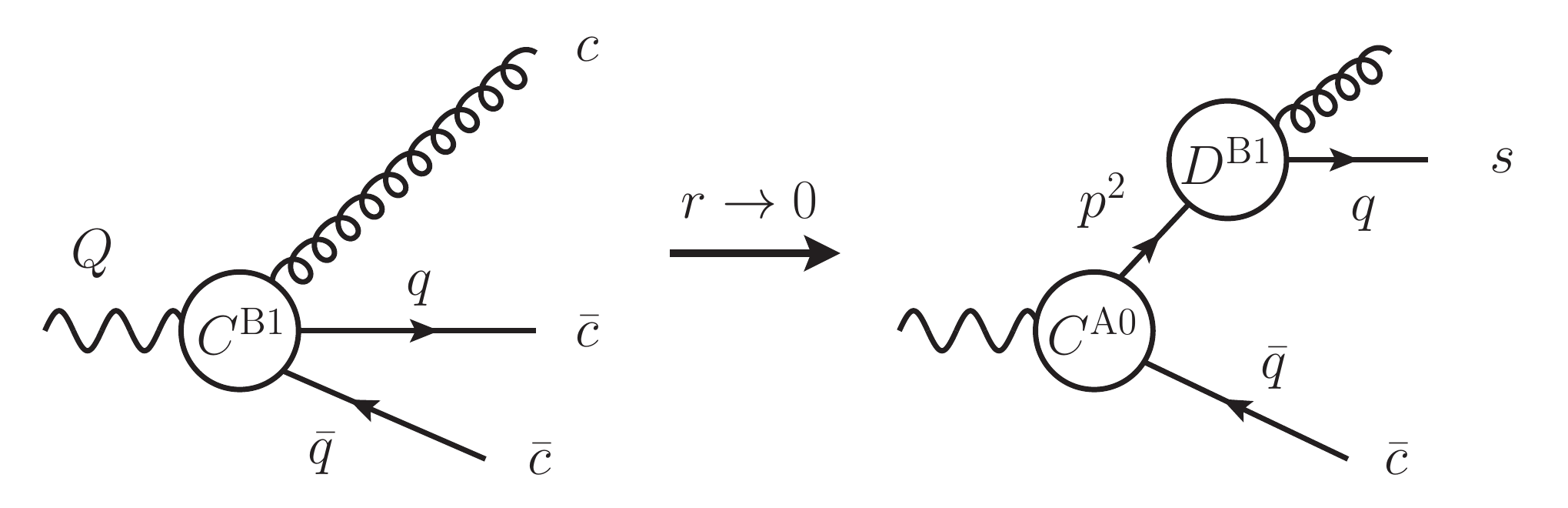}
\caption{
Graphical representation of the factorization \eqref{eq:B1fact1} 
of the B1 matching coefficient in the soft-quark limit.
\label{fig:Cb1refact}}
\end{center}
\end{figure}

By its original definition, the matching coefficient is a 
single-scale, hard function. For $r\to 0$ (soft quark) 
and $r\to 1$ (soft anti-quark), it becomes a two-scale object, 
which can itself be factorized according to 
\cite{Beneke:2020ibj}
\begin{align}
& \hskip0.5cm\displaystyle
C_1^{\rm B1}(Q^2,r) = C^{\rm A0}(Q^2)\times 
\frac{D^{\rm B1}(r Q^2)}{r} 
+\mathcal{O}(r^0)\,,
\label{eq:B1fact1}
\end{align}
where the two scales $Q^2$ and $r Q^2$ are now separated 
into the hard matching coefficient of the leading-power 
A0 SCET current, and a new coefficient $D^{\rm B1}(r Q^2)$, 
which depends only on the endpoint-scale $\sqrt{r} Q$. 
We recall that $C_1^{\rm B1}(Q^2,r)$ is regular in the 
soft anti-quark limit  $r\to 1$. An analogous factorization 
holds for the second B1 operator in the limit $r\to 1$:
\begin{align}
& \hskip0.5cm\displaystyle
C_2^{\rm B1}(Q^2,r) = -C^{\rm A0}(Q^2)\times 
\frac{D^{\rm B1}(\bar{r} Q^2)}{\bar r} +\mathcal{O}(\bar{r}^0)\,.
\label{eq:B1fact2}
\end{align}
Since the two cases are related by $r\leftrightarrow \bar{r}$ 
(plus a global minus sign), the following is phrased for the 
soft-quark limit only. A graphical representation of 
\eqref{eq:B1fact1} is shown in Fig.~\ref{fig:Cb1refact}. 

For $r\to 0$ the outgoing quark represented by 
$\bar\chi_{\bar c}$ field in the second line of 
\eqref{eq:hardmatching} becomes soft-anticollinear. The 
soft-collinear coefficient $D^{\rm B1}(p^2)$, which appears in 
\eqref{eq:B1fact1}, \eqref{eq:B1fact2}, can be defined by matching 
the time-ordered product \cite{Beneke:2020ibj} 
\begin{equation}
\langle g^a_{c}(p_{c}) 
q_{\overline{sc}}(p_{\overline{sc}})| 
\int d^4x \,T\big\{\bar{\chi}_{c}(0),\mathcal{L}_{\xi q}(x) 
\big\} \,|0\rangle=g_s
 \bar{u}(p_{\overline{sc}}) t^a\slashed{\varepsilon}_{c\perp}(p_c) 
\frac{i n_+ p_c}{p^2} \frac{\slashed n_-}{2} \,D^{\rm B1}(p^2)\,,
\label{eq:DB1def}
\end{equation}
with $p^2=(p_{c}+p_{\overline{sc}})^2=r Q^2$ and
\begin{align}
\mathcal{L}_{\xi q}(x)=\bar{q}_{\overline{sc}}(x_{-})
\slashed{\mathcal{A}}_{c\perp}(x)\chi_{c}(x)+\mbox{h.c.}
\end{align}
the leading soft-quark Lagrangian \cite{Beneke:2002ph,Beneke:2002ni} 
with $\bar{q}_s$ replaced by $\bar{q}_{\overline{sc}}$. 
This relation becomes evident by comparing the second diagram in 
Fig.~\ref{fig:SCET} to the right-hand side in 
Fig.~\ref{fig:Cb1refact}. From its definition in terms of a 
time-ordered product it is clear that the matching coefficient 
$D^{\rm B1}(p^2)$ is a universal function that will appear 
in different processes involving soft quark emission.

The leading double-logarithmic resummation of 
$D^{\rm B1}(p^2)$ to all orders in $\alpha_s$ has been derived  
in \cite{Beneke:2020ibj}. It exhibits the all-order 
$(C_A-C_F)^n$ colour coefficient, which seems to be characteristic 
for soft quark 
emission \cite{Vogt:2010cv,Moult:2019uhz,Beneke:2020ibj}. 
Interestingly, the same coefficient 
enters the $ggH$ amplitude with a bottom-quark loop and was 
recently computed at the two-loop level \cite{Liu:2021mac}, 
together with its one-loop evolution kernel, which was obtained 
from renormalization-group invariance of the full $ggH$ amplitude. 

The $D^{\rm B1}(p^2)$ coefficient as well 
as its evolution equation can also be obtained from the 
corresponding B1 operator coefficients and anomalous dimension 
by taking the limit $r\to 0$ and provide the relevant one-loop 
expressions. Since the anti-collinear part of 
the B1 operator has the same field content as the operator 
relevant to light-cone distribution amplitudes, the extraction 
of the anomalous dimension is similar to the derivation 
of the asymptotic kernel for the QED light-meson light-cone 
distribution amplitude \cite{Beneke:2021pkl}, see 
also \cite{Liu:2020wbn}. We refer to App.~\ref{sec:DB1} for 
this derivation, which results in 
\be
D^{\rm B1}(p^2) = 1 + \frac{\as}{4\pi}\,
\big(C_F- C_A \big)
\bigg(\frac{2}{\eps^2} -1 - \frac{\pi^2}{6} \bigg)
\left(\frac{\mu^2}{-p^2-i\varepsilon}\right)^{\!\epsilon}+\ord(\as^2).
\label{eq:DB11loop}
\ee
\be
\frac{d}{d\ln\mu}D^{\rm B1}(p^2) = \int_0^\infty d\hat p^2
\,\gamma_D(\hat p^2,p^2)D^{\rm B1}(\hat p^2)\,,
\label{eq:DBQrge}
\ee
with 
\bea
\gamma_D(\hat p^2,p^2) &=& 
\frac{\alpha_s(C_F-C_A)}{\pi}\,\delta(\hat p^2-p^2)
\ln\left(\frac{\mu^2}{-p^2-i\varepsilon}\right)\nn\\
  && + \,\frac{\alpha_s}{\pi} \left(\frac{C_A}{2}-C_F\right)
p^2 \left[\frac{\theta(\hat p^2 -p^2)}
{\hat p^2 (\hat p^2 -p^2)}
+ \frac{\theta(p^2-\hat p^2 )}{p^2(p^2-\hat p^2 )}\right]_+  
\,,\quad
\label{eq:DB1andim}
\eea
in agreement with \cite{Liu:2021mac}.

When the gluon field is replaced by a photon, 
the corresponding abelian version of $D^{\rm B1}(p^2)$ 
is identical to the jet function, which appears at 
LP in the factorization of the 
radiative semi-leptonic $B$ decay $B\to\gamma\ell\nu$ 
\cite{Bosch:2003fc} and in the NLP endpoint factorization of the 
photonic Higgs decay $H\to\gamma\gamma$ amplitude through 
a bottom-quark loop \cite{Liu:2020wbn}. In the abelian 
case, the two-loop evolution has been inferred from 
renormalization-group consistency of the $B\to\gamma\ell\nu$ 
observable \cite{Liu:2020ydl}. The direct computation of 
the one-loop evolution kernel of this jet function can be 
found in \cite{Bodwin:2021epw}.

\subsection{Endpoint factorization consistency conditions}

As discussed in Sec.~\ref{sec:heuristic} and checked explicitly above 
at $\mathcal{O}(\alpha_s)$, we expect the integrand of the 
A- and B-type terms of the factorization formula to have 
identical asymptotic limits to all orders, which is a 
prerequisite for endpoint factorization. More precisely, 
the limit of anti-collinear momentum component $\nm\cdot p_{\bar c} = r Q, r'Q \to 0$ in the B-type term, 
matches the limit $\nm\cdot k = \omega,\omega' 
\to \infty$ of the corresponding soft momentum component in 
the A-type term.\footnote{There is a similar matching 
of the limit $(1-r)Q,(1-r')Q\to 0$ with the 
soft {\em anti}-quark contribution to the A-type term. 
This contribution can be treated as below, after 
change variables from $r^{(\prime)}\to 
1-r^{(\prime)}$.}

We start from the expressions \eqref{eq:AtypeLaplace}, 
\eqref{eq:BtypeLaplace} in Laplace space, in which the 
integrand 
of the $\omega,\omega'$ and $r,r'$ integrals involves 
no further integrals and the consistency relations are 
algebraic. The coincidence of asymptotic limits implies 
that the following must hold:
\begin{eqnarray}
&&\frac{2C_F}{Q^2} \,f(\epsilon) \,
C^{\rm B1*}_{1}(Q^2,r')C^{\rm B1}_{1}(Q^2,r) \,
\widetilde{\mathcal{J}}_{\bar{c}}^{q\bar{q}(8)}(s_R,r,r^\prime)\,
\widetilde{\mathcal{J}}_{c}^{(g)}(s_L)\,\widetilde{S}^{(g)}(s_R,s_L)\Big|_{
r^{(\prime)}=\frac{\omega^{(\prime)}}{Q}\to \,0}
\nonumber\\
&& \hspace*{0.5cm} 
= \frac{2C_F}{Q} \,f(\epsilon)\,|C^{\rm A0}(Q^2)|^2 \,
\widetilde{\mathcal{J}}_{\bar c}^{(\bar q)}(s_R)
\,\times\,
\bigg\{\,\widetilde{\mathcal{J}}_{c}(s_{L},\omega,\omega')
\,\widetilde{S}_{\rm NLP}(s_R,s_L,\omega,\omega')
\nonumber\\[0.1cm]
&& \hspace*{1cm} +\;\,
\widetilde{\!\!\widehat{\mathcal{J}}}_{\!c}(s_{L},
\omega,\omega')
\,\,\widetilde{\!\widehat{S}}_{\rm NLP}(s_R,s_L,\omega,\omega')\,
\bigg\}\Big|_{\omega,\omega^{\prime} \to\infty}\,.
\label{eq:reff1}
\end{eqnarray} 
The left-hand side of this equation 
represents the leading asymptotics of the 
B-type term \eqref{eq:BtypeLaplace} 
for $r, r' \to 0$, already accounting for the fact that this arises 
only from the square of the amplitude of the first B1 operator, 
since the coefficient of the second is non-singular as $r\to 0$. 
The right-hand side is one half of the total A-type term, which 
we can identify with the soft quark contribution. 
An analogous equation with $r, r' \to \bar r,\bar{r}^\prime$ 
and $C^{\rm B1}_{1} \to C^{\rm B1}_{2}$ applies to the 
limit $r,r'\to 1$. The right-hand side of \eqref{eq:reff1} 
contains two terms in the curly brackets, which 
correspond to different colour structure in the jet 
and soft function, \eqref{eq:defjetNLP} and 
\eqref{eq:NLPsoftfndef}, respectively. We shall now show that 
endpoint factorization consistency implies that the 
hatted term  in \eqref{eq:reff1} is subleading as 
$\omega,\omega^\prime\to \infty$, and can be dropped.

\begin{figure}[t]
\begin{center}
\includegraphics[width=0.8\textwidth]{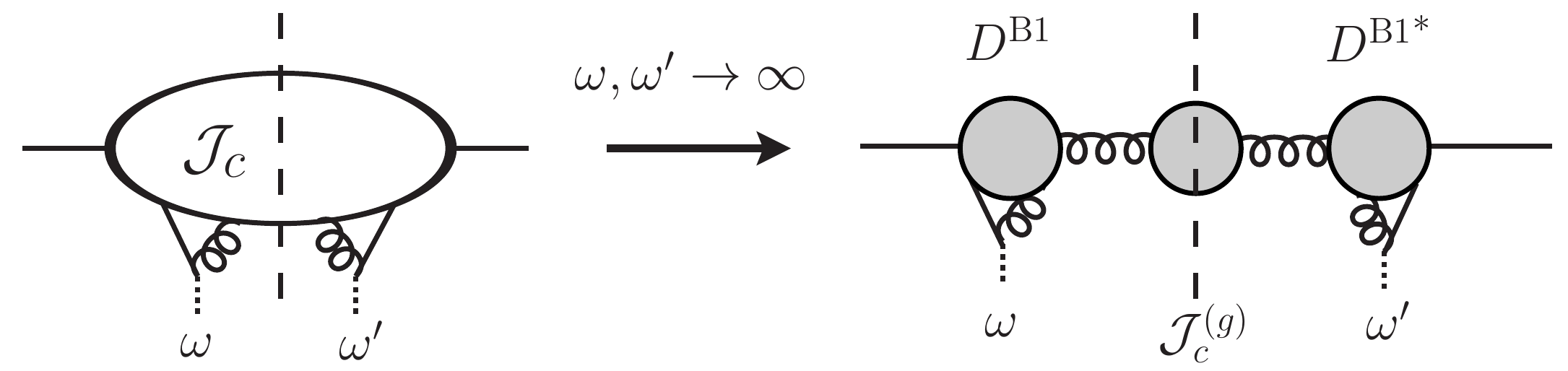}
\caption{Factorization of the radiative collinear function 
$\mathcal{J}_{c}(p^2,\omega,\omega')$, which appears in the 
A-type term, in the limit $\omega, \omega^\prime\to\infty$.
\label{fig:jet1PR}}
\end{center}
\end{figure}

We recall from \eqref{eq:defjetNLP} that 
$\mathcal{J}_{c}(p^2,\omega,\omega')$ is an inclusive 
radiative jet function of the non-local, time-ordered product 
operator defined in \eqref{eq:nonlocaljetfield}. 
Graphically, we may represent it by the diagram to the 
left of the arrow in Fig.~\ref{fig:jet1PR}, 
where the dashed external lines to the left and right 
of the cut denotes the {\em soft} momentum of the 
composite field $\not\hspace*{-1.5mm}\mathcal{A}_{\perp c}\chi_c$ 
in the time-ordered product, which is carried away by the 
soft quark. We now claim that the leading asymptotic 
behaviour proportional to $1/(\omega\omega^\prime)$ for 
large $\omega,\omega^\prime$ requires that the diagram 
is ``one-collinear-particle reducible'', as displayed to 
the right of the arrow in Fig.~\ref{fig:jet1PR}. Then 
colour conservation implies that the collinear line must 
be a colour-octet, that is a gluon, and hence only the 
unhatted colour term \eqref{eq:defjetNLP}, which defines 
$\mathcal{J}_{c}(p^2,\omega,\omega')$ contributes to 
the asymptotic behaviour, as was to be shown.

\begin{figure}[t]
\begin{center}
\includegraphics[width=0.3\textwidth]{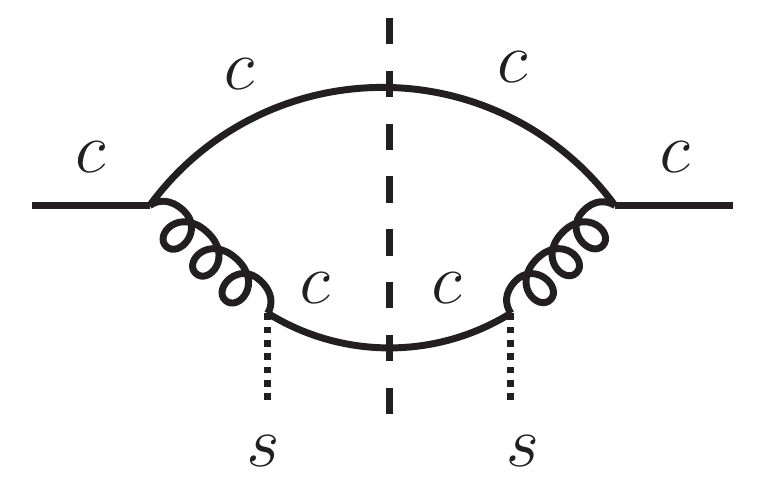}
\caption{Example of a one-collinear-particle reducible diagram, 
which does not contribute to the large-$\omega^{(\prime)}$ 
behaviour of $\mathcal{J}_{c}(p^2,\omega,\omega')$.
\label{fig:jet1PI}}
\end{center}
\end{figure}

Suppose the diagram was not one-collinear-particle reducible,
such as the one-loop example displayed in Fig.~\ref{fig:jet1PI}.
Endpoint factorization consistency requires that the 
large $\omega,\omega^\prime$ must match a corresponding 
B-type contribution, obtained by making the dashed lines 
representing external soft quark anti-collinear. This 
turns internal collinear propagators into hard propagators, 
and the entire diagram into a contribution to the matching 
coefficient of a hard operator with external collinear 
and anti-collinear fields. The key observation is that 
if the diagram was not one-collinear-particle reducible, 
the corresponding hard operator would contain more than 
one external collinear field, in contradiction with the 
B1 operators available at NLP, which contain only a single 
collinear gluon field, see \eqref{eq:hardmatching}. 
Consistency therefore requires that the asymptotic behaviour 
takes the diagrammatic form shown in Fig.~\ref{fig:jet1PR}. 
This proves not only that the hatted jet function is 
irrelevant for endpoint factorization, but further 
that $\mathcal{J}_{c}(p^2,\omega,\omega')$ 
factorizes into the product\footnote{The collinear 
scale on the left-hand side is $\np p \,\omega^{(\prime)} 
=Q  \omega^{(\prime)}$, but we suppressed the factor of 
$Q$ in the respective arguments of 
$\mathcal{J}_{c}\left(p^2,\omega,\omega'\right)$.}
\begin{eqnarray}
\mbox{(I)} &\hskip0.5cm 
\mathcal{J}_{c}\left(p^2,\omega,\omega'\right) = 
\mathcal{J}_{c}^{(g)}(p^2) \,
\displaystyle \frac{D^{\rm B1}(\omega Q)}{\omega}
  \frac{{D^{\rm B1}}^*(\omega^\prime Q)}{\omega^\prime} 
+\mathcal{O}\!\left(\frac{1}{\omega^{(\prime)}}\right)\,,
\label{eq:EPfactI}
\end{eqnarray}
where the function $D^{\rm B1}(p^2)$ is the same as the one  
that appears in the factorization of the hard B1 operator 
coefficient \eqref{eq:B1fact1}. 

Eq.~\eqref{eq:EPfactI} is the first endpoint factorization 
consistency condition. Inserting the Laplace-transformed version 
of this relation together with 
\eqref{eq:B1fact1} for the B1 hard function 
into \eqref{eq:reff1} immediately implies 
a similar consistency relation for the NLP soft quark 
function, 
\begin{eqnarray}
\mbox{(II)} & \hskip-1.5cm 
Q\,\widetilde{\mathcal{J}}^{(\bar{q})}_{\bar{c}}(s_R)\,
\widetilde{S}_{\rm NLP} \left(s_R,s_L,\omega,\omega'\right)\Big|_{\omega^{(\prime)} \to\infty}
\nonumber\\
& \hskip2.5cm \displaystyle
= 
\widetilde{\mathcal{J}}^{q\bar q(8)}_{\bar{c}}\left(s_R,r,r'\right)
\widetilde{S}^{(g)}(s_R,s_L)\Big|_{r^{(\prime)}=\omega^{(\prime)}/Q\to 0}\,.
\end{eqnarray}
The same identity holds 
with $r,r^\prime\to \bar{r},\bar{r}^\prime$. Relations (I) and (II) formalize the heuristic picture developed 
in Sec.~\ref{sec:heuristic}. For large $\omega$, the soft 
quark field in $S_{\rm NLP}$ becomes anti-collinear, 
$\bar{q}_s Y_{n_-} \to \bar{\chi}_{\bar c}$, hence it moves 
from $S_{\rm NLP}$ to the anti-collinear function 
$\mathcal{J}^{q\bar q(8)}_{\bar{c}}$. Removing  $\bar{q}_s Y_{n_-}$ 
from  $S_{\rm NLP}$ leaves the leading-power soft function 
$S^{(g)}$, while adding it as $\bar{\chi}_{\bar c}$ to 
$\mathcal{J}^{(\bar{q})}_{\bar{c}}$ turns the anti-quark 
jet function into $\mathcal{J}^{q\bar q(8)}_{\bar{c}}$. Overall, 
this results in $S_{\rm NLP} \mathcal{J}^{(\bar{q})}_{\bar{c}} \to 
S^{(g)}\mathcal{J}^{q\bar q(8)}_{\bar{c}}$, which is relation 
(II). At the same time, the quark fields in the A-type 
collinear function \eqref{eq:nonlocaljetfield} become highly 
off-shell, which removes them from $\mathcal{J}_{c}(p^2,\omega,\omega^\prime)$, leaving 
only the collinear gluon, and consequently $C^{\rm A0}$ turns 
into $C^{\rm B1}$. Thus, $|C^{\rm A0}|^2 \mathcal{J}_{c} \to 
|C^{\rm B1}|^2 \mathcal{J}_{c}^{(g)}$, which is relation (I).

\subsection{Endpoint factorization formula}
\label{sec:endpointfact}

We are now in the position to derive the endpoint-finite 
factorization formula. For its concise formulation, we employ 
the double-bracket notation introduced in \cite{Liu:2019oav} 
to denote the asymptotic behaviours of the various functions. 
The precise definitions are as follows: In functions of 
$\omega, \omega^\prime$, rescale $\omega \to \kappa 
\omega$, $\omega^\prime \to \kappa 
\omega^\prime$ and take $\kappa\to \infty$. Then 
\begin{eqnarray}
&& \llbracket 
S_{\rm NLP} \left(l_+,l_-,\omega,\omega'\right)
\rrbracket \equiv S_{\rm NLP} \left(l_+,l_-,\omega,\omega'\right)|_{
\mathcal{O}(\kappa^0)}, \label{eq:Sasym}\\
&&\llbracket \mathcal{J}_{c}(p^2,\omega,\omega^\prime)
\rrbracket \equiv \mathcal{J}_{c}(p^2,\omega,\omega^\prime)|_{
\mathcal{O}(\kappa^{-2})}\,,
\end{eqnarray}
where $\delta(\omega-\omega')$ counts as $\kappa^{-1}$. The right-hand 
side of the previous equation equals the right-hand side of 
the consistency relation (I).  
Similarly, in functions of 
$r, r^\prime$, rescale $r \to r \kappa$, $r^\prime \to 
r^\prime \kappa$ and take $\kappa\to 0$, {\em or} the 
corresponding rescaling is applied to $\bar r$, $\bar{r}^\prime$. 
Which of the two is meant, will be indicated by the subscript 
0 or 1 on the double bracket. Then 
\begin{eqnarray}
&& \llbracket C^{\rm B1}_1(Q^2,r) \rrbracket_0 \equiv 
C^{\rm B1}_1(Q^2,r)|_{\mathcal{O}(\kappa^{-1})}\,,
\\[0.1cm]
&& \llbracket C^{\rm B1}_2(Q^2,r) \rrbracket_1 \equiv 
C^{\rm B1}_2(Q^2,r)|_{\mathcal{O}(\kappa^{-1})}\,,
\end{eqnarray}
while $\llbracket C^{\rm B1}_1(Q^2,r) \rrbracket_1 = 
\llbracket C^{\rm B1}_2(Q^2,r) \rrbracket_0 = 0$. With this 
definition the right-hand sides of these equations are 
given by \eqref{eq:B1fact1} and \eqref{eq:B1fact2}, respectively. 
Finally, we have 
\begin{equation}
\llbracket \mathcal{J}^{q\bar q(8)}_{\bar{c}}(p^2,r,r^\prime)
\rrbracket_0 = 
\llbracket \mathcal{J}^{q\bar q(8)}_{\bar{c}}(p^2,r,r^\prime)\rrbracket_1 
\equiv \mathcal{J}^{q\bar q(8)}_{\bar{c}}(p^2,r,r^\prime)|_{
\mathcal{O}(\kappa^0)}
\end{equation}
due to the symmetry $r^{(\prime)}\leftrightarrow \bar{r}^{(\prime)}$.
In the $d$-dimensional expressions there are 
$\epsilon$-dependent powers of $\kappa$, which turn into 
logarithms in the renormalized functions. Thus 
$\epsilon$ counts as an 
infinitesimal variable for the purpose of endpoint 
$\kappa$ power-counting.

\begin{figure}
\begin{center}
\includegraphics[width=0.85\textwidth]{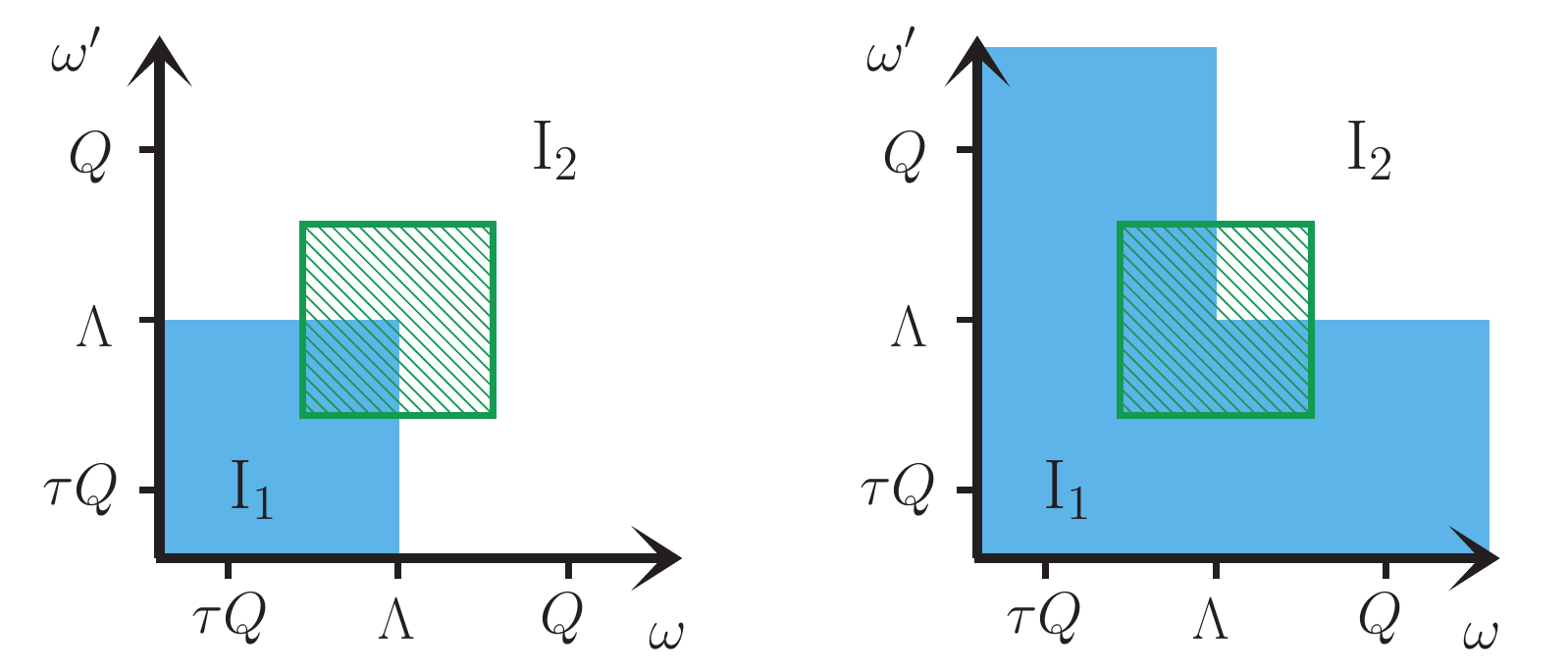}
\caption{\label{fig:overlap}
Version (1) [left] and (2) [right] of the split of 
\eqref{eq:scaleless_integral} into $I_1+I_2$ according to 
the correspondingly indicated regions in the 
$\omega-\omega'$ plane as described in the text 
below \eqref{eq:scaleless_integral}. In the overlap 
region in green the asymptotic behaviour of the A- 
and B-type term must agree.}
\end{center}
\end{figure}

To implement the rearrangement of endpoint-singular terms 
we start with the scaleless integral
\begin{eqnarray}
&&\frac{2 C_F}{Q} \,f(\epsilon)\,|C^{\rm A0}(Q^2)|^2 \widetilde{\mathcal{J}}^{(\bar q)}_{\bar{c}}(s_R) 
\widetilde{\mathcal{J}}_{c}^{(g)}(s_L) \nonumber\\
&&\times \int_0^\infty d\omega d\omega'\,
\frac{D^{\rm B1}(\omega Q)}{\omega}
\frac{D^{\rm B1*}(\omega'Q)}{\omega'} 
\left \llbracket \widetilde{S}_{\rm NLP}(s_R,s_L,\omega,\omega')\right 
\rrbracket,\qquad
\label{eq:scaleless_integral}
\end{eqnarray}
which vanishes in $d$ dimensions. We split this 
integral in two terms $I_{1,2}$, which can be done in two 
ways, illustrated in Fig.~\eqref{fig:overlap}: 
(1) $\omega$ {\em and} $\omega'$ smaller than 
an endpoint factorization parameter $\Lambda$ 
(integral $I_1$), and the complement region $I_2$. 
(2)  $\omega$ {\em or} $\omega'$ smaller $\Lambda$ 
and the complement region. Since the mixed regions 
where one variable is larger and the other smaller than 
$\Lambda$ are not endpoint-singular, both ways could be 
employed for the following considerations, although 
the expressions take a different form. We will use 
the second version. In this case the complement 
region is $\omega,\omega^\prime > \Lambda$ and the 
double-bracket asymptotic behaviour can be used for  
functions of $\omega,\omega'$ in the A-type term.\footnote{
While in version (1), it could be used for functions 
of $r, r'$ in the B-type term with identification 
$r^{(\prime)}=\omega^{(\prime)}/Q$. The resulting 
expressions for this version 
are provided in App.~\ref{version1endpintfact}.} 
In both versions $I_1+I_2=0$. The endpoint rearrangement 
now consists of subtracting $I_1$ from the B-type term 
and $I_2$ from the A-type term. The subtracted 
expressions are now separately endpoint-finite, but 
depend on $\Lambda$. However, as long as no approximations 
are made, the $\Lambda$ dependence cancels exactly between 
the two terms.

We begin with the A-type contribution 
\eqref{eq:AtypeLaplace}, which accounts for soft-quark 
emission, to the factorization theorem and subtract from it the complement region $I_2$, $\omega,\omega^\prime > \Lambda$,  of the integral~(\ref{eq:scaleless_integral}), resulting in
\begin{eqnarray}
\frac{1}{\sigma_0}\frac{\widetilde{d\sigma}}{ds_R ds_L}|_{\rm A-type} &=& 
\frac{2C_F}{Q} \,f(\epsilon)\,|C^{\rm A0}(Q^2)|^2 \,
\widetilde{\mathcal{J}}_{\bar c}^{(\bar q)}(s_R)
\,\int_0^\infty d\omega d\omega'\,
\nonumber\\
&&\hspace*{-3cm}\times\,
\bigg\{\,\widetilde{\mathcal{J}}_{c}(s_{L},\omega,\omega')
\,\widetilde{S}_{\rm NLP}(s_R,s_L,\omega,\omega')
\nonumber\\[-0.15cm]
&&\hspace*{-2.2cm}
-\,\theta(\omega-\Lambda)\theta(\omega'-\Lambda)
\widetilde{\mathcal{J}}_{c}^{(g)}(s_L) 
\frac{D^{\rm B1}(\omega Q)}{\omega}
\frac{D^{\rm B1*}(\omega'Q)}{\omega'} 
\left \llbracket \widetilde{S}_{\rm NLP}(s_R,s_L,\omega,\omega')\right 
\rrbracket
\nonumber\\
&&\hspace*{-2.2cm}
+\;\,\widetilde{\!\!\widehat{\mathcal{J}}}_{\!c}(s_{L},
\omega,\omega') 
\,\,\widetilde{\!\widehat{S}}_{\rm NLP}(s_R,s_L,\omega,\omega')\,\bigg\}\,. \quad
\label{eq:Atype_subtracted_general}
\end{eqnarray}
For $\omega,\omega'\to\infty$, we use the endpoint 
factorization condition (I) from \eqref{eq:EPfactI} to 
rewrite the jet function in the first term of the integrand. 
In this limit it is also justified to replace the soft 
function by its asymptotic form (\ref{eq:Sasym}). Thus the 
first and second term in the integrand cancel. The third 
term is subleading in this limit, and  the convolution 
integrals over $\omega$ and $\omega'$ are now endpoint-finite, as the logarithmically divergent part of the integrand in the large $\omega$ and $\omega'$ region has been removed. 
To make this explicit, we use  \eqref{eq:EPfactI} in 
\eqref{eq:Atype_subtracted_general}, and obtain 
\begin{eqnarray}
\frac{1}{\sigma_0}\frac{\widetilde{d\sigma}}{ds_R ds_L}|_{\rm A-type} &=& 
\frac{2C_F}{Q} \,f(\epsilon)\,|C^{\rm A0}(Q^2)|^2 \,
\widetilde{\mathcal{J}}_{\bar c}^{(\bar q)}(s_R)
\,\int_0^\infty d\omega d\omega'\,
\nonumber\\
&&\hspace*{-3cm}\times\,
\bigg\{\,\widetilde{\mathcal{J}}_{c}(s_{L},\omega,\omega')
\,\widetilde{S}_{\rm NLP}(s_R,s_L,\omega,\omega')
\nonumber\\[0cm]
&&\hspace*{-2.2cm}
-\,\theta(\omega-\Lambda)\theta(\omega'-\Lambda)\,
\llbracket\widetilde{\mathcal{J}}_{c}(s_{L},\omega,\omega')
 \rrbracket
\llbracket \widetilde{S}_{\rm NLP}(s_R,s_L,\omega,\omega')
\rrbracket
\nonumber\\
&&\hspace*{-2.2cm}
+\;\,\widetilde{\!\!\widehat{\mathcal{J}}}_{\!c}(s_{L},
\omega,\omega') 
\,\,\widetilde{\!\widehat{S}}_{\rm NLP}(s_R,s_L,\omega,\omega')\,\bigg\}\,. \quad
\label{eq:Atype_subtracted_general2}
\end{eqnarray}

If we further choose $\Lambda\gg 1/s_{R},1/s_L$, we may simplify 
the previous equation to 
\begin{eqnarray}
\frac{1}{\sigma_0}\frac{\widetilde{d\sigma}}{ds_R ds_L}|_{\rm A-type} &=& 
\frac{2C_F}{Q} \,f(\epsilon)\,|C^{\rm A0}(Q^2)|^2 \,
\widetilde{\mathcal{J}}_{\bar c}^{(\bar q)}(s_R)
\,\int_0^\infty d\omega d\omega'\,
\nonumber\\
&&\hspace*{-2.5cm}\times\,
\bigg\{\,\Big[1-\theta(\omega-\Lambda)\theta(\omega'-\Lambda)\Big]\widetilde{\mathcal{J}}_{c}(s_{L},\omega,\omega')
\,\widetilde{S}_{\rm NLP}(s_R,s_L,\omega,\omega')
\nonumber\\
&&\hspace*{-1.7cm}
+\;\,\widetilde{\!\!\widehat{\mathcal{J}}}_{\!c}(s_{L},
\omega,\omega') 
\,\,\widetilde{\!\widehat{S}}_{\rm NLP}(s_R,s_L,\omega,\omega')\bigg\}\nonumber\\
&&\hspace*{-3cm}=
\,\frac{2C_F}{Q} \,f(\epsilon)\,|C^{\rm A0}(Q^2)|^2 \,
\widetilde{\mathcal{J}}_{\bar c}^{(\bar q)}(s_R)
\,\int d\omega d\omega'\,\Big[1-\theta(\omega-\Lambda)\theta(\omega'-\Lambda)\Big]
\nonumber\\[0.0cm]
&&\hspace*{-2.5cm}\times\,
\bigg\{\,\widetilde{\mathcal{J}}_{c}(s_{L},\omega,\omega')
\,\widetilde{S}_{\rm NLP}(s_R,s_L,\omega,\omega')+\,\widetilde{\!\!\widehat{\mathcal{J}}}_{\!c}(s_{L},
\omega,\omega') 
\,\,\widetilde{\!\widehat{S}}_{\rm NLP}(s_R,s_L,\omega,\omega')\bigg\}\,,\qquad
\label{eq:Atype_subtracted}
\end{eqnarray}
where the equality signs hold up to corrections 
of $\mathcal{O}(1/(s_L\Lambda),1/(s_R\Lambda))$. It is then 
understood that in evaluating \eqref{eq:Atype_subtracted}, 
terms suppressed by powers of $\Lambda$ are dropped.
This is simply \eqref{eq:AtypeLaplace} with integration 
region $\omega,\omega^\prime > \Lambda$ removed, see
Fig.~\ref{fig:overlap}. 

The remaining part $I_1$ of the integral (\ref{eq:scaleless_integral}) can now be combined with  $i=i'=1$ part of the 
B-type term (\ref{eq:BtypeLaplace}). Similarly, 
we proceed for the A-type soft-antiquark contribution and 
the  $i=i'=2$ part of the 
B-type term after using the symmetry under the exchange $r \leftrightarrow \bar{r}$. The term $i\neq i'$ is endpoint-finite so it remains unaffected. 

Beginning with the $i=i'=1$ term, we have
\begin{eqnarray}
\frac{1}{\sigma_0}\frac{\widetilde{d\sigma}}{ds_R ds_L}|_{
\scriptsize \begin{array}{l}
$\rm B--type$\\[-0.1cm] $i=i'=1$\end{array}} 
&=& 
\frac{2 C_F}{Q^2} \, f(\epsilon)\,\bigg[\, 
\widetilde{\mathcal{J}}_{c}^{(g)}(s_L)\,\widetilde{S}^{(g)}(s_R,s_L)
\nonumber\\
&&\hspace*{-1.5cm}\times\,
\int_0^1 dr dr'\,
C^{\rm B1*}_{1}(Q^2,r')C^{\rm B1}_{1}(Q^2,r) 
\,\widetilde{\mathcal{J}}_{\bar{c}}^{q\bar{q}(8)}(s_R,r,r^\prime)
\,\quad 
\nonumber \\ &&\hspace*{-2.5cm}-\,Q\,|C^{A0}(Q^2)|^2 
\widetilde{\mathcal{J}}^{(\bar q)}_{\bar{c}}(s_R) 
\widetilde{\mathcal{J}}_{c}^{(g)}(s_L) 
\int_0^\infty d\omega d\omega'  \,\big[1-\theta(\omega-\Lambda)\theta(\omega'-\Lambda)\big]
\nonumber\\ &&
\hspace*{-1.5cm}\times\,\frac{D^{\rm B1}(\omega Q)}{\omega}
\frac{D^{\rm B1*}(\omega'Q)}{\omega'} 
\left \llbracket \widetilde{S}_{\rm NLP}(s_R,s_L,\omega,\omega')\right 
\rrbracket\bigg]\,.\qquad 
\label{eq:Btype_subtracted}
\end{eqnarray}
We substitute $\omega = r Q$ and $\omega'=r'Q$ in the subtraction term, and employ (\ref{eq:B1fact1}) to express the $D^{\rm B1}$ functions in terms of the asymptotic limits of the B1 matching coefficients. Next, we make use of the endpoint factorization condition (II) to express the asymptotic soft-quark function in terms of the B-type jet function $\widetilde{\mathcal{J}}_{\bar{c}}^{q\bar{q}(8)}(s_R,r,r^\prime)$ in the asymptotic limit to obtain 
\begin{eqnarray}
\frac{1}{\sigma_0}\frac{\widetilde{d\sigma}}{ds_R ds_L}|_{
\scriptsize \begin{array}{l}
$\rm B--type$\\[-0.1cm] $i=i'=1$\end{array}} 
&=& 
\frac{2C_F}{Q^2}\,f(\epsilon) \,\widetilde{\mathcal{J}}_{c}^{(g)}(s_L)\,
\widetilde{S}^{(g)}(s_R,s_L)\, \int_0^\infty dr dr' \nonumber\\ 
&&\hspace*{-2.5cm}\times\, \bigg[ \,
\theta(1-r)\theta(1-r')\,
C^{\rm B1*}_{1}(Q^2,r')C^{\rm B1}_{1}(Q^2,r) 
\,\widetilde{\mathcal{J}}_{\bar{c}}^{q\bar{q}(8)}(s_R,r,r^\prime)
\quad 
\nonumber \\ &&\hspace*{-2cm}-
\big[1-\theta(r-\Lambda/Q)\theta(r'-\Lambda/Q)\big]\,
\nonumber \\ &&\hspace*{-2cm}\times\,
\llbracket C^{\rm B1*}_{1} (Q^2,r')\rrbracket_{0}\, 
\llbracket C^{\rm B1}_{1}(Q^2,r) \rrbracket_{0} \, \llbracket \widetilde{\mathcal{J}}_{\bar{c}}^{q\bar{q}(8)}(s_R,r,r^\prime)\rrbracket_0
\bigg]\,,
\label{eq:Btype_subtracted2}
\end{eqnarray}
where now the integrals over $r,r'$ are evaluated from $0$ to $\infty$. The convolution integrals are now convergent, but they must be evaluated after the integrands are combined.
Provided $\Lambda\ll Q$, this expression can be simplified to
\begin{eqnarray}
\frac{1}{\sigma_0}\frac{\widetilde{d\sigma}}{ds_R ds_L}|_{
\scriptsize \begin{array}{l}
$\rm B--type$\\[-0.1cm] $i=i'=1$\end{array}} 
&=& 
\frac{2C_F}{Q^2}\,f(\epsilon) \,\widetilde{\mathcal{J}}_{c}^{(g)}(s_L)\,
\widetilde{S}^{(g)}(s_R,s_L)
\nonumber\\ 
&&\hspace*{-3cm}\times\,\bigg\{\int_0^1 dr dr' \,\big[1-\theta(\Lambda/Q-r)\theta(\Lambda/Q-r')\big]\,
C^{\rm B1*}_{1}(Q^2,r')C^{\rm B1}_{1}(Q^2,r) 
\,\widetilde{\mathcal{J}}_{\bar{c}}^{q\bar{q}(8)}(s_R,r,r^\prime)
\quad 
\nonumber \\ &&\hspace*{-2.5cm}-
\int_0^\infty dr dr' \,
\big[\theta(r-\Lambda/Q)\theta(\Lambda/Q-r')+
\theta(\Lambda/Q-r)\theta(r'-\Lambda/Q)\big]\,
\nonumber \\ &&\hspace*{-2cm}\times\,
\llbracket C^{\rm B1*}_{1} (Q^2,r')\rrbracket_{0}\, 
\llbracket C^{\rm B1}_{1}(Q^2,r) \rrbracket_{0} \, \llbracket \widetilde{\mathcal{J}}_{\bar{c}}^{q\bar{q}(8)}(s_R,r,r^\prime)\rrbracket_0\,
\bigg\}\,,
\label{eq:Btype_subtracted3}
\end{eqnarray}
up to corrections of $\mathcal{O}(\Lambda/Q)$. 
The first integral in the  curly brackets is 
simply \eqref{eq:BtypeLaplace} with integration 
region $r,r^\prime < \Lambda/Q$ removed, see
Fig.~\ref{fig:overlap}, which is endpoint-finite. 
The second integral is a finite left-over from the 
subtraction term. 

 For completeness, we provide the term $i=i'=2$. In the bare factorization formula (\ref{eq:BtypeLaplace})  we change integration variables to $\bar{r}, \bar{r}'$ to map the singular point $r=r'=1$ to $\bar{r}=\bar{r}'=0$. The subtraction term is then obtained after mapping $\omega =\bar{ r} Q$ and $\omega'=\bar{r}'Q$ and following the same steps as before we find
\begin{eqnarray}
\frac{1}{\sigma_0}\frac{\widetilde{d\sigma}}{ds_R ds_L}|_{
\scriptsize \begin{array}{l}
$\rm B--type$\\[-0.1cm] $i=i'=2$\end{array}} 
&=& 
\frac{2C_F}{Q^2}\, f(\epsilon)\,\widetilde{\mathcal{J}}_{c}^{(g)}(s_L)\,\widetilde{S}^{(g)}(s_R,s_L)\, \int_0^\infty d\bar{r} d\bar{r}' \nonumber\\ 
&&\hspace*{-2.5cm}\times\, \bigg[ \,
\theta(1-\bar r)\theta(1-\bar{r}')\,
C^{\rm B1*}_{2}(Q^2,r')C^{\rm B1}_{2}(Q^2,r) 
\,\widetilde{\mathcal{J}}_{\bar{c}}^{q\bar{q}(8)}(s_R,r,r^\prime)
\quad 
\nonumber \\ &&\hspace*{-2cm}-
\big[1-\theta(\bar r-\Lambda/Q)\theta(\bar{r}'-\Lambda/Q)\big]\,
\nonumber \\ &&\hspace*{-2cm}\times\,
\llbracket C^{\rm B1*}_{2} (Q^2,r')\rrbracket_{1}\, 
\llbracket C^{\rm B1}_{2}(Q^2,r) \rrbracket_{1} \, \llbracket \widetilde{\mathcal{J}}_{\bar{c}}^{q\bar{q}(8)}(s_R,r,r^\prime)\rrbracket_1
\bigg]\,,
\label{eq:Btype_subtractedii2}
\end{eqnarray}
which coincides with \eqref{eq:Btype_subtracted2} up to 
the expected substitutions. Similar simplifications 
hold when $\Lambda\ll Q$ is assumed.
The total endpoint-subtracted B-type is the sum of \eqref{eq:Btype_subtracted2}, \eqref{eq:Btype_subtractedii2} and  
the mixed term $i\not=i'$ in (\ref{eq:B1fact1}). 

By construction, the dependence on $\Lambda$ cancels 
in the sum of the A- and B-type contributions. In 
App.~\ref{version1endpintfact} we provide the 
corresponding version of the 
endpoint-rearranged factorization formulas, when 
\eqref{eq:scaleless_integral} is split according to the 
first version mentioned below this equation, that is, 
for $\omega$ {\em and} $\omega'$ smaller than 
$\Lambda$. 


\section{Renormalization-group equations}
\label{sec:consistency}

In this section we collect the renormalization-group functions, 
which are known for some of the hard, soft and (anti-) collinear 
functions of the two terms in the factorization formula, 
and infer the others from renormalization-group consistency, 
namely the condition that the renormalized {\em integrands} of the 
A- and B-type term must be separately independent of the 
scale $\mu$. The endpoint-factorization conditions then 
provide further relations between the anomalous dimensions 
in the respective large-$\omega$ and small-$r$ limits. We then 
obtain the solutions to leading-logarithmic (LL) accuracy, 
which includes running-coupling effects.

\subsection{RGEs for the A-type functions}
\label{sec:AtypeRGE}

In the A-type term, the hard function and the anti-collinear function are LP objects. Their evolution is therefore well-known. The evolution of the two NLP objects, the collinear function and the soft function, is currently unknown. We cannot reconstruct both RGEs from just consistency arguments since we miss two anomalous dimensions. 

Instead, we will use the endpoint factorization condition (I) from 
\eqref{eq:EPfactI} to derive the evolution of the collinear 
function $\mathcal{J}_c(p^2,\omega,\omega')$ for large 
$\omega,\omega'$ from the known RGEs for the LP gluon jet function 
and the $D^{\rm B1}$ coefficient. Through consistency, we can then 
derive the asymptotic evolution of the soft function. This allows 
us to resum the A-type functions in the endpoint region, which 
is sufficient for LL accuracy. 

\paragraph{Hard function}
The evolution of the LP hard matching coefficient $C^{\rm A0}$ is well known \cite{Becher:2006mr}. It obeys the RGE
\begin{align}
\frac{ d}{ d\ln\mu}C^{\rm A0}\klammer{Q^2,\mu^2}=\left[C_F\gamma_{\text{cusp}}\klammer{\alpha_s}\ln\frac{-Q^2}{\mu^2}+\gamma_{\rm A0}\klammer{\alpha_s}\right]\,C^{\rm A0}\klammer{Q^2,\mu^2},
\label{eq:CA0RGE}
\end{align}
where $-Q^2$ in the logarithm should be read as $-Q^2-i\varepsilon$. The cusp anomalous dimension $\gamma_{\text{cusp}}$ and the hard non-cusp anomalous dimension $\gamma_{A0}$ are given by
\begin{align}
&\gamma_{\text{cusp}}\klammer{\alpha_s}=4\,\frac{\alpha_s}{4\pi}+\mathcal{O}\klammer{\alpha_s^2},\\
\label{eq:A0 non-cusp}&\gamma_{\rm A0}\klammer{\alpha_s}=-6C_F\frac{\alpha_s}{4\pi}+\mathcal{O}\klammer{\alpha_s^2}. 
\end{align}
The one-loop cusp anomalous dimension sums the leading 
logarithms. Non-cusp terms become relevant only at NLL 
accuracy.\footnote{We make an exception for the non-cusp terms 
proportional to $\beta_0$, for example in \eqref{eq:gluonADM} 
below, which determine the scale of the coupling that appears 
in the tree-level $\mathcal{O}(\alpha_s)$ term. See 
Sec.~\ref{sec:resummation} for further discussion.} 
Dropping them, we obtain the LL solution
\begin{align}
\label{eq: hard function general result only cusp}
C_{\text{LL}}^{\rm A0}\klammer{Q^2,\mu^2}=&\exp\left[2C_FS\klammer{\mu_h,\mu}\right]\klammer{\frac{-Q^2}{\mu_h^2}}^{-C_FA_{\gamma_{\text{cusp}}}\klammer{\mu_h,\mu}}C^{\rm A0}\klammer{Q^2,\mu_h^2},
\end{align}
where $\mu_h\sim Q$ denotes the hard initial scale, such that 
$C^{\rm A0}\klammer{Q^2,\mu_h^2}$ free from large logarithms. The functions $S\klammer{\nu,\mu}$ and $A_{\gamma_i}\klammer{\nu,\mu}$ are defined as \cite{Neubert:2004dd}
\begin{align}
\label{eq:Resummation Definition of S}&S\klammer{\nu,\mu}=-\int_{\alpha_s\klammer{\nu}}^{\alpha_s\klammer{\mu}} d\alpha\,\frac{\gamma_{\text{cusp}}\klammer{\alpha}}{\beta\klammer{\alpha}}\int_{\alpha_s\klammer{\nu}}^{\alpha} d\alpha'\frac{1}{\beta\klammer{\alpha'}}\,,\\
\label{eq:Resummation Definition of A}&A_{\gamma_i}\klammer{\nu,\mu}=-\int_{\alpha_s\klammer{\nu}}^{\alpha_s\klammer{\mu}} d\alpha\,\frac{\gamma_i\klammer{\alpha}}{\beta\klammer{\alpha}}\,.
\end{align}
The QCD beta-function is
\be
\beta(\alpha_s) = \frac{ d\alpha_s}{ d\ln\mu}  
= -2\beta_0\,\frac{\alpha_s^2}{4\pi}
+\mathcal{O}\klammer{\alpha_s^3}, \qquad 
\beta_0 = 11-\frac{2n_l}{3}\,.
\ee

\paragraph{Anti-collinear function}

The anti-collinear function appears already at LP. In Laplace space, it obeys the local RGE \cite{Becher:2006mr}
\begin{align}
\frac{d}{d\ln\mu}\widetilde{\mathcal{J}}_{\bar{c}}^{\klammer{\bar{q}}}\klammer{s,Q^2,\mu^2}=-\left[2C_F\gamma_{\text{cusp}}\klammer{\alpha_s}\ln\frac{Q}{se^{\gamma_E}\mu^2}+\gamma_{\mathcal{J}^{\klammer{\bar{q}}}}\klammer{\alpha_s}\right]\widetilde{\mathcal{J}}_{\bar{c}}^{\klammer{\bar{q}}}\klammer{s,Q^2,\mu^2},
\end{align}
where the non-cusp term is given by
\begin{align}
\gamma_{\mathcal{J}^{\klammer{\bar{q}}}}\klammer{\alpha_s}=-6C_F\frac{\alpha_s}{4\pi}+\mathcal{O}\klammer{\alpha_s^2}.
\end{align}
At LL accuracy, the RGE is solved by
\begin{align}
\widetilde{\mathcal{J}}_{\bar{c}, \text{LL}}^{\klammer{\bar{q}}}\klammer{s,Q^2,\mu^2}=\exp\left[-4C_FS\klammer{\mu_{\bar{c}},\mu}\right]\klammer{\frac{Q}{se^{\gamma_E}\mu_{\bar{c}}^2}}^{2C_FA_{\gamma_{\text{cusp}}}\klammer{\mu_{\bar{c}},\mu}}\widetilde{\mathcal{J}}_{\bar{c}}^{\klammer{\bar{q}}}\klammer{s,Q^2,\mu_{\bar{c}}^2},
\end{align}
where $\mu_{\bar{c}}$ denotes the anti-collinear initial scale.

\paragraph{Collinear function}

The collinear function  $\mathcal{J}_c(p^2,\omega,\omega')$ is a new NLP object and its anomalous dimension is currently unknown. In the asymptotic region of large $\omega, \omega'$, it factorizes into the LP gluon jet function and two $D^{B1}$ matching coefficients as shown in \eqref{eq:EPfactI}. We therefore derive its asymptotic evolution from the LP gluon jet function,  which renormalizes in Laplace space as \cite{Becher:2009th}
\begin{align}
\frac{d}{d\ln\mu}\widetilde{\mathcal{J}}_c^{(g)}(s,Q^2,\mu^2)=
-\left[2C_A\gamma_{\text{cusp}}\klammer{\alpha_s}\ln\frac{Q}{se^{\gamma_E}\mu^2}+\gamma_{\mathcal{J}^{\klammer{g}}}\klammer{\alpha_s}\right]\,\widetilde{\mathcal{J}}_{c}^{(g)}(s,Q^2,\mu^2),
\label{eq:gluonjetRGE}
\end{align}
with the  gluon non-cusp term given by
\begin{align}
\gamma_{\mathcal{J}^{\klammer{g}}}\klammer{\alpha_s}=-2\beta_0\frac{\alpha_s}{4\pi}+\mathcal{O}\klammer{\alpha_s^2}, 
\label{eq:gluonADM}
\end{align}
and the evolution of the $D^{\rm B1}$ coefficient. Making the cusp 
term explicit, \eqref{eq:DBQrge}, \eqref{eq:DB1andim} or 
\eqref{eq:DB1RGEinomega}, \eqref{eq:DB1RGEinomegagamma} 
imply the non-local RGE 
\begin{eqnarray}
\frac{ d}{ d\ln\mu}D^{\rm B1}(\omega,\mu^2)
&=& -\klammer{C_F-C_A}\gamma_{\text{cusp}}\klammer{\alpha_s}\ln\frac{-\omega Q-i\varepsilon}{\mu^2} \,D^{\rm B1}(\omega,\mu^2)
\nonumber\\
&& +\,\int_0^\infty d\hat{\omega}\,\hat{\gamma}_D\klammer{\hat{\omega},\omega} D^{\rm B1}(\hat{\omega},\mu^2),
\label{eq:DB1evolution}
\end{eqnarray}
where the non-cusp anomalous dimension $\hat{\gamma}_D\klammer{\hat{\omega},\omega}$ is given by the second line of \eqref{eq:DB1RGEinomegagamma}. 
The first endpoint factorization consistency condition \eqref{eq:EPfactI} now implies the following asymptotic RGE for the collinear function $\mathcal{J}_c$,
\begin{eqnarray}
\frac{d}{d\ln\mu}\left\llbracket\widetilde{\mathcal{J}}_c\klammer{s,\omega,\omega',Q^2,\mu^2}\right\rrbracket 
&=&-\bigg[2C_A\gamma_{\text{cusp}}\klammer{\alpha_s}\ln\frac{Q}{se^{\gamma_E}\mu^2}
\nonumber\\
&& \hspace*{-3cm} +\,\klammer{C_F-C_A}\gamma_{\text{cusp}}\klammer{\alpha_s}\klammer{\ln\frac{\omega Q}{\mu^2}+\ln\frac{\omega' Q}{\mu^2}}\bigg]\,
\left\llbracket\widetilde{\mathcal{J}}_c\klammer{s,\omega,\omega',Q^2,\mu^2}\right\rrbracket
\nonumber\\
&& \hspace*{-4cm}-\gamma_{\mathcal{J}^{\klammer{g}}}\klammer{\alpha_s}\left\llbracket\widetilde{\mathcal{J}}_c\klammer{s,\omega,\omega',Q^2,\mu^2}\right\rrbracket
+\int_0^\infty d\hat{\omega}\,\frac{\hat{\omega}}{\omega}\,\hat{\gamma}_D\klammer{\hat{\omega},\omega}\left\llbracket\widetilde{\mathcal{J}}_c\klammer{s,\hat{\omega},\omega',Q^2,\mu^2}\right\rrbracket
\nonumber\\
&& \hspace*{-4cm}+\,\int_0^\infty d\hat{\omega}'\,\frac{\hat{\omega}'}{\omega'}\,\hat{\gamma}^*_D\klammer{\hat{\omega}',\omega'}\left\llbracket\widetilde{\mathcal{J}}_c\klammer{s,\omega,\hat{\omega}',Q^2,\mu^2}\right\rrbracket.
\label{eq:Atype_collinearRGE}
\end{eqnarray}
There is no mixing of the unhatted collinear function into the hatted collinear function, which is irrelevant asymptotically. At LL accuracy, we drop the non-cusp terms (except for those proportional to $\beta_0$) and obtain the solution
\begin{align}
\left\llbracket\widetilde{\mathcal{J}}_{c,\text{LL}}\klammer{s,\omega,\omega',Q^2,\mu^2}\right\rrbracket=
\nonumber&\frac{\alpha_s\klammer{\mu_c}}{\alpha_s\klammer{\mu}}\exp\left[-4C_AS\klammer{\mu_c,\mu}-4\klammer{C_F-C_A}S\klammer{\mu_{c\Lambda},\mu}\right]\\
\nonumber&\hspace*{-3cm}\times\klammer{\frac{Q}{se^{\gamma_E}\mu_c^2}}^{2C_A A_{\gamma_{\text{cusp}}}\klammer{\mu_c,\mu}}\klammer{\frac{\omega Q}{\mu_{c\Lambda}^2}}^{\klammer{C_F-C_A} A_{\gamma_{\text{cusp}}}\klammer{\mu_{c\Lambda},\mu}}\\
&\hspace*{-3cm}\times\klammer{\frac{\omega' Q}{\mu_{c\Lambda}^2}}^{\klammer{C_F-C_A} A_{\gamma_{\text{cusp}}}\klammer{\mu_{c\Lambda},\mu}}\left\llbracket\widetilde{\mathcal{J}}_c\klammer{s,\omega,\omega',Q^2,\mu_c^2,\mu_{c\Lambda}^2}\right\rrbracket\,.
\end{align}
We introduce two initial scales, $\mu_c$ and $\mu_{c\Lambda}$, 
connected to the two different cusp logarithms in 
\eqref{eq:Atype_collinearRGE}, allowing for the possibility to choose 
different initial scales for the gluon jet function and 
$D^{\rm B1}$ coefficient, which will be used in Sec.~\ref{sec:logsandscales}.

\paragraph{Soft function}

The soft quark function $S_{\rm NLP}(l_+,l_-,\omega,\omega')$ 
appears first at NLP and its complete RGE is currently unknown 
even at lowest order. We can obtain its asymptotic evolution 
for large $\omega, \omega'$ from RGE consistency. For 
fixed values of $\omega, \omega'$ and $r,r'$, respectively, 
the integrands of the A- and B-type term must be individually 
RGE-invariant. Imposing this requirement on the A-type term, 
we obtain the asymptotic RGE for the soft function,
\begin{eqnarray}
\frac{ d}{ d\ln\mu}\left\llbracket\widetilde{S}_{\text{NLP}}(s_R,s_L,\omega,\omega',\mu^2)\right\rrbracket
&=&\bigg[2C_A\gamma_{\text{cusp}}(\alpha_s)\ln\frac{1}{s_Le^{\gamma_E}s_Re^{\gamma_E}\mu^2}
\nonumber\\
&& \hspace*{-5cm} 
+\,\klammer{C_F-C_A}\gamma_{\text{cusp}}(\alpha_s)\klammer{\ln\frac{\omega }{s_Re^{\gamma_E}\mu^2}+\ln\frac{\omega'}{s_Re^{\gamma_E}\mu^2}}\bigg] 
\left\llbracket\widetilde{S}_{\text{NLP}}(s_R,s_L,\omega,\omega',\mu^2)\right\rrbracket
\nonumber\\[0.1cm]
&& \hspace*{-5cm}+\,\Big[\gamma_{\mathcal{J}^{\klammer{\bar{q}}}}(\alpha_s)+\gamma_{\mathcal{J}^{(g)}}(\alpha_s)-2\gamma_{A0}\klammer{\alpha_s}\Big]\left\llbracket\widetilde{S}_{\text{NLP}}\klammer{s_R,s_L,\omega,\omega',\mu^2}\right\rrbracket
\nonumber\\
&& \hspace*{-5cm}-\,\int_0^\infty d\hat{\omega}\,\frac{\omega}{\hat{\omega}}\,\hat{\gamma}_D(\omega,\hat{\omega})\left\llbracket\widetilde{S}_{\text{NLP}}(s_R,s_L,\hat{\omega},\omega',\mu^2)\right\rrbracket
\nonumber\\
&& \hspace*{-5cm}-\,\int_0^\infty d\hat{\omega}'\,\frac{\omega'}{\hat{\omega}'}\,\hat{\gamma}^*_D(\omega',\hat{\omega}')\left\llbracket\widetilde{S}_{\text{NLP}}(s_R,s_L,\omega,\hat{\omega}',\mu^2)\right\rrbracket .
\end{eqnarray}
Consistency requires that there is no mixing of the unhatted soft function into the hatted soft function. At LL accuracy, the asymptotic RGE is solved by
\begin{eqnarray}
\left\llbracket\widetilde{S}_{\text{NLP,LL}}(s_R,s_L,\omega,\omega',\mu^2)\right\rrbracket
&=&\frac{\alpha_{s}(\mu)}{\alpha_s(\mu_s)}\exp\left[4C_AS\klammer{\mu_s,\mu}+4\klammer{C_F-C_A}S\klammer{\mu_{s\Lambda},\mu}\right]
\nonumber\\
&&\hspace*{-4cm}\times\,
\klammer{\frac{1}{s_Le^{\gamma_E}s_Re^{\gamma_E}\mu_s^2}}^{\!-2C_AA_{\gamma_{\text{cusp}}}\klammer{\mu_s,\mu}}
\,\klammer{\frac{\omega }{s_Re^{\gamma_E}\mu_{s\Lambda}^2}}^{\!-\klammer{C_F-C_A}A_{\gamma_{\text{cusp}}}\klammer{\mu_{s\Lambda},\mu}}
\nonumber\\
&&\hspace*{-4cm}\times\,\klammer{\frac{\omega' }{s_Re^{\gamma_E}\mu_{s\Lambda}^2}}^{\!-\klammer{C_F-C_A}A_{\gamma_{\text{cusp}}}\klammer{\mu_{s\Lambda},\mu}}\,
\left\llbracket\widetilde{S}_{\text{NLP}}(s_R,s_L,\omega,\omega',\mu_s^2,\mu_{s\Lambda}^2)\right\rrbracket,
\end{eqnarray}
where again we allow for two separate initial soft scales, 
$\mu_s$ and $\mu_{s\Lambda}$. The lowest-order initial condition is proportional to  $\alpha_s$, which we choose to evaluate at $\mu_s$. Together with the prefactor $\alpha_s(\mu)/\alpha_s(\mu_s)$, which arises from the $\beta_0$-term in $\gamma_{\mathcal{J}^{(g)}}$, this results in an overall factor of $\alpha_s(\mu)$. 
The point of keeping the non-cusp $\beta_0$-term in $\gamma_{\mathcal{J}^{(g)}}$ is that it renders the result independent of the 
initial scale -- choosing to evaluate $\alpha_s$ at $\mu_{s\Lambda}$ instead, we would obtain the same resummed soft function.   

\subsection{RGEs for the B-type functions}
\label{sec:BtypeRGE}

In the B-type contribution, the collinear function and the soft function are known LP objects while the hard functions $C_i^{\rm B1}$ and the anti-collinear function $\mathcal{J}_{\bar c}^{q\bar q(8)}(p^2,r,r')$ are NLP objects. The evolution equation of the hard matching coefficients is of the form discussed in \cite{Beneke:2017ztn,Beneke:2018rbh} for B1 SCET operators, hence we can derive 
the anomalous dimensions for the new anti-collinear function from 
consistency, i.e.~the scale-independence of the entire B-type 
term for given $r, r'$. However, the leading logarithms to gluon 
thrust arise only from the endpoint region $r,r'\to 0, 1$, 
and therefore we will only need the equations for the 
asymptotic hard and anti-collinear functions. 

\paragraph{Collinear function} 

The  anomalous dimension of the collinear function is well-known 
since it appears in LP factorization theorems. The RGE equation and 
anomalous dimension have already been given in 
\eqref{eq:gluonjetRGE} and \eqref{eq:gluonADM}, respectively.
The LL solution to the RGE reads
\begin{align}
\widetilde{\mathcal{J}}_{c,\text{LL}}^{\klammer{g}}\klammer{s,\mu^2}=\frac{\alpha_s\klammer{\mu_c}}{\alpha_s\klammer{\mu}}\exp\left[-4C_AS\klammer{\mu_c,\mu}\right]\klammer{\frac{Q}{se^{\gamma_E}\mu_c^2}}^{2C_AA_{\gamma_{\text{cusp}}}\klammer{\mu_c,\mu}}\widetilde{\mathcal{J}}_c^{\klammer{g}}\klammer{s,\mu_c^2},
\end{align}
where $\mu_c$ denotes the collinear initial scale. 

\paragraph{Soft function}

The LP two-hemisphere soft function for gluons, which is defined 
with adjoint Wilson lines, obeys the evolution equation
\begin{eqnarray}
\label{eq:LPGluonSoftFunctionADM}
\frac{d}{d\ln\mu}\widetilde{S}^{(g)}(s_R,s_L,\mu^2)=\left[2C_A\gamma_{\text{cusp}}(\alpha_s)\ln\frac{1}{s_L s_Re^{2\gamma_E}\mu^2}+\gamma_{S^{(g)}}(\alpha_s)\right]\widetilde{S}^{\klammer{g}}(s_R,s_L,\mu^2)\;\;
\end{eqnarray}
in Laplace space. 
The non-cusp part of the anomalous dimension is given by \cite{Berger:2010xi}
\begin{align}
\gamma_{S^{(g)}}(\alpha_s)=0+\mathcal{O}(\alpha_s^2)\,.
\end{align}
At LL accuracy, the solution to the RGE reads
\begin{align}
\widetilde{S}_{\text{LL}}^{\klammer{g}}(s_R,s_L,\mu^2)=&\,
\exp\left[4C_AS\klammer{\mu_s,\mu}\right]\klammer{\frac{1}{s_L s_Re^{2\gamma_E}\mu_s^2}}^{\!-2C_AA_{\gamma_{\text{cusp}}}\klammer{\mu_s,\mu}}\\
\nonumber&\times\widetilde{S}^{\klammer{g}}(s_R,s_L,\mu_s^2),
\end{align}
where $\mu_s$ denotes the soft initial scale.

\paragraph{Hard function} 
Following the calculations of \cite{Beneke:2017ztn,Beneke:2018rbh}, the specific anomalous dimension matrix for 
the two coefficients $C^{\rm B1}_i$, $i=1,2$
needed here can be inferred from App.~\ref{app:asympDB1RGE}. For 
the flavour non-singlet case, which is assumed here, $\Delta_F$ 
can be set to zero there, and the two coefficients evolve 
independently,  
\bea
\frac{ d}{ d\ln\mu}C^{\rm B1}_i(Q^2,r) &=& \int_0^1  d\hat r\, \gamma_{\rm B1}(\hat r,r)C^{\rm B1}_i(Q^2,\hat r)\,,
\eea
with the same anomalous dimension (see~\eqref{eq:gamma}) 
\bea
\gamma_{\rm B1}(r,s) &=& -\frac{\alpha_s}{\pi}\,\delta(r-s)\left[C_A\ln\left(\frac{\mu^2}{-Q^2}\right)-\frac{C_A}{2}\ln(r\bar r)+\frac{3C_F}{2}\right]+2\gamma(r,s)\,,\qquad
\eea
where
\bea
\gamma(r,s) &=& \frac{\alpha_s}{2\pi}\left(\frac{C_A}{2}-C_F\right)\Bigg[ \frac{\bar r}{\bar s}\left(\left(\frac{\theta(r-s)}{r-s}\right)_+ + \theta(r-s)\right)\nn\\
&& {} + \frac{r}{ s}\left(\left(\frac{\theta(s-r)}{s-r}\right)_+ + \theta(s-r)\right) \Bigg]\,.
\eea 

The non-cusp part of this RGE contains a hidden large 
logarithm of $r$ ($\bar r$), which must be extracted and 
combined with the cusp part when one wants to sum the hard 
function for small $r$ or $\bar{r}$, see 
App.~\ref{app:asympDB1RGE}. This 
results in the evolution equation \eqref{eq:DBQrge} for 
the $D^{\rm B1}$ coefficient. The evolution equation 
for the asymptotic $C^{\rm B1}_i$ coefficients then 
follows from \eqref{eq:B1fact1}, \eqref{eq:B1fact2}. Since the 
leading logarithms to gluon thrust arise only from these 
asymptotic regions, we provide the RGEs for the asymptotic 
coefficients in the following.

The asymptotic B1 coefficients factorize into a product of A0 
coefficient and the $D^{\rm B1}$ coefficient. Combining 
\eqref{eq:CA0RGE} and \eqref{eq:DB1evolution}, we find 
\begin{eqnarray}
\frac{ d}{ d\ln\mu}\left\llbracket C^{\rm B1}_1\klammer{Q^2,r,\mu^2}\right\rrbracket_0
&=&\bigg[\,C_F\gamma_{\text{cusp}}\klammer{\alpha_s}\ln\frac{-Q^2}{\mu^2}
\nonumber\\
&&\hspace*{-3cm}-\,\klammer{C_F-C_A}\gamma_{\text{cusp}}\klammer{\alpha_s}\ln\frac{-rQ^2}{\mu^2}+\gamma_{\rm A0}\klammer{\alpha_s}\bigg]\left\llbracket C^{\rm B1}_1\klammer{Q^2,r,\mu^2}\right\rrbracket_0
\nonumber\\
&&\hspace*{-3cm}+\,Q\int_{0}^\infty d\hat{r}\,\frac{\hat{r}}{r}\,
\hat{\gamma}_D\klammer{\hat{r}Q,rQ}\left\llbracket C^{\rm B1}_1\klammer{Q^2,\hat{r},\mu^2}\right\rrbracket_0\,.
\end{eqnarray}
The LL solution reads 
\begin{eqnarray}
\left\llbracket C_1^{\rm B1}\klammer{Q^2,r,\mu^2}\right\rrbracket_0
&=&\exp\left[2C_FS\klammer{\mu_h,\mu}-2\klammer{C_F-C_A}S\klammer{\mu_{h\Lambda},\mu}\right]
\nonumber\\
&&\hspace*{-2cm}\times
\left(\frac{-Q^2}{\mu_h^2}\right)^{\!-C_FA_{\gamma_{\text{cusp}}}\klammer{\mu_h,\mu}}\klammer{\frac{-rQ^2}{\mu_{h\Lambda}^2}}^{\klammer{C_F-C_A}A_{\gamma_{\text{cusp}}}\klammer{\mu_{h\Lambda},\mu}}
\nonumber\\
&& \hspace*{-2cm}\times\left\llbracket C_1^{\rm B1}\klammer{Q^2,r,\mu_h^2,\mu_{h\Lambda}^2}\right\rrbracket_0,\quad
\label{eq:CB1LLsol}
\end{eqnarray}
where as in the case of similar A-type RGE solutions, we allowed for two different initial scales, $\mu_h$ and $\mu_{h\Lambda}$, 
connected with the different cusp logarithms. 

For each of the two endpoint limits $r\rightarrow0,1$, we only 
need one of the two B1 coefficients since the other one is regular. 
The RGE equation and solution for 
$C_2^{\rm B1}\klammer{Q^2,r,\mu^2}$ is obtained from the above 
by replacing $r\to \bar{r}$, and similarly for $\hat{r}$ including 
the integration measure $ d\hat{r}$.

\paragraph{Anti-collinear function}

The evolution of the hard, collinear, and soft functions is known,  
hence we can derive the evolution of the anti-collinear function through consistency. We demand that the B-type contribution before integration over $r,r'$ is RGE-invariant on its own, which allows us to derive the RGE and anomalous dimension of the anti-collinear 
function in the form
\begin{eqnarray}
\frac{ d}{ d\ln\mu}\widetilde{\mathcal{J}}_{\bar{c}}^{q\bar{q}\klammer{8}}\klammer{s_R,r,r'}&=&\bigg[-2C_A\gamma_{\text{cusp}}\klammer{\alpha_s}\ln\frac{Q}{s_Re^{\gamma_E}\mu^2}
\nonumber\\
&& \hspace*{-3cm}
+\,\gamma_{\mathcal{J}^{\klammer{g}}}(\alpha_s)-\gamma_{S^{\klammer{g}}}\klammer{\alpha_s}\bigg]\,\widetilde{\mathcal{J}}_{\bar{c}}^{q\bar{q}\klammer{8}}\klammer{s_R,r,r'}\nonumber\\
&&\hspace*{-3cm}
 -\,\int_0^1 d\hat{r}\,\hat{\gamma}_{\rm B1}\klammer{r,\hat{r}}\widetilde{\mathcal{J}}_{\bar{c}}^{q\bar{q}\klammer{8}}\klammer{s_R,\hat{r},r'}-\int_0^1 d\hat{r}'\,\hat{\gamma}^*_{\rm B1}\klammer{r',\hat{r}'    }\widetilde{\mathcal{J}}_{\bar{c}}^{q\bar{q}\klammer{8}}\klammer{s_R,r,\hat{r}'}\,,\qquad
\end{eqnarray}
where the non-cusp part of the B1 anomalous dimension $\hat{\gamma}_{\rm B1}$ is given by
\begin{align}
\hat{\gamma}_{\rm B1}\klammer{r,s}=\frac{\alpha_s}{\pi}\delta\klammer{r-s}\left[\frac{C_A}{2}\ln\klammer{r\bar{r}}-\frac{3C_F}{2}\right]+2\gamma\klammer{r,s}\,.
\end{align}
As for the hard function, the non-cusp part of the full anomalous dimension contains hidden leading logarithms coming from the endpoint region. We therefore use asymptotic anomalous dimensions for the LL resummation. RGE-consistency yields in the two endpoint regions
\begin{eqnarray}
\frac{ d}{ d\ln\mu}\left\llbracket\widetilde{\mathcal{J}}_{\bar{c}}^{q\bar{q}\klammer{8}}\klammer{s,r,r',\mu^2}\right\rrbracket_0&=&-\bigg[2C_F\gamma_{\text{cusp}}\klammer{\alpha_s}\ln\frac{Q}{se^{\gamma_E}\mu^2}\nonumber\\
&&\hspace*{-3cm}-\klammer{C_F-C_A}\gamma_{\text{cusp}}\klammer{\alpha_s}\left(\ln\frac{rQ}{se^{\gamma_E}\mu^2}+\ln\frac{r'Q}{se^{\gamma_E}\mu^2}\right)\nonumber\\
&&\hspace*{-3cm}+\left[2\gamma_{A0}(\alpha_s)+\gamma_{{S}^{\klammer{g}}}(\alpha_s)-\gamma_{\mathcal{J}^{\klammer{g}}}(\alpha_s)\right]\bigg]\left\llbracket\widetilde{\mathcal{J}}^{q\bar{q}\klammer{8}}_{\bar{c}}\klammer{s,r,r',\mu^2}\right\rrbracket_0\nonumber\\
&&\hspace*{-3cm}-Q\int_0^\infty d\hat{r}\,\frac{r}{\hat{r}}\,\hat{\gamma}_D\klammer{rQ,\hat{r}Q}\left\llbracket\widetilde{\mathcal{J}}^{q\bar{q}\klammer{8}}_{\bar{c}}\klammer{s,\hat{r},r',\mu^2}\right\rrbracket_0\nonumber\\
&&\hspace*{-3cm}-Q\int_0^\infty d\hat{r}'\,\frac{r'}{\hat{r}'}\,\hat{\gamma}^*_D\klammer{r'Q,\hat{r}'Q}\left\llbracket\widetilde{\mathcal{J}}^{q\bar{q}\klammer{8}}_{\bar{c}}\klammer{s,r,\hat{r}',\mu^2}\right\rrbracket_0\,.
\end{eqnarray}
At LL accuracy, the asymptotic RGE is solved by
\begin{eqnarray}
\left\llbracket\widetilde{\mathcal{J}}_{\bar{c},\text{LL}}^{q\bar{q}\klammer{8}}\klammer{s,r,r',\mu^2}\right\rrbracket_0&=&\frac{\alpha_s(\mu)}{\alpha_s(\mu_{\bar{c}})}
\exp\left[-4C_FS\klammer{\mu_{\bar{c}},\mu}+4\klammer{C_F-C_A}S\klammer{\mu_{\bar{c}\Lambda},\mu}\right]\nonumber\\
&&\hspace*{-3cm}\times\,\klammer{\frac{Q}{se^{\gamma_E}\mu_{\bar{c}}^2}}^{\!2C_FA_{\gamma_{\text{cusp}}}\klammer{\mu_{\bar{c}},\mu}}\klammer{\frac{rQ}{se^{\gamma_E}\mu_{\bar{c}\Lambda}^2}}^{\!-\klammer{C_F-C_A}A_{\gamma_{\text{cusp}}}\klammer{\mu_{\bar{c}\Lambda},\mu}}\nonumber\\
&&\hspace*{-3cm}
\times\,\klammer{\frac{r'Q}{se^{\gamma_E}\mu_{\bar{c}\Lambda}^2}}^{\!-\klammer{C_F-C_A}A_{\gamma_{\text{cusp}}}\klammer{\mu_{\bar{c}\Lambda},\mu}}\left\llbracket\widetilde{\mathcal{J}}_{\bar{c}}^{q\bar{q}\klammer{8}}\klammer{s,r,r',\mu_{\bar{c}}^2,\mu_{\bar{c}\Lambda}^2}\right\rrbracket_0,
\end{eqnarray}
where $\mu_{\bar{c}}$ and $\mu_{\bar{c}\Lambda}$ denote the 
anti-collinear initial scales. As for the A-type soft function, we may evaluate the factor of $\alpha_s$ in the fixed-order initial condition either at $\mu_{\bar{c}}$ or $\mu_{\bar{c}\Lambda}$, 
and $\mu_{\bar{c}}$ has been chosen above. This choice does not affect the overall resummed anti-collinear function since the ratio of couplings in front always guarantees a global factor of $\alpha_s\klammer{\mu}$. Once again the RGE and solution 
for the anti-collinear function for $r\to 1$, $\llbracket\widetilde{\mathcal{J}}_{\bar{c},\text{LL}}^{q\bar{q}\klammer{8}}\klammer{s,r,r',\mu^2}\rrbracket_1$, are obtained from the above by replacing 
 $r\to \bar{r}$, and similarly for $\hat{r}$ including 
the integration measure $ d\hat{r}$ as well as for the 
corresponding primed quantities.


\section{Resummation} 
\label{sec:resummation}
 
With the individual functions resummed to LL accuracy, we 
proceed to the resummation of the leading logarithms of 
gluon thrust at NLP. All functions are renormalized, so 
from now on we set $d=4$. 

We begin by rearranging the endpoint-subtracted expressions
\eqref{eq:Atype_subtracted_general2} and 
\eqref{eq:Btype_subtracted2} for the A- and B-type term, 
respectively. In \eqref{eq:Atype_subtracted_general2}, 
we rewrite
\begin{eqnarray}
&&\widetilde{\mathcal{J}}_c\klammer{s_L,\omega,\omega'}\widetilde{S}_{\text{NLP}}\klammer{s_R,s_L,\omega,\omega'}
\nonumber\\
&&\hspace*{0.5cm}=\,
\widetilde{\mathcal{J}}_c\klammer{s_L,\omega,\omega'}\widetilde{S}_{\text{NLP}}\klammer{s_R,s_L,\omega,\omega'}
-\sigma(\omega,\omega')\left\llbracket\widetilde{\mathcal{J}}_c\klammer{s_L,\omega,\omega'}\right\rrbracket\left\llbracket\widetilde{S}_{\text{NLP}}\klammer{s_R,s_L,\omega,\omega'}\right\rrbracket
\nonumber\\
&&\hspace*{1cm}+\,
\sigma(\omega,\omega')\left\llbracket\widetilde{\mathcal{J}}_c\klammer{s_L,\omega,\omega'}\right\rrbracket\left\llbracket\widetilde{S}_{\text{NLP}}\klammer{s_R,s_L,\omega,\omega'}\right\rrbracket,
\end{eqnarray}
where $\sigma(\omega,\omega')$ is an auxiliary function, 
which equals unity whenever $\omega^{(\prime)}s_R, 
\omega^{(\prime)}s_L \gg 1$ and which goes to zero 
for  $\omega^{(\prime)}\to 0$. We find 
\begin{eqnarray}
\frac{1}{\sigma_0}\frac{\widetilde{ d\sigma}}{ ds_R ds_L}|_{\rm A-type}&=&
\frac{2C_F}{Q}\,\left|C^{\rm A0}\klammer{Q^2}\right|^2\widetilde{\mathcal{J}}^{\klammer{\bar{q}}}_{\bar{c}}\klammer{s_R}
\int_0^\infty d\omega d\omega'
\times\Bigg\{
\nonumber\\
&&\hspace*{-3cm}
\left[\sigma(\omega,\omega')-\theta\klammer{\omega-\Lambda}\theta\klammer{\omega'-\Lambda}\right]\left\llbracket\widetilde{\mathcal{J}}_c\klammer{s_L,\omega,\omega'}\right\rrbracket\left\llbracket\widetilde{S}_{\text{NLP}}\klammer{s_R,s_L,\omega,\omega'}\right\rrbracket
\nonumber\\
&&\hspace*{-3cm}+\,\bigg[\widetilde{\mathcal{J}}_c\klammer{s_L,\omega,\omega'}\widetilde{S}_{\text{NLP}}\klammer{s_R,s_L,\omega,\omega'}
-\sigma(\omega,\omega')\left\llbracket\widetilde{\mathcal{J}}_c\klammer{s_L,\omega,\omega'}\right\rrbracket\!\left\llbracket\widetilde{S}_{\text{NLP}}\klammer{s_R,s_L,\omega,\omega'}\right\rrbracket\bigg]
\nonumber\\
&&\hspace*{-3cm}+\,\widetilde{\widehat{\mathcal{J}}}_c\klammer{s_L,\omega,\omega'}\widetilde{\widehat{S}}_{\text{NLP}}\klammer{s_R,s_L,\omega,\omega'}\Bigg\}\,.
\label{eq:Atyperearranged}
\end{eqnarray}
In this expression, the first line in curly brackets is the 
subtraction term, which contains the endpoint divergence, 
if $\Lambda\to\infty$, and therefore results in a 
logarithmically enhanced contribution from the 
$\omega,\omega'$ convolution. The next term in square 
brackets resembles a ``plus-distribution'' type term -- 
this combination has no support for large $\omega,\omega'$ 
and does not produce an extra logarithm. Finally, the 
last term in curly brackets involving the hatted soft 
function is regular by itself and does not require 
subtraction. After this rewriting, only the first 
line in curly brackets needs to be kept at LL accuracy. 
The introduction of the auxiliary function $\sigma$ 
is necessary to eliminate a spurious singularity 
for small $\omega,\omega'$, which arises if one wants 
to integrate the subtraction and plus-distribution-like 
term separately, and which 
cancels between the two.

Similarly, we write \eqref{eq:Btype_subtracted2} in the form
\begin{eqnarray}
\frac{1}{\sigma_0}\frac{\widetilde{ d\sigma}}{ ds_R ds_L}|_{
\scriptsize \begin{array}{l}
$\rm B--type$\\[-0.1cm] $i=i'=1$\end{array}} &=&
\frac{2C_F}{Q^2}\,\widetilde{J}_c^{\klammer{g}}\klammer{s_L}\widetilde{S}^{\klammer{g}}\klammer{s_R,s_L}\times\Bigg\{
\nonumber\\
&&\hspace*{-2cm}\int_0^\infty dr dr'\,\left[\theta\klammer{1-r}\theta\klammer{1-r'}-1+\theta\klammer{r-\frac{\Lambda}{Q}}\theta\klammer{r'-\frac{\Lambda}{Q}}\right]
\nonumber\\
&&\hspace*{-1.5cm}\times\left\llbracket C_1^{\rm B1*}\klammer{Q^2,r'}\right\rrbracket_0 \left\llbracket C_1^{\rm B1}\klammer{Q^2,r}\right\rrbracket_0 \left\llbracket \widetilde{\mathcal{J}}_{\bar{c}}^{q\bar{q}\klammer{8}}\klammer{s_R,r,r'}\right\rrbracket_0
\nonumber\\
&&\hspace*{-2cm}+
\int_0^1 dr dr'\,\bigg[C_1^{\rm B1*}\klammer{Q^2,r'}C_1^{\rm B1}\klammer{Q^2,r}\widetilde{\mathcal{J}}_{\bar{c}}^{q\bar{q}\klammer{8}}\klammer{s_R,r,r'}
\nonumber\\[-0.15cm]
&&\hspace*{-1.5cm}-\left\llbracket C_1^{\rm B1*}\klammer{Q^2,r'}\right\rrbracket_0 \left\llbracket C_1^{\rm B1}\klammer{Q^2,r}\right\rrbracket_0 \left\llbracket \widetilde{\mathcal{J}}_{\bar{c}}^{q\bar{q}\klammer{8}}\klammer{s_R,r,r'}\right\rrbracket_0\bigg]\,\Bigg\}\,.
\label{eq:Btyperearranged}
\end{eqnarray}
The first integral in curly brackets represents the 
subtraction term, which contributes to the leading 
logarithms and cancels the $\Lambda$-dependence of the 
corresponding A-type term. The second integral with 
support in $[0,1]$ is of the plus-distribution type 
and can be safely integrated to $r,r'=0$. This term 
can be dropped at LL accuracy. An auxiliary function 
is not required for the B-type term.

We remind the reader that there is an 
identical soft anti-quark $A$-type term and the 
$i=i'=2$ term, which is identical to the 
$i=i'=1$ term due to the symmetry in $r^{(\prime)}
\leftrightarrow \bar{r}^{(\prime)}$. The $i\not= i'$ 
interference terms can be dropped at this point 
for LL accuracy. 

The previous two equations are general and provide a 
suitable starting point for the resummation of 
NLP logarithms beyond the leading ones. Since some of 
the non-asymptotic anomalous dimensions in the 
plus-distribution-like terms are not yet available, 
the following focuses on the leading logarithms, 
including a new set of next-to-leading logarithms 
associated with the running coupling in the leading 
terms.

\subsection{Counting of logarithms and proper 
choice of scales}
\label{sec:logsandscales}

Standard Sudakov exponentiation puts an observable into 
the form 
\begin{equation}
I \sim F(\alpha_s) \exp\left(\ln\tau \,g_0(\lambda) + 
g_1(\lambda) +\alpha_s g_2(\lambda)+\ldots\right)\,,
\end{equation}
where $\lambda = \alpha_s\ln\tau$ counts as 
$\mathcal{O}(1)$, while $\tau\ll 1$. The function 
$g_0$ is needed to sum the leading logarithms, 
$g_1$ is NLL etc. $F$ collects the initial condition 
and has an expansion without large logs. The 
$\mathcal{O}(\alpha_s)$ correction to the leading 
approximation of $F$ is already a next-to-next-to-leading 
logarithmic (NNLL) effect. The large-logarithm 
counting primarily refers to the {\em exponent}.

The {\em integrands} of \eqref{eq:Atyperearranged}, 
\eqref{eq:Btyperearranged} are of this form when the 
solutions to the RGEs from Sec.~\ref{sec:consistency} 
are used. However, one now has to take integrals of 
the form
\begin{equation}
\int_{\tau Q}^\Lambda \frac{d\omega}{\omega}\,I(\omega),
\quad 
\int^1_{\Lambda/Q} \frac{dr}{r}\,I(r),
\label{eq:logints}
\end{equation}
and the assumption underlying standard Sudakov resummation 
that $\omega =\mathcal{O}(\tau)$ and $r=\mathcal{O}(1)$ 
is not justified, since either $\Lambda\gg\tau Q$ 
or $\Lambda\ll Q$, or both. The fact that the integrals 
are logarithmic for large $\omega$ and small $r$ implies 
two modifications of the standard resummation scheme.

First, the logarithmic integrals combine to an extra 
logarithm promoting their log-counting by one order. 
Thus, inserting the LL-resummed RGE functions in the 
subtraction terms in \eqref{eq:Atyperearranged}, 
\eqref{eq:Btyperearranged} will result in leading 
logarithms for the observables. On the other hand, 
inserting them into the plus-distribution 
type terms will result only in next-to-leading logarithms, 
since these terms are not sensitive to the large 
$\omega$ and small $r$ region by construction 
and therefore lack one power of logarithm. This counting 
extends to higher logarithmic order in an obvious 
manner and justified the previous assertion that 
to LL accuracy, it suffices to focus on the subtraction 
terms in  \eqref{eq:Atyperearranged}, 
\eqref{eq:Btyperearranged}.

\begin{figure}
\begin{center}
\includegraphics[width=0.65\textwidth]{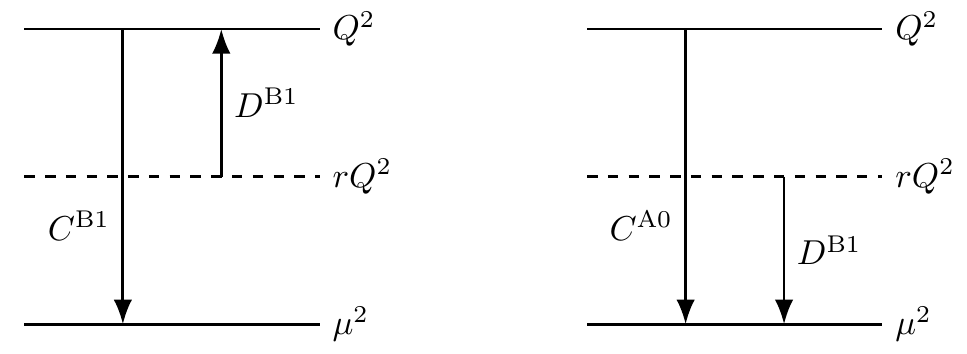}
\caption{
Graphical illustration of the resummation of the initial condition 
of the B1 operator for $r\ll 1$ (left) and the LL-accurate 
approximate procedure employed here.
\label{fig:ICres}}
\end{center}
\end{figure}

Second, the fact that $\omega =\mathcal{O}(\tau)$ 
and $r=\mathcal{O}(1)$ cannot be assumed can be seen 
as the need for resumming the initial condition of 
the RGE evolution. Consider for example the hard function 
$C^{\rm B1}_1(Q^2,r,\mu^2)$. Standard RGE summation 
implies that in the initial condition, one chooses 
the scale $\mu_h\sim Q$, at which there are no large 
logarithms and evolves to $\mu\ll Q$. However, the 
initial condition for $\mu_h\sim Q$ may still contain 
an infinite series of $\ln r$ terms. While these 
normally count as $\mathcal{O}(1)$, they lead to a 
breakdown of the perturbative expansion in integrals 
such as \eqref{eq:logints}, which receive contributions 
from $r\ll 1$. These logarithms are summed by evolving 
with asymptotic evolution kernel for the $D^{\rm B1}$ 
coefficient from $\mu^2\sim r Q^2$ to $\mu_h^2$. At LL 
accuracy, 
combining this evolution with the standard evolution from 
$\mu_h$ to $\mu$, amounts to the expression 
\eqref{eq:CB1LLsol} with split initial scales -- 
$\mu_h\sim Q$ in $C^{\rm A0}$, and the running scale 
$\mu_{h\Lambda}^2\sim r Q^2$ for the part that arises 
from the $D^{\rm B1}$ coefficient, whose natural 
scale is $r Q^2$. The procedure is illustrated in 
Fig.~\ref{fig:ICres}. A similar reasoning applies 
to the choice of initial conditions for the anti-collinear 
function $\widetilde{\mathcal{J}}_{\bar{c}}^{q\bar{q}\klammer{8}}\klammer{s_R,r,r'}$ and the functions 
$\widetilde{\mathcal{J}}_c\klammer{s_L,\omega,\omega',Q^2,\mu^2}$, 
$\widetilde{S}_{\text{NLP}}\klammer{s_R,s_L,\omega,\omega',\mu^2}$, which appear in the A-type term. Hence, we choose 
the initial scales in the hard, soft and (anti-) collinear 
functions to be of order of 
\begin{align}
&\mu_h^2\sim Q^2\,,&&\mu_c^2\sim \frac{Q}{s_L}\,,&&\mu_{\bar{c}}^2\sim \frac{Q}{s_R}\,,&&\mu_s^2\sim \frac{1}{s_Ls_R}\,,
\nonumber\\
&\mu_{h\Lambda}^2\sim rQ^2\,,&&\mu_{c\Lambda}^2\sim \omega Q\,,&&\mu_{\bar{c}\Lambda}^2\sim \frac{rQ}{s_R}\,,&&\mu_{s\Lambda}^2\sim \frac{\omega}{s_R}\,.
\label{eq:initialscales}
\end{align}
To ensure the cancellation of the rearrangement scale 
$\Lambda$ between the A- and B-type term, the integrands 
of the subtractions terms must match, which requires
\begin{equation}
\label{eq:intermediatescaleconstraints}
\mu_{s\Lambda}\sim\mu_{\bar{c}\Lambda}\quad\text{and}
\quad\mu_{h\Lambda}\sim\mu_{c\Lambda}\,,
\end{equation}
which is satisfied by \eqref{eq:initialscales} with 
the identification $rQ\leftrightarrow\omega$. 

We now insert the resummed functions from Sec.~\ref{sec:consistency}, 
evolved from their respective initial scales to a common scale $\mu$,  
into the {\em integrands} of \eqref{eq:Atyperearranged},  
\eqref{eq:Btyperearranged}, and simplify the expressions employing 
the identities  
\begin{eqnarray}
\label{eq:ASidentities}
&& A_{\gamma_i}\klammer{\mu_1,\mu_2}+A_{\gamma_i}\klammer{\mu_2,\mu_3}=A_{\gamma_i}\klammer{\mu_1,\mu_3}\,,\\
&&S\klammer{\mu_1,\mu_2}+S\klammer{\mu_2,\mu_3}=S\klammer{\mu_1,\mu_3}+\ln\frac{\mu_1}{\mu_2}\,A_{\gamma_{\text{cusp}}}\klammer{\mu_2,\mu_3}\,,
\end{eqnarray} 
and the special case $\mu_3=\mu_1$. 
We also abbreviate $A\equiv A_{\gamma_{\text{cusp}}}$. 
We find 
\begin{eqnarray}\label{eq:LLintegrandA}
&& \left|C^{\rm A0}(Q^2)\right|^2\widetilde{\mathcal{J}}_{\bar{c}}^{\klammer{\bar{q}}}\klammer{s_R,Q^2}
\left\llbracket\widetilde{\mathcal{J}}_{c}\klammer{s_L,\omega,\omega'}\right\rrbracket\left\llbracket\widetilde{S}_{\text{NLP}}\klammer{s_R,s_L,\omega,\omega'}\right\rrbracket
\nonumber\\[0.1cm]
&&\hspace*{0.5cm} 
=\,\exp\left[4C_FS\klammer{\mu_h,\mu_{\bar{c}}}+4C_AS\klammer{\mu_s,\mu_c}+4\klammer{C_F-C_A}S\klammer{\mu_{s\Lambda},\mu_{c\Lambda}}\right]
\nonumber\\[0.1cm]
&&\hspace*{1cm} 
\times\klammer{\frac{Q^2}{\mu_h^2}}^{\!-2C_FA\klammer{\mu_h,\mu_{\bar{c}}}}\klammer{\frac{1}{s_Ls_Re^{2\gamma_E} \mu_s^2}}^{\!-2C_A A\klammer{\mu_s,\mu_c}}\klammer{\frac{\omega }{s_R e^{\gamma_E}\mu_{s\Lambda}^2}}^{\!-2\klammer{C_F-C_A}A\klammer{\mu_{s\Lambda},\mu_{c\Lambda}}}
\nonumber\\
&&\hspace*{1cm} \times\klammer{s_R e^{\gamma_E} Q}^{2C_FA\klammer{\mu_{c\Lambda},\mu_{\bar{c}}}+2C_A A\klammer{\mu_c,\mu_{c\Lambda}}}\left[\frac{\alpha_s(\mu_{c})}{4\pi}\,\delta\klammer{\omega-\omega'}\frac{1}{\omega s_R}\right]
\end{eqnarray}
for the A-type integrand, and for the B-type, we find
\begin{eqnarray}\label{eq:LLintegrandB}
	&&\left\llbracket C_{1}^{\rm B1}\klammer{Q^2,r}\right\rrbracket_0\left\llbracket C_{1}^{\rm B1*}\klammer{Q^2,r'}\right\rrbracket_0\widetilde{\mathcal{J}}_{c}^{\klammer{g}}\klammer{s_L}\left\llbracket\widetilde{\mathcal{J}}_{\bar{c}}^{q\bar{q}\klammer{8}}\klammer{s_R,r,r'}\right\rrbracket_0\widetilde{S}^{\klammer{g}}\klammer{s_R,s_L}\nonumber\\[0.1cm]
	&&\hspace*{0.5cm} 
	=\,\exp\left[4C_FS\klammer{\mu_h,\mu_{\bar{c}}}
	+4C_AS\klammer{\mu_s,\mu_c}
	+4\klammer{C_F-C_A}S\klammer{\mu_{\bar{c}\Lambda},\mu_{h\Lambda}}\right]
	\nonumber\\[0.1cm]
	&&\hspace*{1cm} \times\klammer{\frac{Q^2}{\mu_h^2}}^{\!-2C_F A\klammer{\mu_h,\mu_{\bar{c}}}}
	\klammer{\frac{1}{s_Ls_R e^{2\gamma_E}\mu_s^2}}^{\!-2C_A A\klammer{\mu_s,\mu_c}}\klammer{\frac{rQ}{s_R e^{\gamma_E}\mu_{\bar c\Lambda}^2}}^{\!-2\klammer{C_F-C_A}A\klammer{\mu_{\bar{c}\Lambda},\mu_{h\Lambda}}}
	\nonumber\\
	&&\hspace*{1cm}\times\klammer{s_R e^{\gamma_E} Q}^{2C_F A\klammer{\mu_{h\Lambda},\mu_{\bar{c}}}+2C_A A\klammer{\mu_c,\mu_{h\Lambda}}}\left[\frac{\alpha_s(\mu_c)}{4\pi}\delta\klammer{r-r'}\frac{Q}{rs_R}\right].
\end{eqnarray}
As it should be with all functions evaluated consistently to LL accuracy the above expressions are manifestly independent of the arbitrary renormalization scale $\mu$. 
Note that if we would not have included the terms proportional to $\beta_0$ in the non-cusp parts of the anomalous dimensions, the only change would be that the overall prefactor of $\alpha_s$ in the square bracket in~\eqref{eq:LLintegrandA}
would be evaluated at scale $\mu_s$ (or $\mu_{s\Lambda}$) instead of $\mu_c$. Similarly, in~\eqref{eq:LLintegrandB} the prefactor would be evaluated at  $\mu_{\bar{c}}$ (or $\mu_{\bar{c}\Lambda}$) instead of $\mu_c$. The scale at which the prefactor is evaluated formally contributes at the same order in the logarithmic counting as the LL terms in the argument of the exponential factors, when including the running coupling. Therefore, it is necessary to take into account the terms proportional to $\beta_0$ in the non-cusp parts of the anomalous dimensions to obtain a consistent LL result. In addition, we note that this also ensures that the final result is insensitive to a possible redefinition $\widetilde{\mathcal{J}}_{\bar{c}}^{q\bar{q}\klammer{8}}\to \widetilde{\mathcal{J}}_{\bar{c}}^{q\bar{q}\klammer{8}}/\alpha_s$ and $C_{1}^{\rm B1}\to C_{1}^{\rm B1}\times\sqrt{\alpha_s}$, that corresponds to a reshuffling of the overall factor of $\alpha_s$ from the anti-collinear to the hard function, and similar ambiguities in the 
definitions of the functions in the A-type term.

We note the similarity of the two expressions. In fact, 
enforcing the relations \eqref{eq:intermediatescaleconstraints}, both expressions 
are exactly equal up to an overall factor of $Q$ with the 
identification $\omega^{(\prime)}=r^{(\prime)} Q$.
We thus sum the subtraction terms from
\eqref{eq:Atyperearranged},  \eqref{eq:Btyperearranged}, 
and change variables from $r$ to $\omega/Q$ in the 
B-type contribution \eqref{eq:Btyperearranged}. We also 
choose the auxiliary function to be of the form 
$\sigma(\omega,\omega) = \theta(\omega-\sigma)$, where 
$\sigma$ is of order $1/s_R$, or $1/s_L$. The two integrals 
to be added then combine as 
\begin{equation}
\int_0^\infty\frac{d\omega}{\omega}
\Big[\,\theta(\omega-\sigma)-\theta(\omega-\Lambda)\,\Big] 
+\int_{\Lambda/Q}^1 \frac{dr}{r} 
=\int_\sigma^Q\frac{d\omega}{\omega}\,,
\end{equation}
and we obtain 
\begin{eqnarray}
	\frac{1}{\sigma_0}\frac{\widetilde{ d\sigma}}{ ds_R ds_L}|_{\rm LL}&=&
	2\cdot \frac{2 C_F}{Q s_R}\frac{\alpha_s(\mu_c)}{4\pi}\,
	\,\exp\left[4C_FS\klammer{\mu_h,\mu_{\bar{c}}}
	+4C_AS\klammer{\mu_s,\mu_c}\right]
	\nonumber\\
	&&\hspace*{-2cm}
	\times\klammer{\frac{Q^2}{\mu_h^2}}^{\!-2C_F A\klammer{\mu_h,\mu_{\bar{c}}}}
	\klammer{\frac{1}{s_Ls_R e^{2\gamma_E}\mu_s^2}}^{\!-2C_A A\klammer{\mu_s,\mu_c}}
	\nonumber\\
	&&\hspace*{-2cm}
	\times\int_\sigma^Q\frac{d\omega}{\omega}\,
	\exp\left[4\klammer{C_F-C_A}S\klammer{\mu_{s\Lambda},\mu_{h\Lambda}}\right]
	\klammer{\frac{\omega }{s_R e^{\gamma_E} \mu_{s\Lambda}^2}}^{\!-2\klammer{C_F-C_A}A\klammer{\mu_{s\Lambda},\mu_{h\Lambda}}}
	\nonumber\\
	&&\hspace*{-0.8cm}\times\,\klammer{s_R e^{\gamma_E} Q}^{2C_FA\klammer{\mu_{h\Lambda},\mu_{\bar{c}}}+2C_A A\klammer{\mu_c,\mu_{h\Lambda}}}
	\,,
	\label{eq:mainLLresult}
\end{eqnarray}
where $\mu_{h\Lambda}^2\sim \omega Q$ and 
$\mu_{s\Lambda}^2\sim \omega/s_R$ according to
\eqref{eq:initialscales}. We also added an overall factor 
of 2 to account for the identical soft anti-quark and 
$i=i'=2$ contribution. This is our main result for the 
LL resummed two-hemisphere invariant mass distribution 
in the NLP gluon thrust region. The Laplace-space 
gluon thrust distribution 
itself is obtained as
\begin{equation}
\frac{1}{\sigma_0}\frac{\widetilde{ d\sigma}}{ ds} = 
\frac{1}{\sigma_0}\frac{\widetilde{ d\sigma}}{ ds_R ds_L}|_{s_R=s_L=s},
\end{equation}
which follows from \eqref{eq:taudist} and the definition of 
the Laplace transformation.

\subsection{Double-logarithmic limit}

The double-logarithmic limit corresponds to evaluating 
the renormalization group functions at leading order 
in an expansion in $\alpha_s$:
\begin{eqnarray}
A(\nu,\mu)&=&-\frac{\alpha_s}{2\pi}\ln\frac{\mu^2}{
\nu^2}\,,
\\
S(\nu,\mu)&=&-\frac{\alpha_s}{8 \pi}\ln^2\frac{\mu^2}{
\nu^2}\,.
\end{eqnarray}
Setting scales as prescribed by \eqref{eq:initialscales}, 
the LL result \eqref{eq:mainLLresult} simplifies to 
\begin{eqnarray}
\label{eq:dbllog}
\frac{1}{\sigma_0}\frac{\widetilde{ d\sigma}}{ ds_R ds_L}&=&
\frac{\alpha_s C_F}{\pi}\frac{1}{Q s_R}
\,e^{-\frac{\alpha_s C_A}{\pi}\ln\klammer{s_L e^{\gamma_E} Q}\ln\klammer{s_R e^{\gamma_E} Q}}\int_{\sigma}^Q\frac{ d\omega}{\omega}\,
\klammer{\frac{\omega}{Q}}^{\!\frac{\alpha_s}{\pi}\klammer{C_F-C_A}\ln\klammer{s_R e^{\gamma_E} Q}
}
\nonumber\\
&=&
\frac{C_F}{C_F-C_A}\frac{1}{Q s_R  \ln(s_R e^{\gamma_E} Q)}
\exp\,\bigg[-\frac{\alpha_s C_A}{\pi}\ln\klammer{s_L e^{\gamma_E} Q}\ln\klammer{s_R e^{\gamma_E} Q}\bigg]
\nonumber\\
&&\times\bigg\{
1-\klammer{\frac{\sigma}{Q}}^{\!\frac{\alpha_s}{\pi}\klammer{C_F-C_A}\ln\klammer{s_R e^{\gamma_E} Q}} \bigg\}\,.
\end{eqnarray}
The scale of $\alpha_s$ remains undetermined in the 
double-logarithmic approximation.

To obtain the thrust distribution we set $s_R=s_L=s$ and 
choose $\sigma=1/(s e^{\gamma_E})$. 
For the inverse Laplace-transformation from $s$ to $\tau=1-T$ we use that the Laplace transform of $\tau^a$ is $\Gamma(1+a)/(sQ)^{1+a}$. Taylor-expanding this relation in $a$ gives 
\be
  \frac{1}{sQ}\ln^n\left(\frac{1}{s e^{\gamma_E} Q}\right) \rightarrow \ln^n\tau-\frac{\pi^2}{12}n(n-1)\ln^{n-2}\tau+{\cal O}(\ln^{n-3}\tau)\,,
\ee
where the arrow denotes inverse Laplace-transformation.
At the double-logarithmic level, the inverse 
Laplace-transformation is therefore simply accomplished by 
multiplying with $sQ$ and
substituting $s e^{\gamma_E}\to 1/(Q\tau)$ resulting in 
\begin{eqnarray}
\frac{1}{\sigma_0}\frac{ d\sigma}{ d\tau}|_{\rm DL} &=&
\frac{C_F}{C_F-C_A}\frac{1}{\ln(1/\tau)}
\,e^{-\frac{\alpha_s C_A}{\pi}\ln^2\tau}\,
\left\{
1-e^{-\frac{\alpha_s}{\pi}\klammer{C_F-C_A}\ln^2\tau} 
\right\}\,.
\label{eq:doublelogresult}
\end{eqnarray}
This agrees with the previous 
result \cite{Moult:2019uhz,Beneke:2020ibj}, which, 
however, was obtained through $d$-dimensional factorization 
and consistency relations.\footnote{The overall sign of (E.13) 
in \cite{Beneke:2020ibj} is given incorrectly. The signs of all 
previous equations leading to (E.13) are, however, correct.}

\subsection{Leading logarithms and running coupling effects}

The present approach based on renormalized functions 
satisfying standard RGEs allow us to go beyond the 
double-logarithmic limit. As a first application, we
consider the LL approximation, which includes 
one-loop running-coupling effects. The RGE functions 
read
\begin{align}
&A\klammer{\nu,\mu}=\frac{2}{\beta_0}\ln\frac{\alpha_s\klammer{\mu}}{\alpha_s\klammer{\nu}},\\
\nonumber&S\klammer{\nu,\mu}=\frac{4\pi}{\beta_0^2\alpha_s\klammer{\nu}}\left[1-\frac{\alpha_s\klammer{\nu}}{\alpha_s\klammer{\mu}}-\ln\frac{\alpha_s\klammer{\mu}}{\alpha_s\klammer{\nu}}\right],
\end{align}
and the strong coupling $\alpha_s$ at a scale $\nu$ is 
given by
\begin{align}
\alpha_s\klammer{\nu}=\frac{\alpha_s(\mu)}{1+\frac{\beta_0}{2\pi}\alpha_s\klammer{\mu}\ln\frac{\nu}{\mu}}
\label{eq:onelooprunningalphas}
\end{align}
in terms of the coupling at a reference scale $\mu$. 
In the following $\alpha_s$ without argument 
refers to $\alpha_s(\mu)$.

These expressions are now to be used in 
\eqref{eq:mainLLresult}, which upon 
implementing \eqref{eq:initialscales} (with ``$\sim$'' replaced 
by ``='' and $s_i$ by $s_i e^{\gamma_E}$)
reads\footnote{In an abuse of notation, we here 
give the scale arguments of $\alpha_s$, $A$ and $S$ as 
$\mu^2$ rather than $\mu$.}
\begin{eqnarray}
\frac{1}{\sigma_0}\frac{\widetilde{ d\sigma}}{ ds_R ds_L}|_{\rm LL}&=&
\frac{\alpha_s(Q/(s_L e^{\gamma_E})) C_F}{\pi}\,\frac{1}{Q s_R}
\nonumber\\
&&\hspace*{-2cm}
\times\,
\exp\left[4C_F S(Q^2,\frac{Q}{s_R e^{\gamma_E}})
+4C_A S(\frac{1}{s_L s_R e^{2\gamma_E}},\frac{Q}{s_L e^{\gamma_E}})\right]
\nonumber\\
&&\hspace*{-2cm}
\times\int_\sigma^Q\frac{d\omega}{\omega}\,
\exp\left[-4\klammer{C_F-C_A}S(\omega Q,\frac{\omega}{s_R e^{\gamma_E}})\right]
\nonumber\\
&&\hspace*{-1.2cm}\times\,
\klammer{s_R e^{\gamma_E} Q}^{
2C_F A\klammer{\omega/s_R  e^{\gamma_E},Q/s_R  e^{\gamma_E}}
+2C_A A\klammer{Q/s_L  e^{\gamma_E},\omega/s_R  e^{\gamma_E}}}
\,.
\label{eq:LLresult}
\end{eqnarray}
This is our main result for the resummed two-hemisphere 
invariant mass distribution in the gluon-thrust region. 
The integral over $\omega$ can be performed only numerically. 

An approximate analytic expression can be obtained by expanding 
the one-loop running coupling to linear order in $\beta_0$ in 
the exponent and prefactor, resulting in 
\begin{eqnarray}
\frac{1}{\sigma_0}\frac{\widetilde{ d\sigma}}{ ds_R ds_L}|_{\rm LL}
&=& 
\frac{\alpha_s C_F}{\pi}\,\frac{1}{Q s_R}\,
\,\bigg(1-\frac{\beta_0\alpha_s}{4\pi}
\ln\frac{Q}{s_L e^{\gamma_E} \mu^2}\bigg)
\nonumber\\
&&\hspace*{-1.5cm}\times\,
\exp\Bigg[-\frac{\alpha_s C_A}{\pi}\ln\klammer{s_L e^{\gamma_E} Q}\ln\klammer{s_R e^{\gamma_E} Q}
\bigg(1-\frac{\beta_0\alpha_s}{4\pi}
\ln\frac{Q}{\sqrt{s_R s_L} e^{\gamma_E} \mu^2}\bigg)\,\Bigg]
\nonumber\\
&&\hspace*{-1.5cm}\times\,\int_{\sigma}^Q\frac{ d\omega}{\omega}
\klammer{\frac{\omega}{Q}}^{\!\frac{\alpha_s}{\pi}\klammer{C_F-C_A}\ln\klammer{s_R e^{\gamma_E} Q}
\,\big(1-\frac{\beta_0\alpha_s}{4\pi}
\ln\frac{Q\sqrt{\omega}}{\sqrt{s_R e^{\gamma_E}}\mu^2}\big)}
\nonumber\\[0.1cm]
&&\hspace*{-2cm}=\,\frac{C_F}{Q s_R}\exp\Bigg[-\frac{\alpha_s}{\pi}C_A\ln\klammer{s_L e^{\gamma_E} Q}\ln\klammer{s_R e^{\gamma_E} Q}\klammer{1-\frac{\beta_0\alpha_s}{4\pi}\ln\frac{Q}{\sqrt{s_L s_R} e^{\gamma_E}\mu^2}}\Bigg]
\nonumber\\
&&\hspace*{-1.5cm}\times\,
\frac{\sqrt{2\pi}\left(1-\frac{\beta_0\alpha_s}{4\pi}\ln\frac{Q}{s_L e^{\gamma_E} \mu^2}\right)}
{\sqrt{\beta_0}\sqrt{C_A-C_F}\sqrt{\ln(s_R e^{\gamma_E} Q})}
\klammer{s_R e^{\gamma_E}  Q}^{\frac{2\klammer{C_F-C_A}}{\beta_0}\klammer{1-\frac{\beta_0\alpha_s}{4\pi}\ln\frac{Q^{\frac{3}{2}}}{\sqrt{s_R e^{\gamma_E}}\mu^2}}^2}\nonumber\\[0.2cm]
&&\hspace*{-1.5cm}\times\,\Bigg[\,
\mathrm{erfi}\klammer{
\sqrt{2 \,\frac{C_A-C_F}{\beta_0} \ln(s_R e^{\gamma_E} Q)}
\,\bigg(1-\frac{\beta_0\alpha_s}{4\pi}\ln\frac{\sigma\sqrt{Q}}{\sqrt{s_R e^{\gamma_E}}\mu^2}\bigg)}\nonumber\\
&&\hspace*{-0.8cm}-\,\mathrm{erfi}\klammer{
\sqrt{2 \,\frac{C_A-C_F}{\beta_0} \ln(s_R e^{\gamma_E} Q)}
\,\bigg(1-\frac{\beta_0\alpha_s}{4\pi}\ln\frac{Q^{\frac{3}{2}}}{\sqrt{s_R e^{\gamma_E}}\mu^2}\bigg)}\Bigg]\,,
\label{eq:linearbeta0LLresults}
\end{eqnarray}
where 
\begin{equation}
\mathrm{erfi}(x) = -\frac{2 i}{\sqrt{\pi}}\int_0^{i x} dt\,e^{-t^2}\,.
\end{equation}
Once again, the thrust distribution in Laplace-space is obtained 
by identifying $s_R=s_L\equiv s$, in which case we also 
choose $\sigma=1/(s e^{\gamma_E}) $. 

\begin{figure}[p]
\begin{center}
\includegraphics[width=0.7\textwidth]{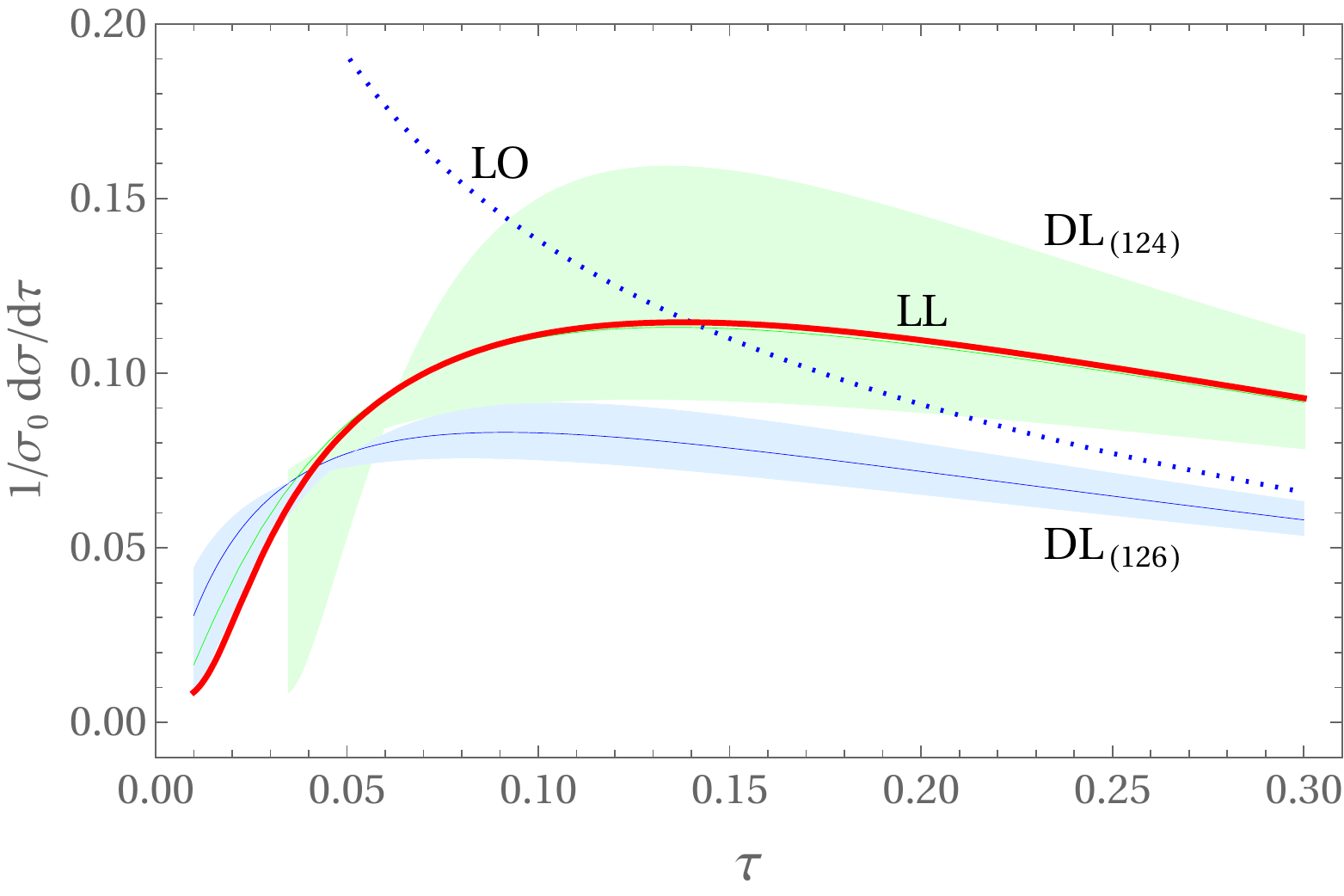}\\[0.7cm]
\includegraphics[width=0.7\textwidth]{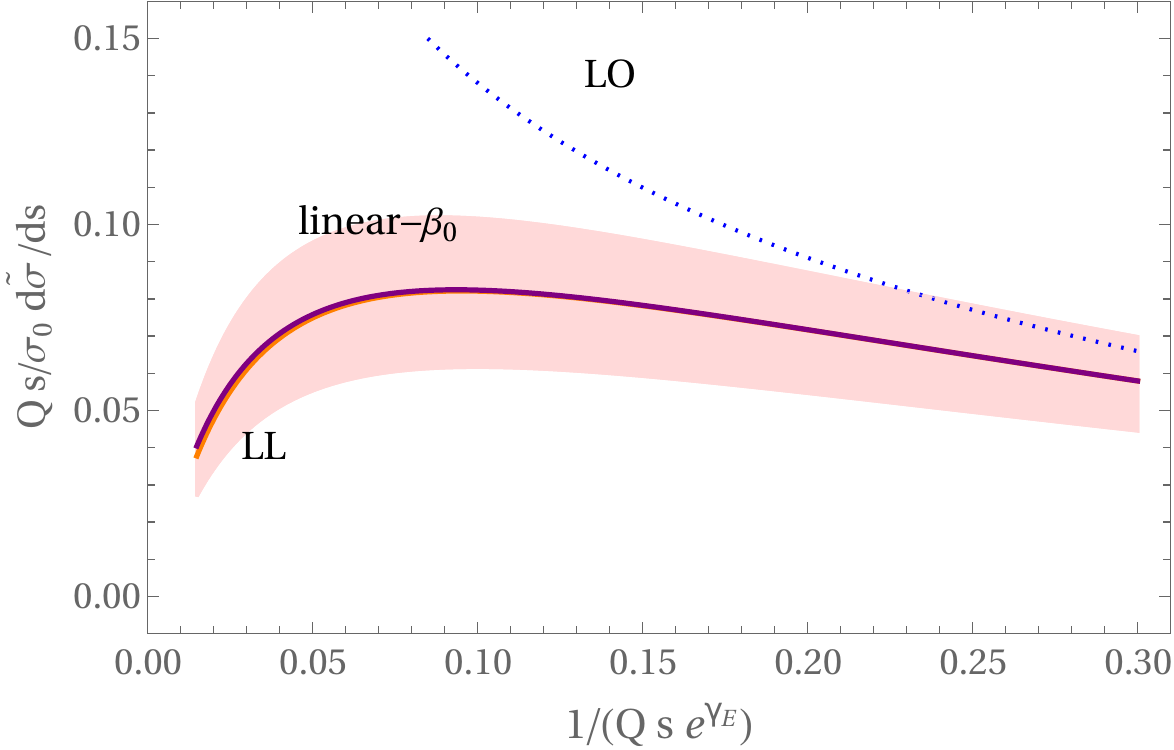}\\[0.2cm]
\caption{\label{fig:numerics}
Upper panel: Gluon thrust distribution 
$\frac{1}{\sigma_0}\frac{ d\sigma}{ d\tau}$ in the double-logarithmic 
approximation including scale variation (light-blue band from 
\eqref{eq:doublelogresult} and light-green band from the numerical 
inverse Laplace transform of \eqref{eq:dbllog}), and the 
LL  approximation (red line). 
The tree-level result is shown in dotted-blue.
Lower panel: Laplace-space gluon thrust distribution  (times $Q s$),
$\frac{Q s}{\sigma_0}\frac{\widetilde{ d\sigma}}{ ds}$, 
in the LL approximation (red line), allowing for a variation of the 
initial scales as described in the text (light-red band). Also shown is the 
truncation of the LL result to linear order in $\beta_0$ from 
\eqref{eq:linearbeta0LLresults} (blue), but the difference is hardly 
visible.}
\end{center}
\end{figure}

To display the effect of resummation and of running coupling 
effects, we show in the upper panel of 
Fig.~\ref{fig:numerics} the gluon-thrust 
distribution $\frac{1}{\sigma_0}\frac{ d\sigma}{ d\tau}$ in 
the leading $\mathcal{O}(\alpha_s)$ approximation 
(blue-dotted line) and in the resummed double-logarithmic 
approximation \eqref{eq:doublelogresult} (light-blue band). 
We adopt $Q=m_Z=91.1876\,$GeV corresponding to final states from 
hadronic $Z$-boson decay, and $\alpha_s(m_Z)=0.1179$. The coupling 
is evolved with one-loop accuracy and $n_l=5$ according to 
\eqref{eq:onelooprunningalphas}. In the leading-order approximation 
we choose $\alpha_s$ at the collinear scale $\sqrt{\tau} Q$. 
Since the scale of $\alpha_s$ is not fixed in the double-logarithmic 
approximation, we vary it from the soft scale $\tau Q$ to the 
hard scale $Q$, which produces the shown band and a considerable 
uncertainty. For comparison, we also show (light-green band) 
the numerical inverse Laplace transform of the double-logarithmic 
approximation \eqref{eq:dbllog} in Laplace space. 
To this end, we compute 
\begin{equation}
\frac{1}{\sigma_0}\frac{ d\sigma}{ d\tau} = \frac{Q}{2\pi i}
\int_{c-i\infty}^{c+i\infty} \!\! ds\, e^{s\tau Q}\,
\frac{1}{\sigma_0}\frac{\widetilde{ d\sigma}}{ ds_R ds_L}|_{s_R=s_L=s},
\label{eq:inverseLaplacedist}
\end{equation}
where $c>0$ is chosen to enclose the singularities of 
the Laplace-space distribution, excluding the one from the 
Landau pole of the running coupling, which corresponds to  
the  ``minimal prescription'' of \cite{Catani:1996yz}.
We also set 
$\sigma = 1/(s e^{\gamma_E})$ in \eqref{eq:dbllog}. We observe 
a significant difference between the two bands, which must 
be attributed to formally next-to-leading logarithmic effects 
in the relation between Laplace and momentum space. 

Finally, the 
upper plot shows (red line) the 
full leading-logarithmic result \eqref{eq:LLresult} 
(with the $\omega$ integral 
evaluated numerically for $\sigma = 1/(s e^{\gamma_E})$) after 
taking the inverse Laplace transformation 
\eqref{eq:inverseLaplacedist} numerically. The LL result
turns out to be very close to the DL one when the 
inverse Laplace transformation is computed in the same way. 
The green band ends at $\tau\approx 0.03$, 
since the DL result at the soft scale 
becomes sensitive to the strong coupling regime 
at smaller values $\tau$. 
We also recall that perturbative 
resummation requires 
$Q\tau\gg \Lambda_{\rm QCD}\approx 0.5\,$GeV and $\tau\ll 1$.

We can gain some insight on the importance of intrinsic 
{\em next}-to-leading logarithms by varying in the LL result the 
various matching scales around the 
values adopted in \eqref{eq:LLresult}. For this purpose, we 
vary the three pairs of scales $(\mu_h, \mu_{h\Lambda})$, 
$(\mu_c,\mu_{\bar c})$, $(\mu_s, \mu_{s\Lambda})$ by a factor of $1/2$ 
and 2 around their default scales. We then take the minimum 
and maximum values 
of the 15 possible combinations of $(1,\frac{1}{2},\frac{1}{2})$ 
etc. to compute the scale variation, 
excluding the cases where different scales are 
varied in different directions. For simplicity, we show this 
result for the normalized Laplace-space distribution 
$\frac{Q s}{\sigma_0}\frac{\widetilde{ d\sigma}}{ ds}$ in the 
lower panel of Fig.~\ref{fig:numerics} as the light-red 
band around the red curve (LL) that represents 
\eqref{eq:LLresult}. For comparison the tree-level (LO) and 
linear-$\beta_0$ truncation \eqref{eq:linearbeta0LLresults} 
of the LL expression are displayed, but the difference between 
the latter approximation and the full result is hardly visible. 
The sizeable scale variation seen in the figure emphasizes the need for 
NLL resummation. The renormalized and endpoint-rearranged 
factorization formula derived in this work provides the 
starting point for this systematic improvement. However, the 
actual implementation requires the calculation of the presently 
unknown non-cusp parts of the $\mathcal{O}(\alpha_s)$ anomalous 
dimension of 
$\mathcal{J}_c(p^2,\omega,\omega')$ and $S_{\rm NLP}(l_+, l_-,
\omega,\omega')$  
for general $\omega^{(\prime)}$, which require 
two-loop calculations beyond the scope of this work, as 
well as the  $\mathcal{O}(\alpha_s^2)$ cusp parts of the anomalous 
dimensions of the NLP objects.


\section{Conclusion}
\label{sec:conclusion}

The lack of convergence of the convolution integrals appearing in subleading-power factorization theorems is currently the biggest obstacle to resummation of the power-suppressed large logarithmic corrections in collider physics. This applies in particular to the classic 
$1\to 2$ and $2\to 1$ processes, such as event shapes in 
$e^+ e^-$ annihilation,  deep-inelastic scattering for $x\to 1$, 
and the Drell-Yan threshold. 

In earlier work \cite{Beneke:2020ibj} on off-diagonal deep-inelastic scattering for $x\to 1$, we showed that the hard function of the subleading power B-type operator factorizes in the limit when the collinear momentum fraction carried by one of the quark fields tends to zero. To avoid dealing with the non-perturbative parton distributions, in this work we considered the analogue of the off-diagonal channel for $e^+ e^-$ annihilation, 
``gluon thrust''. Earlier results on the double-logarithmic 
resummation of this next-to-leading power kinematic configuration 
employed $d$-dimensional factorization and consistency relations
\cite{Moult:2019uhz,Beneke:2020ibj}, which do not readily generalize 
beyond the leading logarithms.

In this paper, we derived a novel endpoint factorization relation for 
the  NLP thrust distribution in the two-jet region for the 
flavour-nonsinglet off-diagonal 
contribution, where a gluon-initiated jet recoils against 
a quark-antiquark pair, which involves sub\-leading-power soft and jet 
functions defined at the cross-section level. The above-mentioned 
factorization of the subleading power B-type operator plays again 
a crucial role for the consistency of the rearrangement that renders 
the expressions free from endpoint divergences. The framework 
developed in the present paper allows for the first time to remove 
systematically endpoint divergences in the convolution integrals of the SCET$_{\rm I}$ factorization theorems and hence opens the path 
to systematic next-to-leading-power resummation for collider 
observables involving soft quark emission. Employing the subtraction 
of endpoint divergences with the help of operatorial endpoint
 factorization conditions, we were able to reshuffle the factorization 
theorem such that the individual terms are free from 
endpoint divergences in convolutions and can be expressed in 
terms of renormalized hard, soft and collinear functions in four 
dimensions. At this point, standard renormalization-group 
techniques can be used to obtain the resummed integrands. The 
basic structure of the rearrangements turns out to be surprisingly 
similar to the one for Higgs decay through 
light-quark loops into two photons  \cite{Liu:2019oav,Liu:2020wbn}, 
which however is a SCET$_{\rm II}$ process, where factorization 
applies to the amplitude. For the $1\to 2$ and $2\to 1$ processes 
considered here, there are no rapidity divergences releated to momentum modes of the same virtuality, and 
dimensional regularization suffices to make the convolutions 
well-defined \cite{Beneke:2018gvs,Beneke:2019oqx}, yet their 
divergence as $\epsilon\to 0$ requires rearrangement and 
refactorization as discussed here.

For the NLP thrust distribution in the power-suppressed two-jet 
region, where a gluon-initiated jet recoils against a quark-antiquark 
pair, we derived the necessary anomalous dimension of the NLP jet 
and soft 
functions using renormalization-group consistency and endpoint 
factorization relations. We also explicitly computed the one-loop anomalous dimension for the hard matching coefficients. These ingredients allowed us to perform the first resummation of the endpoint-divergent 
SCET$_{\rm I}$ observables at the LL accuracy using exclusively renormalization-group methods. After resummation of the integrands, a judicious 
choice of initial conditions is necessary for the endpoint-subtraction terms, requiring a scale setting depending on the value of the 
convolution variable. We verified that our results simplified to double logarithmic accuracy agrees with the earlier results \cite{Moult:2019uhz,Beneke:2020ibj} and evaluated the numerical impact of the new 
LL corrections. Our main technical result for gluon thrust is 
\eqref{eq:LLresult}, which provides a concise expression for 
the two-hemisphere invariant mass distribution in Laplace space.

The presented method relies on universal properties of the soft and 
collinear limits and can be applied to other SCET$_{\rm I}$ problems. 
An immediate target for the further exploration of NLP resummation 
is the extension of the present work for gluon thrust and related 
soft-quark emission processes to NLL accuracy, which requires 
the computation of renormalization kernels for the NLP soft and jet functions at order $\mathcal{O}(\alpha_s)$, i.e., at the two-loop level 
(counting phase-space integrals as loops). 
For the {\em diagonal} processes, the leading NLP logarithms 
have been considered for the thrust distribution and the 
DY threshold in \cite{Moult:2018jjd,Beneke:2018gvs,Bahjat-Abbas:2019fqa,Beneke:2019mua}, and they turn out to be almost trivial in 
comparison with the ones for the off-diagonal processes, as there 
are no endpoint divergences. This is no longer expected beyond the 
LL accuracy \cite{Beneke:2019oqx}. The present framework supplies a 
method to address the endpoint divergences that appear in the soft 
gluon limit at NLP, which is relevant to the diagonal processes. 

\subsubsection*{Acknowledgements} 

This work has been supported in part by the Excellence Cluster 
ORIGINS funded by the Deutsche Forschungsgemeinschaft 
(DFG, German Research Foundation) under Ger\-many's Excellence Strategy --EXC-2094 --390783311. S.J. is supported by the UK Science and Technology Facilities Council (STFC) under grant ST/T001011/1. R.S. is supported by the United States Department of Energy under Grant Contract DE-SC0012704. L.V. is supported by Fellini Fellowship for Innovation at INFN, funded by the European Union's Horizon 2020 research programme under the Marie Sk\l{}odowska-Curie Cofund Action, grant agreement no. 754496. J.W. was supported in part by the National Natural Science Foundation of China (No.~12005117) 
and the Taishan Scholar Foundation of Shandong province (tsqn201909011).
Figures were drawn with \texttt{Jaxodraw}~\cite{Binosi:2008ig}. 


\begin{appendix}

\section{\boldmath $C^{\rm B1}$ and $D^{\rm B1}$}
\label{sec:DB1}

In this appendix we provide explicit results for the
hard matching coefficient $C^{\rm B1}_{i}(Q^2,r)$ associated to
the SCET operators defined in the second line of 
\eqref{eq:hardmatching} and \eqref{eq:BtypeDirac}. In 
momentum space, 
\be
J^{B1,\mu}_i(r) = \bar\chi_{\bar c}( r)\Gamma_i^{\mu\nu}\mathcal{A}_{\nu\perp c}\chi_{\bar c}(\bar r),\quad i=1,2\,,
\label{eq:B1momspace}
\ee
where $r$ and $ \bar r=1-r$ denote the momentum fractions of the anti-collinear quark and anti-quark, respectively, and
\be
  \Gamma_1^{\mu\nu} = \frac{\slashed{n}_-}{2}\gamma_\perp^\nu\gamma_\perp^\mu,\qquad
  \Gamma_2^{\mu\nu} = \frac{\slashed{n}_-}{2}\gamma_\perp^\mu\gamma_\perp^\nu\,.  
\ee
We then obtain the one-loop correction and one-loop evolution equation 
of the refactorization coefficient $D^{\rm B1}(p^2)$ by taking 
the $r\to 0$ limit of the corresponding results for  
$C^{\rm B1}_{1}(Q^2,r)$ and \eqref{eq:B1fact1},
\bea
\label{eq:DB1def2}
C^{\rm B1}_1(Q^2,r) &=& C^{\rm A0}(Q^2)\,\frac{D^{\rm B1}(rQ^2)}{r}
+\text{endpoint-regular}\,.
\eea
Here endpoint-regular contributions diverge at most logarithmically 
in the limit $r\to 0$.


\subsection{\boldmath One-loop results}
\label{sec:CB1}

Allowing for the one-loop correction to \eqref{eq:CB11tree}, 
we write 
\be\label{CB1loopgeneralform}
C_1^{\rm B1}(Q^2,r) =
\frac{1}{r} \left(1+\frac{\as}{4\pi} \,\Delta^{\rm B1}(Q^2,r)
+\mathcal{O}(\alpha_s^2)\right).
\ee
We recall that charge conjugation implies 
$C_2^{\rm B1}(Q^2,r) = -C_1^{\rm B1}(Q^2,\bar{r})$, 
hence it suffices to 
discuss the case $i=1$. A standard matching calculation of 
the $\gamma^*\to q\bar{q}g$ amplitude results in 
\bea\label{deltaB1full} \nn
\Delta^{\rm B1}(Q^2,r) &=&
\bigg[\bigg(C_F-\frac{C_A}{2}\bigg)
\bigg(- \frac{2}{\eps^2} + \frac{2}{\bar r \eps}
+ \frac{4 r}{\bar r} +\frac{\pi^2}{6} \bigg)
+C_F \frac{2}{\bar r} \bigg]
\bigg(-\frac{\mu^2}{Q^2}\bigg)^{\eps} \\ \nn
&&\hspace{-2.0cm}+\,\bigg[C_F \bigg(\frac{2}{\eps^2} - \frac{2 r}{\bar r \eps}
- \frac{1 + 5 r}{\bar r} -\frac{\pi^2}{6} \bigg)
+C_A \bigg(- \frac{2}{\eps^2} + \frac{r}{\bar r \eps}
+ \frac{1 + r}{\bar r} + \frac{\pi^2}{6}\bigg)\bigg]
\bigg(-\frac{\mu^2}{r Q^2}\bigg)^{\eps} \\
&&\hspace{-2.0cm}+\,\bigg[C_F \bigg(- \frac{5}{\eps} -10 \bigg)
+C_A \bigg(- \frac{1}{\eps^2} + \frac{1}{\eps} + \frac{\pi^2}{12}\bigg)\bigg]
\bigg(-\frac{\mu^2}{\bar r Q^2}\bigg)^{\eps}
+\ord(\eps)\,.
\eea
The limit $r \to 0$ is of particular interest, and results in
\be
\Delta^{\rm B1}(Q^2,r) \stackrel{r\to 0}{=} C_F \, \Delta^{A0}
\bigg(-\frac{\mu^2}{Q^2}\bigg)^{\eps}
+\big(C_F- C_A \big)
\bigg(\frac{2}{\eps^2} -1 - \frac{\pi^2}{6} \bigg)
\bigg(-\frac{\mu^2}{r Q^2}\bigg)^{\!\eps}
+\ord(\eps,r),
\ee
which shows that the B1 coefficient depends on two scales, 
$Q^2$ and $rQ^2$. We can identify 
\be
\Delta^{A0} = - \frac{2}{\eps^2} - \frac{3}{\eps}
-8 +\frac{\pi^2}{6} + \ord(\eps)
\ee
with the one-loop contribution to the leading-power 
hard matching coefficient coefficient $C^{\rm A0}$:
\be
C^{\rm A0}(Q^2) = 1 + \frac{\as C_F }{4\pi} \Delta^{A0}
\left(-\frac{\mu^2}{Q^2}\right)^{\!\eps} +\ord(\as^2).
\ee
Comparing to \eqref{eq:B1fact1}, we find 
\be
D^{\rm B1}(p^2) = 1 + \frac{\as}{4\pi}\,
\big(C_F- C_A \big)
\bigg(\frac{2}{\eps^2} -1 - \frac{\pi^2}{6} \bigg)
\left(-\frac{\mu^2}{p^2}\right)^{\!\epsilon} +\ord(\as^2).
\label{eq:DB11loopapp}
\ee
For completeness, we check that the $r\to 1$ limit
of $\Delta^{\rm B1}(Q^2,r)$ does not contain $1/\bar{r}$ terms:
\bea \nonumber
\Delta^{\rm B1}(Q^2,r) &\stackrel{r\to 1}{=}&
\bigg[C_F \bigg(\frac{2}{\eps} -1 \bigg)
+C_A \bigg(-\frac{1}{\eps^2} - \frac{1}{\eps}
+2 + \frac{\pi^2}{12} \bigg) \bigg]
\bigg(-\frac{\mu^2}{Q^2}\bigg)^{\eps} \\
&&\hspace{-2.0cm}+\,\bigg[C_F \bigg(- \frac{5}{\eps} -10 \bigg)
+C_A \bigg(- \frac{1}{\eps^2} + \frac{1}{\eps} + \frac{\pi^2}{12}\bigg)\bigg]
\bigg(-\frac{\mu^2}{\bar r Q^2}\bigg)^{\eps}
+\ord(\eps,\bar r)\,.
\label{eq:DeltaB1xto1limit}
\eea


\subsection{\boldmath Derivation of the evolution equation of the 
asymptotic refactorization coefficient $D^{\rm B1}$}
\label{app:asympDB1RGE}

Our goal is to show that the RGE \eqref {eq:DBQrge} for the 
matching coefficient $D^{\rm B1}(r Q^2)$ can be derived 
from the standard evolution equation of the hard-matching 
coefficient $C_i^{\rm B1}(Q^2,r)$ for
generic value of $r$. We note that this finding is distinct 
from the case of B1 operators containing a quark and a
gluon in the same collinear direction, where the 
soft-gluon limit needs to be treated separately from the
endpoint-regular piece~\cite{Beneke:2019kgv}.

The renormalization of SCET operators of the B1 type has 
been studied in detail 
in~\cite{Beneke:2017ztn,Beneke:2018rbh}, except for 
the fermion-number zero case, to which the operators 
$J^{B1,\mu}_i(r)$ in \eqref{eq:B1momspace} belong.
The anomalous dimension matrix is determined by computing the $\overline{\text{MS}}$ renormalization factors 
$Z_{ij}^{\mu\nu}(r,s)=\delta_{ij}g^{\mu\nu}_\perp\delta(r-s)+\delta Z_{ij}^{\mu\nu}(r,s)$ defined via
\be
  J^{B1,\mu}_{i,\text{ren}}(r) = \int ds\, Z_{ij}^{\mu\nu}(r,s) J^{B1,\nu}_{j,\text{bare}}(s)\,.
\ee
Following the lines of~\cite{Beneke:2017ztn,Beneke:2018rbh}, we find at one-loop order $\delta Z_{ij}^{\mu\nu}(r,s)=g^{\mu\nu}_\perp \delta Z_{ij}(r,s)$ with 
\bea
  \delta Z_{11}(r,s) &=& \delta Z_{22}(r,s) = -\frac{\alpha_s}{2\pi}\delta(r-s)\left[\frac{C_A}{\epsilon^2}+\frac{C_A}{\epsilon}\ln\left(\frac{\mu^2}{-Q^2-i\varepsilon}\right)-\frac{C_A}{2\epsilon}\ln(r\bar r)+\frac{3C_F}{2\epsilon}\right]\nn\\
  && {} +\frac{\gamma(r,s)}{\epsilon}\,,\\
  \delta Z_{12}(r,s) &=& \delta Z_{21}(r,s) = \frac{\alpha_s}{2\pi}\Delta_F\frac{2T_F}{\epsilon}r\bar r \,,
\eea
where
\bea\label{eq:gamma}
  \gamma(r,s) &=& \frac{\alpha_s}{2\pi}\,\Bigg\{ \left(\frac{C_A}{2}-C_F\right)\Bigg[ \frac{\bar r}{\bar s}\left(\left(\frac{\theta(r-s)}{r-s}\right)_+ + \theta(r-s)\right)\nn\\
  && {} + \frac{r}{ s}\left(\left(\frac{\theta(s-r)}{s-r}\right)_+ + \theta(s-r)\right) \Bigg]
   +2\Delta_F T_F r\bar r\Bigg\} \,.
\eea
The momentum fractions have support in the interval $0\leq r,s\leq 1$.
Furthermore, $T_F=1/2$, and $\Delta_F=1\;(0)$ in the flavour-singlet (non-singlet) projection of the anti-collinear 
$\bar\chi_{\bar c}\chi_{\bar c}$ fields in $J^{B1,\mu}_i$. The expression 
in square brackets defines a symmetric distribution, which applies 
to integration over test functions of $r$ or $s$, 
with definition 
\bea
\int dr\, f(r) \,[D(r-s)]_+ &=& 
\int dr\, D(r-s) \,(f(r)-f(s))\,,
\\
\int ds\, [D(r-s)]_+ \,g(s) &=& 
\int ds\, D(r-s) \,(g(s)-g(r))\,.
\eea 
The definition applies to any integration range, which is left unspecified here. In~\eqref{eq:gamma} the integrations are limited to
$0\leq r,s\leq 1$, but we will also consider the case $0\leq r,s\leq \infty$ below.

The anomalous dimension can then be obtained as
\be
  {\bf \Gamma} = {\bf Z}\frac{d}{d\ln\mu}{\bf Z}^{-1}\,,
\ee
giving the RGEs
\be
  \frac{d}{d\ln\mu} C^{\rm B1}_j(Q^2,s) = \int_0^1 dr \, C^{\rm B1}_i(Q^2,r) \Gamma_{ij}(r,s) \,.
\ee
Here we already used that the current renormalization for the specific operators $J^{\rm B1,\mu}_i(r)$ considered here is diagonal in Lorentz indices, such that
they can be dropped in the Wilson coefficients and anomalous dimension matrix.

We checked that convolving the tree-level coefficients $C^\text{B1}_i(Q^2,r)$ with $\delta Z_{ij}(r,s)$ agrees with the
divergent part of the explicit one-loop result for $C^\text{B1}_j(Q^2,s)$ as given in the previous subsection for the flavour non-singlet projection, including both the contributions that are singular as well as those that are regular with respect to the dependence on $s$. This confirms the observation that the standard evolution equation for the B1 current can be used in this context, despite of the soft (anti-)quark singularity for $r\to 0$ ($\bar r\to 0$). In the following we are interested in extracting from
the evolution equation the part that describes the renormalization of the endpoint-singular piece captured by $D^\text{B1}$. 

The all-order identity $C^{\rm B1}_2(Q^2,r) = 
- C^{\rm B1}_1(Q^2,\bar r)$ is consistent with the relations 
$\delta Z_{11}(r,s)=\delta Z_{22}(r,s)=\delta Z_{22}(\bar r,\bar s)$ and $\delta Z_{12}(r,s)=\delta Z_{21}(r,s)=\delta Z_{21}(\bar r,\bar s)$. Furthermore it can be checked that $\delta Z_{12}(r,s)$ does not contribute to the renormalization of $D^{\rm B1}(r Q^2)$, due to the independence on $s$ (see below).
Therefore it is sufficient to consider  $Z(r,s)\equiv \delta(r-s)+\delta Z_{11}(r,s)$ in the following.

The matching coefficient $C^{\rm B1}_1(Q^2,r)$ for the problem at hand features a power-like endpoint singularity $\propto 1/r$ for $r\to 0$, that may be
modulated by logarithmic corrections, and is captured by the $D^{\rm B1}$ contribution in~\eqref{eq:B1fact1}.
In order to extract the renormalization factor for $D^{\rm B1}$, we are interested in the convolution with $\Gamma_{11}(r,s)$ (or equivalently $Z(r,s)$ at order $\alpha_s$).
The result of the convolution can be split into an endpoint divergent piece $\propto 1/s$, and a regular part. The former describes the renormalization of $D^{\rm B1}$, and the latter yields a mixing of $D^{\rm B1}$ into the endpoint-finite part of $C^{\rm B1}_1(Q^2,r)$. Here we are interested only in the endpoint-singular contribution. In order to isolate it, we consider test functions that
represent the most general possible dependence of the Wilson coefficient on the momentum fraction, and investigate the integral
\be\label{eq:fzF}
  \int dr \,\frac{f(r)}{r} Z(r,s) = \frac{F(s)}{s}+\text{endpoint-regular}\,.
\ee
Here $f(r)$ is an arbitrary test function defined for $r\geq 0$, that depends logarithmically on $r$, and represents the $D^{\rm B1}$ coefficient. 
For example, $f(r)$ could be given by some power of $\ln(r)$ or a more general polylogarithmic dependence.
The right-hand side states that the result can be written as an endpoint-singular piece, with some function $F(s)$ that also depends logarithmically on $s$, and
an endpoint-regular contribution. Below we show that this is indeed the case, and that $F(s)$ can be obtained via
\be
  F(s) = \int dr f(r) \,Z_{A0A0}\times Z_{D^{\rm B1}D^{\rm B1}}(r,s)\,,
\ee
with a renormalization factor $Z_{D^{\rm B1}D^{\rm B1}}(r,s)$, and 
$Z_{A0A0}$ the renormalization constant for the leading-power 
coefficient $C^{\rm A0}(Q^2)$, being given at order $\alpha_s$ by
\be
  Z_{A0A0}=1-\frac{\alpha_s}{2\pi}\left[\frac{C_F}{\epsilon^2}+\frac{C_F}{\epsilon}\ln\left(\frac{\mu^2}{-Q^2-i\varepsilon}\right)+\frac{3C_F}{2\epsilon}\right]\,.
\ee
The strategy for obtaining $Z_{D^{\rm B1}D^{\rm B1}}$ is to apply the method of regions \cite{Beneke:1997zp} to~\eqref{eq:fzF}, and expanding the integrand such that a homogeneous scaling in the limit $r,s\to 0$ is obtained.

As a first step, we need to check whether the integral over $r$ exists, despite the endpoint-singular factor $1/r$.
This is trivially the case for the local contribution to $Z(r,s)$, and we therefore only need to check the piece
containing $\gamma(r,s)$,
\bea\label{eq:fg}
  \int dr \,\frac{f(r)}{r} \gamma(r,s) &=& \frac{\alpha_s}{2\pi}\left(\frac{C_A}{2}-C_F\right)  \Bigg[ \int_s^1 dr  \left(\frac{\frac{\bar r}{\bar s} \frac{f(r)}{r}-\frac{f(s)}{s}}{r-s} + \frac{\bar r}{\bar s} \frac{f(r)}{r}\right)\nn\\
  && \hspace*{-2cm} + \int_0^s dr  \left(\frac{\frac{r}{ s}\frac{f(r)}{r}-\frac{f(s)}{s}}{s-r} + \frac{r}{s}\frac{f(r)}{r}\right) \Bigg]
  + \frac{\alpha_s}{\pi}\Delta_F T_F \int_0^1 dr\, r\bar r \,\frac{f(r)}{r}\,.\quad
\eea
The integrals in the second line exist due to the explicit factor $r$ in the second line of~\eqref{eq:gamma}, that cancels the endpoint-singularity $1/r$. The first line has no endpoint singularity due to the lower integration boundary. 

As the next step, we need to expand the integrand assuming a homogeneous scaling for $r,s\to 0$.
The first integral in the second line of~\eqref{eq:fg} has already a homogeneous scaling, and only the first term in the round bracket contributes to $F(s)/s$, while the second term yields an endpoint-regular contribution $\int_0^s dr f(r)/s$. 
In addition, the integral in the ``singlet'' term proportional to $\Delta_F$ is independent of $s$ and can therefore not contribute to $F(s)$ in~\eqref{eq:fzF}.

Let us now turn to the first line of~\eqref{eq:fg}. Expanding in the region of small $r,s$ allows us to replace $\bar r/\bar s\to 1$. Next, we see that the integral $\int_s^1 dr f(r)/r$ contributes only to the endpoint-finite part
on the right-hand side of~\eqref{eq:fzF}, and can therefore be dropped in the present discussion. However, the first term in the round bracket in the first line of~\eqref{eq:fg} yields contributions that
scale as $1/s$. To isolate them, we replace the upper integration boundary $1$ by a Heaviside function $\theta(1-r)$ within the integrand. The method of regions instructs us to homogenize also the Heaviside function, corresponding to extending the upper integration boundary to $+\infty$. At this point, we use that the test function $f(r)$ depends logarithmically on $r$. It can therefore be extended over the interval $0\leq r <\infty$, in accordance with the method-of-region expansion.
However, this procedure is too naive, since it implies that the integral diverges for $r\to\infty$. This problem is linked to the observation that the respective integral yields contributions that are logarithmically enhanced as $\ln(s)/s$
for $s\to 0$. They can in turn be isolated by the rewriting
\be
  \int_s^1 dr  \,\frac{f(r)/r-f(s)/s}{r-s} = \int_s^1 dr  \, 
\frac{f(r)-f(s)}{r(r-s)} + \frac{f(s)}{s}\ln(s)\,.
\ee
After this rewriting we can extend the integration range of the remaining integral to $s\leq r<\infty$, yielding the desired homogeneous scaling.\footnote{We note the similarity of this 
procedure to the extraction of the ``asymptotic kernel'' for 
the QED-generalized light-cone distribution amplitude of 
a light meson in \cite{Beneke:2021pkl}, which involves 
a colour-singlet but electrically charged operator.}

Note that the relation above could equivalently be written at the level of plus-distributions, but we prefer to keep the explicit form here in order to be clear about the treatment of the integration domain in the various steps.

Altogether, this implies the contributions that scale
as $1/s$ up to logarithmic corrections have been isolated, and are given by
\bea
  \int dr \,\frac{f(r)}{r} \gamma(r,s)\Big|_{\propto \ln()/s} &=& \frac{\alpha_s}{2\pi}\left(\frac{C_A}{2}-C_F\right)  \Bigg[ \int_s^\infty dr  \left(\frac{f(r)-f(s)}{r(r-s)} \right)+ \frac{f(s)}{s}\ln(s)\nn\\
  && {} +  \int_0^s dr  \left(\frac{f(r)-f(s)}{s(s-r)} \right) \Bigg]\,.
\eea
By comparing with~\eqref{eq:fzF} we can read off the renormalization coefficient of $D^{\rm B1}$,
\bea
  Z_{D^{\rm B1}D^{\rm B1}}(r,s) &=& \frac{\alpha_s(C_F-C_A)}{2\pi}\,\delta(r-s)\left[\frac{1}{\epsilon^2}+\frac{1}{\epsilon}\ln\left(\frac{\mu^2}{-rQ^2-i\varepsilon}\right)\right]+\frac{\hat\gamma(r,s)}{\epsilon}\,,\quad
\eea
where
\bea
  \hat\gamma(r,s) &=& \frac{\alpha_s}{2\pi} \left(\frac{C_A}{2}-C_F\right) 
s \left[\frac{\theta(r-s)}{r(r-s)} 
  + \frac{\theta(s-r)}{s(s-r)}\right]_+\,,
\eea
and the plus distributions are defined with respect to the interval $0\leq r,s<\infty$ here.
The anomalous dimension is, at order $\alpha_s$, given by
\be
  \Gamma_{D^{\rm B1}D^{\rm B1}}(r,s) = -\frac{d}{d\ln \mu}Z_{D^{\rm B1}D^{\rm B1}}(r,s) = 2\epsilon Z_{D^{\rm B1}D^{\rm B1}}(r,s)-\frac{\partial Z_{D^{\rm B1}D^{\rm B1}}}{\partial\ln\mu}\,,
\ee
which gives
\bea
  \Gamma_{D^{\rm B1}D^{\rm B1}}(r,s) &=&\frac{\alpha_s(C_F-C_A)}{\pi}\,\delta(r-s)\ln\left(\frac{\mu^2}{-rQ^2-i\varepsilon}\right)
   + 2\hat\gamma(r,s)\,.
\eea
We can express the result equivalently in terms of $\omega$ and $\hat\omega$ using $s=\omega/Q, r=\hat \omega/Q$.
This yields
\be
  \frac{d}{d\ln\mu}D^{\rm B1}(\omega) = \int_0^\infty \!d\hat\omega 
\,\gamma_D(\hat\omega,\omega)D^{\rm B1}(\hat\omega)\,,
\label{eq:DB1RGEinomega}
\ee
with
\bea
  \gamma_D(\hat\omega,\omega) &=& \frac{\alpha_s(C_F-C_A)}{\pi}\delta(\omega-\hat\omega)\ln\left(\frac{\mu^2}{-\omega Q-i\varepsilon}\right)\nn\\
  && +\,\frac{\alpha_s}{\pi} \left(\frac{C_A}{2}-C_F\right)
 \omega \left[\frac{\theta(\hat\omega -\omega)}{\hat\omega (\hat\omega -\omega)}
  + \frac{\theta(\omega-\hat\omega )}{\omega(\omega-\hat\omega )}\right]_+\,.
\label{eq:DB1RGEinomegagamma}
\eea
This result is consistent with Eq.~(3.2) in~\cite{Liu:2021mac} obtained by a different method, when rewriting $\hat\omega\to x \omega$ and $\omega Q\to p^2$ in terms of the variables
$x$ and $p^2$ used there, and accounting for the different convention of the order of arguments of $\gamma_D$ in \eqref{eq:DB1RGEinomega}.


\section{Alternative version of the endpoint-finite 
factorization formula}
\label{version1endpintfact}

In this appendix consider the alternative split of the integration regions mentioned in Sec.~\ref{sec:endpointfact}, i.e. we present the factorization formula for the version (1), 
corresponding to $\omega$ {\em and} $\omega'$ smaller than 
an endpoint factorization parameter $\Lambda$ 
(integral $I_1$), and the complement region $I_2$, see Fig.~\ref{fig:overlap}.

Subtracting the complement region $I_2$, the A-type 
term takes the from 
\begin{eqnarray}
\frac{1}{\sigma_0}\frac{\widetilde{d\sigma}}{ds_R ds_L}|_{\rm A-type} &=& 
\frac{2C_F}{Q} \,f(\eps)\,|C^{\rm A0}(Q^2)|^2 \,
\widetilde{\mathcal{J}}_{\bar c}^{(\bar q)}(s_R)
\,\int d\omega d\omega'\,
\nonumber\\
&&\hspace*{-3cm}\times\,
\bigg\{\,\widetilde{\mathcal{J}}_{c}(s_{L},\omega,\omega')
\,\widetilde{S}_{\rm NLP}(s_R,s_L,\omega,\omega')
\nonumber\\
&&\hspace*{-2cm}- 
\left[1-\theta(\Lambda-\omega)\theta(\Lambda-\omega')\right]
\llbracket \widetilde{\mathcal{J}}_{c}(s_{L},\omega,\omega') \rrbracket
\left \llbracket S_{\rm NLP}(s_R,s_L,\omega,\omega')\right 
\rrbracket
\nonumber\\
&&\hspace*{-2cm}+\;\,\widetilde{\!\!\widehat{\mathcal{J}}}_{\!c}(s_{L},
\omega,\omega') 
\,\,\widetilde{\!\widehat{S}}_{\rm NLP}(s_R,s_L,\omega,\omega')\,\bigg\}\,. \quad
\label{eq:Atype_subtracted_not_expanded}
\end{eqnarray}
If we further assume that $\Lambda \gg 1/s_R, 1/s_L$, then we can keep only the leading terms in $1/(s_R\Lambda)$ and $1/(s_L\Lambda)$ and the previous equation simplifies to
\begin{eqnarray}
\frac{1}{\sigma_0}\frac{\widetilde{d\sigma}}{ds_R ds_L}|_{\rm A-type} &=& 
\frac{2C_F}{Q} \,f(\eps)\,|C^{\rm A0}(Q^2)|^2 \,
\widetilde{\mathcal{J}}_{\bar c}^{(\bar q)}(s_R)
\,\int d\omega  d\omega'\,
\nonumber\\
&&\hspace*{-3cm}\times\,\bigg\{\,
\theta(\Lambda-\omega)\theta(\Lambda-\omega') \Big[\widetilde{\mathcal{J}}_{c}(s_{L},\omega,\omega')
\,\widetilde{S}_{\rm NLP}(s_R,s_L,\omega,\omega')
\nonumber\\
&&\hspace*{-2.3cm}+\;\,\widetilde{\!\!\widehat{\mathcal{J}}}_{\!c}(s_{L},
\omega,\omega') 
\,\,\widetilde{\!\widehat{S}}_{\rm NLP}(s_R,s_L,\omega,\omega')\,\Big]
\\
&&\hspace*{-2.7cm}+\,
\big[\theta(\omega- \Lambda)\theta(\Lambda-\omega')+\theta(\omega'- \Lambda)\theta(\Lambda-\omega)\big]
\;\Big[\,\widetilde{\mathcal{J}}_{c}(s_{L},\omega,\omega')
\,\widetilde{S}_{\rm NLP}(s_R,s_L,\omega,\omega')
\nonumber\\
&&\hspace*{-2.3cm}\,- 
\llbracket \widetilde{\mathcal{J}}_{c}(s_{L},\omega,\omega') \rrbracket
\left \llbracket S_{\rm NLP}(s_R,s_L,\omega,\omega')\right 
\rrbracket +\;\,\widetilde{\!\!\widehat{\mathcal{J}}}_{\!c}(s_{L},
\omega,\omega') 
\,\,\widetilde{\!\widehat{S}}_{\rm NLP}(s_R,s_L,\omega,\omega')\Big]\,\bigg\}\,. \quad
\nonumber
\label{eq:Atype_subtracted_not_expanded2}
\end{eqnarray}

The remaining part $I_1$ of the scaleless integral \eqref{eq:scaleless_integral} must be combined withe the B-type term. To make notation more concise, we introduce $r_\Lambda \equiv \Lambda/Q$. The B-type term for $i=i'=1$ is 
\begin{eqnarray}
\frac{1}{\sigma_0}\frac{\widetilde{d\sigma}}{ds_R ds_L}|_{ \begin{subarray}{c}\text{B-type}\\
 \text{ i=i'=1}\end{subarray}} 
&=& 
\frac{2C_F}{Q^2}\,f(\eps) \,\mathcal{J}_{c}^{(g)}(s_L)\,S^{(g)}(s_R,s_L)\, \int_0^{\infty} dr dr' \nonumber\\ &&\hspace*{-3cm}\times\,
 \bigg\{\,
\theta(1-r)\theta(1-r')C^{\rm B1*}_{1}(Q^2,r')C^{\rm B1}_{1}(Q^2,r) 
\,\mathcal{J}_{\bar{c}}^{q\bar{q}(8)}(s_R,r,r^\prime)
\quad 
\nonumber \\ &&\hspace*{-2.3cm}-
\left[\theta(r_\Lambda-r)\theta(r_\Lambda-r')\right]
\llbracket C^{\rm B1*}_{1} (Q^2,r')\rrbracket_{0} \llbracket C^{\rm B1}_{1}(Q^2,r) \rrbracket_{0} \, \llbracket \mathcal{J}_{\bar{c}}^{q\bar{q}(8)}(s_R,r,r^\prime)\rrbracket
\,\bigg\}\,.\qquad 
\label{eq:Btype_subtracted_11_expanded}
\end{eqnarray}
If we choose $r_\Lambda \ll 1$, then the B-type term simplifies to
\begin{eqnarray}
\frac{1}{\sigma_0}\frac{\widetilde{d\sigma}}{ds_R ds_L}|_{ \begin{subarray}{c}\text{B-type}\\
 \text{ i=i'=1}\end{subarray}} &=& 
\frac{2C_F}{Q^2} \,f(\eps)\,\mathcal{J}_{c}^{(g)}(s_L)\,S^{(g)}(s_R,s_L)\, \int_0^1 \!dr \int_0^1\!dr' 
\left[1-\theta(r_\Lambda-r)\theta(r_\Lambda-r')\right]
\nonumber\\ &\times&\,
 \bigg\{\,
C^{\rm B1*}_{1}(Q^2,r')C^{\rm B1}_{1}(Q^2,r) 
\,\mathcal{J}_{\bar{c}}^{q\bar{q}(8)}(s_R,r,r^\prime)
\nonumber\\
&&\hspace*{0.5cm}-\,
\llbracket C^{\rm B1*}_{1} (Q^2,r')\rrbracket_{0} \llbracket C^{\rm B1}_{1}(Q^2,r) \rrbracket_{0} \, \llbracket \mathcal{J}_{\bar{c}}^{q\bar{q}(8)}(s_R,r,r^\prime)\rrbracket\,
\bigg\}
\label{eq:Btype_subtracted_11_expanded2}
\end{eqnarray}
up to corrections of 
$\mathcal{O}(\Lambda/Q)$. Just as for version (2) given in the main text, we can obtain the $i=i'=2$ case by $C_1^{\rm B1(*)}\to C_2^{\rm B1(*)} $, $\llbracket \ldots \rrbracket_0\to \llbracket \ldots \rrbracket_1$ and $r^{(')} \to \bar{r}^{(')}$.

\end{appendix}

\bibliography{NLPthrust}

\end{document}